\documentclass[aps,prd,floatfix,showpacs,10pt,nofootinbib]{revtex4-1}

\usepackage{amsmath,amssymb,amsfonts,amsthm} %usepackage{amsthm,amsfonts,amssymb,amsmath,amsbsy}
\usepackage{color}
\usepackage{float}
\usepackage{graphicx}
\usepackage{pstricks,pstricks-add}
\usepackage{pst-plot}
\usepackage[scriptsize,nooneline,hang]{caption}
\usepackage[hang,nooneline,scriptsize]{subfigure}
\definecolor{dark-green}{rgb}{0,0.7,0}
\definecolor{dark-blue}{rgb}{0,0.2,0.5}
\definecolor{med-blue}{rgb}{0,0.7,1}
\definecolor{mblue}{rgb}{0,0.2,1}
\definecolor{cnc}{rgb}{0.8,0,0}
\definecolor{light-red}{rgb}{1,0.8,0.8}
\definecolor{dark-yellow}{rgb}{1,0.8,0}
\definecolor{light-blue}{rgb}{0.8,0.9,1}
\definecolor{grey}{rgb}{0.211,0.211,0.211}
\definecolor{verylight-blue}{rgb}{0.93,0.95,1}
\definecolor{light-yellow}{rgb}{1,0.9,0.8}
\newcommand{\weglassen}[1]{}

\begin{document}

\title{Charged $Q$-balls and boson stars and dynamics of charged test particles}

\author{Yves Brihaye $^{(1)}$ }
\email{yves.brihaye@umons.ac.be}

\author{Valeria Diemer \footnote{n\'ee Kagramanova} $^{(2)}$}
\email{valeriya.diemer@uni-oldenburg.de}

\author{Betti Hartmann $^{(3), (4)}$ }
\email{b.hartmann@jacobs-university.de}
%\affiliation{School of Engineering and Science, Jacobs University Bremen, 28759 Bremen, Germany}

\affiliation{
$(1)$ Facult\'e de Sciences, Universit\'e de Mons, 7000 Mons, Belgium\\
$(2)$ Institut f\"ur Physik, Universit\"at Oldenburg, 26111 Oldenburg, Germany\\
$(3)$ School of Engineering and Science, Jacobs University Bremen, 28759 Bremen, Germany\\
$(4)$ Universidade Federal do Esp\'irito Santo (UFES), 
Departamento de F\'isica, Vit\'oria (ES), Brazil}
\date{\today}

\date\today

\begin{abstract}
We construct electrically charged $Q$-balls and boson stars in a model with a scalar self-interaction
potential resulting from gauge mediated supersymmetry breaking. We discuss the properties of
these solutions in detail and emphasize the differences to the uncharged case. We observe that $Q$-balls can only be constructed
up to a maximal value of the charge of the scalar field, while for boson stars the interplay between the attractive
gravitational force and the repulsive electromagnetic force determines their behaviour. We find that the vacuum
is stable with respect to pair production in the presence of our charged boson stars.
We also study the motion of charged, massive test particles in the space-time of boson stars. We find that
in contrast to charged black holes the motion of charged test particles in charged boson star space-times
is planar, but that the presence of the scalar field plays a crucial r\^ole for the qualitative features of the
trajectories. Applications of this test particle motion can be made in the study of extreme-mass 
ratio inspirals (EMRIs) as well as astrophysical plasmas relevant 
e.g. in the formation of 
accretion discs and polar jets of compact objects. 
\end{abstract}

\maketitle

\section{Introduction}
Non-topological solitons in contrast to topological solitons \cite{ms} are not stable due to
a topological charge, but due to a globally conserved Noether charge that results from a continuous symmetry 
existing in the
system \cite{fls,lp}. The non-topological soliton most frequently discussed is the $Q$-ball \cite{coleman} and its gravitating counterpart, the boson
star \cite{kaup,misch,flp,jetzler,new1,new2,Liddle:1993ha}. These are solutions appearing in complex scalar
field models that possess a continuous U(1) symmetry. Due to the recent confirmation of the existence of 
a fundamental scalar field in nature \cite{cern_lhc} these models seem to be very appealing. 
When the U(1) symmetry is global, the Noether charge
corresponds to the particle number. When the U(1) symmetry is gauged the product between the Noether charge
and the constant defining the coupling between the gauge field and the scalar field can be interpreted as
total charge of the soliton. 

As is well known, soliton solutions exist in models with a subtle interplay between non-linearity and dispersion
and as such specific potentials of the scalar field are necessary to obtain $Q$-ball solutions. It was shown in \cite{coleman}
that a potential of at least 6th order in the scalar field is necessary to obtain solutions and these
solutions have been constructed in \cite{vw,kk1,kk2,bh}. Note that when
gravity is added to the system this restriction is not present, however, using the same scalar field potential
in the gravitating case leads to a proper flat space-time limit of the solutions \cite{kk1,kk2,bh2}.
Unfortunately, 6th order potentials
are non-renormalizable and hence the question arises whether appropriate potentials can be derived from
some particle physics models. Indeed, it is well known that in supersymmetric extensions of the
Standard Model $Q$-ball solutions exist \cite{kusenko} and these have been discussed with respect to their astrophysical
applications and possible r\^ole in dark matter models \cite{dm, implications}. In \cite{cr,ct} a potential
arising in gauge-mediated supersymmetry breaking has been discussed and the corresponding $Q$-ball
solutions have been constructed. The gravitating counterparts were discussed in \cite{hartmann_riedel2,hartmann_riedel}.

Charged $Q$-balls and boson stars have also been studied previously \cite{jetzler2,Arodz:2008nm, Kleihaus:2009kr,Pugliese:2013gsa,Tamaki:2014nua}. 
In particular, a detailed study in models with a V-shaped potential (in which compact objects 
arise \cite{Arodz:2008jk,Arodz:2008nm}) has been presented in \cite{Kleihaus:2009kr}.
In these latter mentioned models with a V-shaped potential boson stars with a well-defined outer
surface can be considered as toy models for or even alternatives to compact objects such as neutron stars
or even supermassive black holes \cite{schunck_liddle} residing
e.g. in the center of galaxies. Hence, the applications of boson stars is two-fold: 
\begin{itemize}
 \item They can be considered
as toy models for neutron stars. Neutron stars are very difficult to model due to a lacking general
equation of stay governing the matter inside the star.  Boson stars of stellar size masses might be interesting
in this respect. And then, since very compact objects accrete matter around them, it is interesting to
study the accretion of matter to boson stars as well. This can e.g. be done by studying the geodesic motion of test 
particles in the space-time of boson stars -- in analogy to the studies done in black hole space-times
\cite{tejeda1,tejeda2}. In fact, the assumption made in this so-called {\it ballistic limit} is that
of non-interacting, free test particles. The dynamics of these type of particles in uncharged boson star space-times 
has been discussed in \cite{Eilers:2013lla}. However, since in astrophysical settings we would expect the accreted matter to consist of
a plasma, i.e. charged constituents, one of the aims of this paper is to study the motion of charged particles
in boson star space-times. 
\item While objects with a  ``hard core" as alternatives to supermassive black holes in the center of our galaxy
have been ruled out \cite{Broderick:2005xa}, the objects constructed in this paper do not have a well-defined
surface outside which the energy density is strictly vanishing. To our knowledge no observation yet
clearly rules out a non-compact horizonless object as a possible center of our galaxy. In fact,
recent observations of radio waves with wavelength of $1.3$ mm emitted from the galactic center suggest
that the size of this object ({\it Sagittarius A$^*$}) is smaller than the expected apparent event horizon 
size of the assumed black hole \cite{Doeleman:2008qh}.
The question that we try to address then is how a possible charge of such an object could influence rotation curves of stars
around the galactic center and whether any signal of the charge of either the object or the accreted matter
would show up in the radio wave signals. 
\end{itemize}

Our paper is organised as follows: in Section II, we discuss the field theoretical model of electrically charged
$Q$-balls and boson stars. In Section III we present numerical solutions to the coupled system of ordinary, non-linear differential
equations. In particular, we will put emphasize on the r\^ole of the charge on the stability and existence of
the solutions. In Section IV, we discuss the motion of charged, massive test particles in the space-time
of electrically charged boson stars and also comment on possible astrophysical applications.

\section{Charged  $Q$-balls and boson stars}
In the following we will discuss the field theoretical model to describe the 
space-time of charged, non-spinning, non-compact $Q$-balls and boson stars in which the
test particles will move.

The action $\tilde{S}$ of the field theoretical model reads:
\begin{equation}
 \tilde{S}=\int \sqrt{-g} d^4 x \left( \frac{R}{16\pi G} + {\cal L}_{m}\right)
\end{equation}
where $R$ is the Ricci scalar, $G$ denotes Newton's constant and
the matter Lagrangian is given by
\begin{equation}
\label{lag}
 {\cal L}_{m}=-\frac{1}{4}F_{\mu\nu}F^{\mu\nu}-D_{\mu} \Phi D^{\mu} \Phi^*
 - U(\vert\Phi\vert)
\end{equation}
where $\Phi$ denotes a complex scalar field and we choose as signature of the metric
$(-+++)$. $F_{\mu\nu}=\partial_{\mu} A_{\nu} - \partial_{\nu} A_{\mu}$ is the field strength tensor
of a U(1) gauge field $A_{\mu}$ and $D_{\mu} \Phi = (\partial_{\mu} - i \tilde{e} A_{\mu})\Phi$ denotes the covariant
derivative with gauge coupling $\tilde{e}$. $U(\vert\Phi\vert)$ is the scalar field potential 
that arises in gauge-mediated supersymmetry breaking (GMSB) in Supersymmetric extensions of the
Standard Model \cite{cr,ct}:
\begin{equation}
\label{potential}
 U(\vert\Phi\vert)=m^2\eta_{\rm susy}^2 \left(1-\exp\left(-\frac{\vert\Phi\vert^2}{\eta_{\rm susy}^2}\right)\right) \ . 
\end{equation}
In GMSB the supersymmetry (SUSY) breaking happens in a hidden sector and the breaking is mediated via gauge fields
to the visible sector of the Standard Model \cite{Giudice:1998bp}. These fields are the so-called messenger fields with corresponding
messenger mass scale $\eta_{\rm susy}$. These are coupled to the gauge multiplets with coupling constant
$g\sim 10^{-2}$. The soft breaking mass scale is then given by $m$ and is on the order of TeV.
The lightest supersymmetric particle (LSP) predicted by GMSB is the gravitino with mass $m_{3/2}$. Hence, limits of the parameter
$\eta_{\rm susy}$ are typically given in terms of this mass $m_{3/2}$ as follows
\begin{equation}
 \frac{\rm 1 TeV}{\sqrt{m}} \leq \sqrt{\eta_{\rm susy}} \leq \frac{\sqrt{g}}{4\pi} \frac{\sqrt{m_{3/2} M^r_{\rm pl}}}{\sqrt{m}}   \ ,
\end{equation}
where $M^r_{\rm pl}=(8\pi G)^{-1/2}\sim 2.4 \cdot 10^{18} GeV$ is the reduced Planck mass. The recent results from the ATLAS and CMS detectors at LHC 
seem to indicate that the minimal GMSB is ruled out since it predicts a mass of the Brout-Englert-Higgs (BEH) boson
of less than 122 GeV which is clearly in disagreement with the confirmed value \cite{Arbey:2011ab}. However,
extended GMSB models still allow for the measured value of the BEH boson and new models have been discussed 
(see e.g. \cite{CahillRowley:2012gu,CahillRowley:2012cb}). Current bounds on $m_{3/2}$ depend on the next-to-lightest supersymmetric
particle (NLSP) and the decay channels. Mostly the gravitino is assumed to have masses above roughly 1 keV and not larger than
1 GeV (limits exist also from Cosmology and Astrophysics). The messenger scale $\eta_{\rm susy}$  could then be between roughly 1 TeV and
up to $10^8$ TeV.

The matter Lagrangian ${\cal L}_{m}$ (\ref{lag}) is invariant under a local U(1) transformation.
As such the locally conserved Noether
current $j^{\mu}$, $\mu=0,1,2,3$, associated to this symmetry is given by
\begin{equation}
j^{\mu}
 = -i \left(\Phi^* D^{\mu} \Phi - \Phi (D^{\mu} \Phi)^*\right) \  \ {\rm with} \ \ \
j^{\mu}_{; \mu}=0  \ .
\end{equation}
The globally conserved Noether charge $N$ of the system then reads
\begin{equation}
\label{charge}
 N= -\int \sqrt{-g} j^0 d^3 x  \  ,
\end{equation}
which corresponds to a particle number such that $Q\equiv \tilde{e}N$ is the total charge. Note that
since $N$ is positive the sign of $e$ will determine the sign of the charge of the $Q$-balls and boson
stars. 
Finally, the energy-momentum tensor is given by
\begin{eqnarray}
T_{\mu\nu} &=& g_{\mu\nu}{L}_M
-2 \frac{\partial {L}_M}{\partial g^{\mu\nu}}
=
    ( F_{\mu\alpha} F_{\nu\beta} g^{\alpha\beta}
   -\frac{1}{4} g_{\mu\nu} F_{\alpha\beta} F^{\alpha\beta})
\nonumber\\
&-& 
   \frac{1}{2} g_{\mu\nu} \left(     (D_\alpha \Phi)^* (D_\beta \Phi)
  + (D_\beta \Phi)^* (D_\alpha \Phi)    \right) g^{\alpha\beta}
  + (D_\mu \Phi)^* (D_\nu \Phi) + (D_\nu \Phi)^* (D_\mu \Phi) - 
    U(\Phi|)   \ .
  \label{tmunu}
\end{eqnarray}
The coupled system of ordinary differential equations is given by the Einstein
equations
\begin{equation}
\label{einstein}
 G_{\mu\nu}=8\pi G T_{\mu\nu}
\end{equation}
with $T_{\mu\nu}$ as in (\ref{tmunu}) and the matter field equations
\begin{eqnarray}
\partial_\mu \left ( \sqrt{-g} F^{\mu\nu} \right) =
   \sqrt{-g} e \Phi^* D^\nu \Phi  \ \ \ , \ \ \ D_\mu D^\mu \Phi - \frac{\partial U}{\partial \vert\Phi\vert^2}\Phi = 0
 . \label{feq_matter} \end{eqnarray}
In the following, we want to study non-spinning $Q$-balls and boson stars. 
For the metric we use the following Ansatz in isotropic coordinates
\begin{equation}
\label{metric}
 ds^2=-f(r) dt^2 + \frac{l(r)}{f(r)}\left[dr^2 + r^2 d\theta^2 + r^2 \sin^2\theta d\varphi^2\right]  \ .
\end{equation}
For the matter fields we choose
\begin{equation}
\label{ansatz1}
\Phi=\phi(r) e^{i\omega t}  \ , \ A_\mu d x^\mu = A(r) dt \ ,
\end{equation}
such that our solutions will carry only electric charge and $\omega$ denotes the internal frequency of the
scalar field. The concrete expression for the charge of the solution
(\ref{charge}) is then
\begin{equation}
 Q= 8\pi \tilde{e} \int\limits_{0}^{\infty} \frac{\sqrt{l^3}}{f^2} r^2 \left(\omega - \tilde{e} A\right)\phi^2 dr 
\end{equation}
The initially complex scalar field can be chosen to be real. The reason is the U(1) gauge symmetry
that allows us to gauge away any non-vanishing phase. We hence apply the following U(1) gauge transformation
\begin{equation}
 \Phi \rightarrow \Phi e^{-i\omega t} \ \ , \ \   A\rightarrow A + \frac{\omega}{\tilde{e}}  \ .
\end{equation}
In order to be able to use dimensionless quantities we introduce the following rescalings
\begin{equation}
\label{rescale}
 r\rightarrow \frac{r}{m} \ \ , \ \ \omega \rightarrow m\omega \ \ ,  \
 \  \phi\rightarrow \eta_{\rm susy} \phi \ \  , \ \ A\rightarrow \eta_{\rm susy} A 
\end{equation}
and find that the equations depend only on the dimensionless coupling constants  
\begin{equation}
 e = \frac{\eta_{\rm susy}}{m} \tilde{e} \ \ \ , \ \ \ \alpha=8\pi G\eta_{\rm susy}^2 = 8\pi \frac{\eta_{\rm susy}^2}{M_{\rm pl}^2}  \ ,
\end{equation}
where $M_{\rm pl}$ is the Planck mass. Note that with these rescalings the scalar boson mass $m$ 
becomes equal to unity. The charge of the scalar field will be given in multiples of $\eta_{\rm susy}/m$ which
can be between unity and $10^{8}$. $\alpha$ on the other hand is in general very small, but in this paper we will also consider
larger values of $\alpha$ in order to understand the qualitative behaviour of the boson star solutions.

We can also write
\begin{equation}
 \frac{\alpha}{e^2} = 8\pi \frac{m^2 G}{\tilde{e}^2} = 2 \frac{m^2 G}{\alpha_{\rm fs}} = 274 \cdot m^2 G
\end{equation}
where in our natural unit system ($\epsilon_0=\hbar=c\equiv 1$)  $\alpha_{\rm fs}$ denotes the fine structure
constant with $1/\alpha_{\rm fs}=137$. Since stable boson stars should only exist when the gravitational
attraction compensates the electromagnetic repulsion (see e.g. \cite{Jetzer:1993nk}) we would expect that they 
exist only for $m^2 G > \alpha_{\rm fs}$, i.e.
for $\alpha/e^2 \gtrsim 2$.

The Lagrangian gets rescaled like ${\cal L} \rightarrow {\cal L}/(m^2 \eta_{\rm susy}^2)$ such that
the masses we will give in our numerical analysis below are measured in units of $\eta_{\rm susy}^2/m$ which is 
(with the given values as above) between roughly 1 TeV and $10^{16}$ TeV. The upper limit here corresponds to the Planck mass which - of course -
we would not be able to reach for boson stars. However, these are just rough approximations and we note that boson stars
in supersymmetric extensions of the Standard Model can be quite heavy.

Using these rescalings the field equations of motion read
\begin{equation}
\label{phi_eq}
  \phi''=- \frac{r l' + 4l}{2rl} \phi'  -(\omega-e A)^2 \frac{l}{f^2} \phi + \frac{l}{f}\phi \exp(-\phi^2)   \ ,
\end{equation}
\begin{equation}
\label{a_eq}
 A''=-\left(\frac{f'}{f} + \frac{l'}{2l} + \frac{2}{r}\right)A' + \frac{2l}{f} e(eA-\omega) \phi^2
\end{equation}
for the matter field functions and
\begin{equation}
 f'' = f'\left(-\frac{l'}{2l} + \frac{f'}{f} - \frac{2}{r}\right) + 2\alpha l 
\left(\frac{2(\omega-eA)^2 \phi^2}{f} - 1 + \exp(-\phi^2)\right)   \ ,
\end{equation}
\begin{equation}
\label{l_eq}
 l''=l'\left(\frac{l'}{2l} - \frac{3}{r}\right) + 4\alpha l^2 \left(\frac{(\omega-eA)^2 \phi^2}{f^2} - \frac{1-\exp(-\phi^2)}{f}\right)
\end{equation}
for the metric functions. The prime now and in the following denotes the derivative with respect to $r$. 
Note that the equations are unchanged under $(e,A) \leftrightarrow (-e,-A)$.
The coupled set of non-linear ordinary differential equations (\ref{phi_eq}) - (\ref{l_eq}) has to be solved subject to appropriate boundary conditions. 
These are given by the requirement
of regularity at the origin
\begin{equation}
\label{bc0}
 f'\vert_{r=0} = 0 \ \ , \ \ l'\vert_{r=0} = 0 \ \ , \ \ \phi'\vert_{r=0} = 0 \ \ , \ \ A'(0)=0 
\end{equation}
and by the requirement of finite energy, asymptotically flat solutions
\begin{equation}
\label{bcinf}
 f(r=\infty)=1 \ \ , \ \ l(r=\infty)=1 \ \ , \ \ \phi(r=\infty)=0 \ \ , \ \ A(r=\infty)=-\frac{\omega}{e} \ .
\end{equation}
For numerical calculations, we have found it more convenient, however, to adapt the boundary conditions.
As can be easily verified, the equations depend only on the combination $(\omega-eA)$. We take advantage of this
and impose $A(0)=0$ instead of $A(r=\infty)=-\frac{\omega}{e}$. The asymptotic behaviour of the matter fields
reads 
\begin{equation}
 A(r \rightarrow \infty) \sim \mu + \frac{Q}{r} \ \ , \ \  \phi(r\rightarrow\infty) \sim 
\frac{1}{r}\exp\left(- \sqrt{1-(\omega-e\mu)^2} r\right)  \ ,
\end{equation}
where $\mu$ is the value of $A(r)$ at $r\rightarrow \infty$, which
will be computed numerically by the program. Furthermore, we impose an additional condition on
$\phi(r)$ in order to avoid the trivial solution $\phi(r)\equiv 0$. For that we add an equation
$\omega''=0$, hence making $\omega$ a function. By imposing the additional boundary conditions
$\phi(0)=\phi_0$ and $\omega'\vert_{r=0}=0$ we find a consistent solution with $\omega$ constant. 

The set of coupled, nonlinear ordinary differential equations (\ref{einstein})-(\ref{feq_matter}) has been
studied with a V-shaped potential in \cite{Kleihaus:2009kr}, however in Schwarzschild-like coordinates. 
In \cite{hartmann_riedel,Eilers:2013lla} uncharged boson stars with an exponential potential of the form (\ref{potential})
have been considered. 

\section{Soliton solutions}
We have solved the set of differential equations numerically using the ODE solver Colsys \cite{colsys}.
The relevant parameters are $\alpha$ and the electric coupling constant $e$.
Once choosing these constants, families of solutions labeled by $\phi(0)$ (and hence $\omega$) can be constructed.

\subsection{Charged $Q$-balls}
\begin{figure*}[h!]
\begin{center}
\subfigure[][$M$ and $N$ as function of $\omega$]{\label{M_N_om}\includegraphics[width=7.2cm]{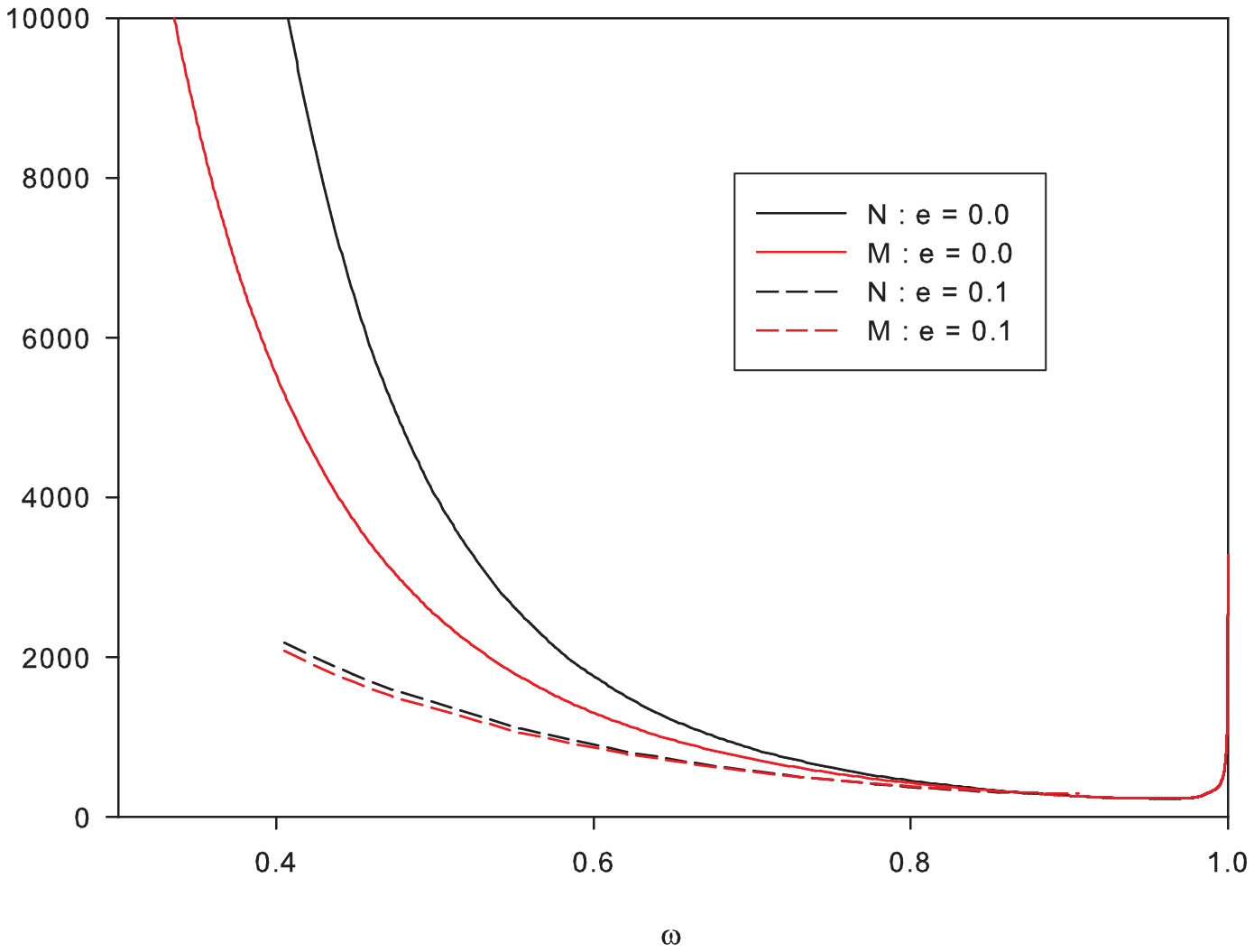}}
\subfigure[][$M$ as function of $N$]{\label{M_N}\includegraphics[width=8cm]{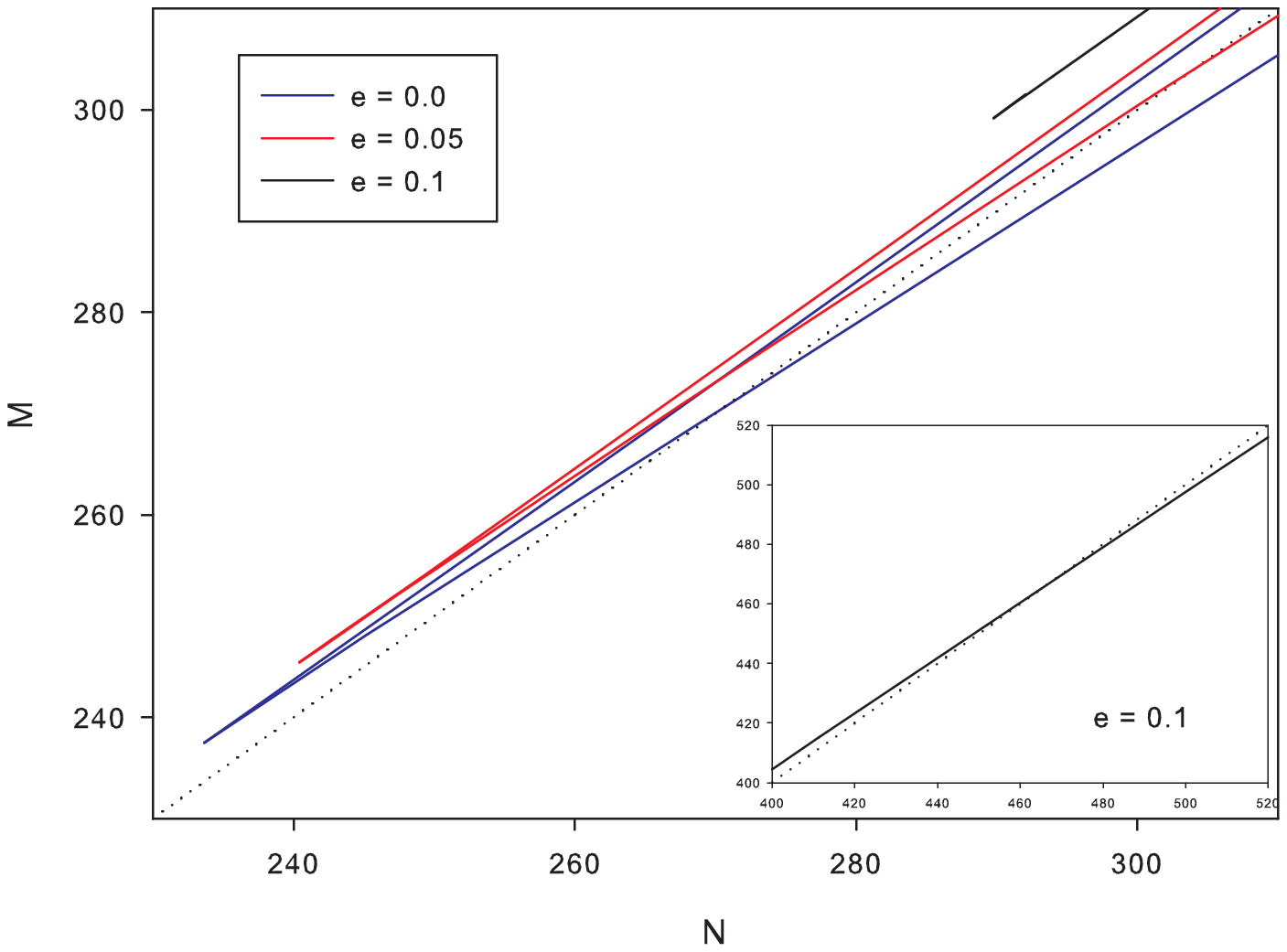}}
\end{center}
\caption{
(Left:) The mass $M$ (red) and particle number $N$ (black) of uncharged $Q$-balls (solid, $e=0.0$) and charged
$Q$-balls (dashed, $e = 0.1$), respectively, as function of $\omega$. (Right:) The mass $M$ as function of the
particle number $N$ for uncharged $Q$-balls (blue, $e=0.0$) and two differently charged $Q$-balls
(red for $e=0.05$, black for $e=0.1$). The insert shows the region around the point where
$M=N$ for $Q$-balls with $e=0.1$.\label{   }}
\end{figure*}

This corresponds to the case $\alpha=0$, i.e. to solitons in flat space-time. In this case
we have $f=l\equiv 1$. These solutions are called ``$Q$-balls''
in the literature and we hence refer to them as such here. 
Let us first discuss the restrictions on the parameter $\omega$ that we have. From \cite{vw} we know that in the uncharged
case, i.e. for $e=0$ the frequency $\omega$ is bounded from above and below, i.e. $\omega_{\rm min} \le \omega \le \omega_{\rm max}$. 
To see how this constraint changes in the presence of an electromagnetic field note that we can rewrite (\ref{phi_eq}) 
in analogy to the
uncharged case as follows
\begin{equation}
 \frac{1}{2} \phi'^2 + \frac{1}{2}\left(\omega-eA\right)^2 \phi^2 -\frac{1}{2}U(\phi) = 
E - 2\int\limits_0^r \frac{\phi'^2}{r} dr  \ ,
\end{equation}
where $E$ is an integration constant. 
This describes the frictional motion of a particle with ``coordinate'' $\phi$ at ``time'' $r$ in an effective 
potential of the form
\begin{equation}
 V(\phi)= \frac{1}{2}\left(\omega-eA\right)^2 \phi^2 -\frac{1}{2}U(\phi)  \ .
\end{equation}
The requirements then are that $V''(\phi=0) < 0$ and $V(\phi) > 0$ for some $\phi\neq 0$ \cite{vw}.
In our case with the potential (\ref{potential}) this leads to the following conditions
\begin{equation}
\label{bounds}
(e \mu-\omega)^2 < 1 \ \ , \ \ (e\mu-\omega)^2 > 0 \ ,
\end{equation}
where the latter condition is not really an extra condition since all quantities are assumed to be real. 
Keeping these conditions in mind we have studied the case of $e$ fixed and varying $\phi(0)$ in order
to understand how the dependence of $\omega$ on $\phi(0)$ changes in the presence of an electromagnetic field. 
In the case $e=0$, i.e. for uncharged $Q$-balls, the limit $\omega\rightarrow \omega_{\rm max}$ corresponds
to $\phi(0)\rightarrow 0$. This is the so-called {\it thick wall limit}. On the other hand, $\omega\rightarrow \omega_{\rm min}$
corresponds to $\phi(0)\rightarrow \infty$, the {\it thin wall limit}. In both cases, the mass $M$ and the charge $Q$ 
of the solutions diverges. This is shown in Fig.\ref{M_N_om}, where we give the mass $M$
and particle number $N$ of the $Q$-balls in dependence on $\omega$ for $e=0$ and for $e=0.1$. 
Let us first discuss the case $e=0$ again in detail:
in the limit $\phi(0)\to 0$ (which corresponds to $\omega \to 1$)
the $Q$-balls have $M > N$ and are hence unstable to decay into $N$ individual quanta of mass $m\equiv 1$. 
Increasing the parameter $\phi(0)$
progressively we can construct a family of stable solutions with $M< N$ for $\phi(0)>1.75$
(this corresponds to $\omega < 0.88$). This is illustrated in Fig. \ref{M_N}.
Increasing $\phi(0)$ further the frequency $\omega$ approaches zero, the mass and particle number
of the $Q$-ball both tend to infinity.

\begin{figure*}[h!]
\begin{center}
\subfigure[][$\Omega$ and $\omega$ as function of $\phi(0)$]{\label{om_phi0}\includegraphics[width=7.8cm]{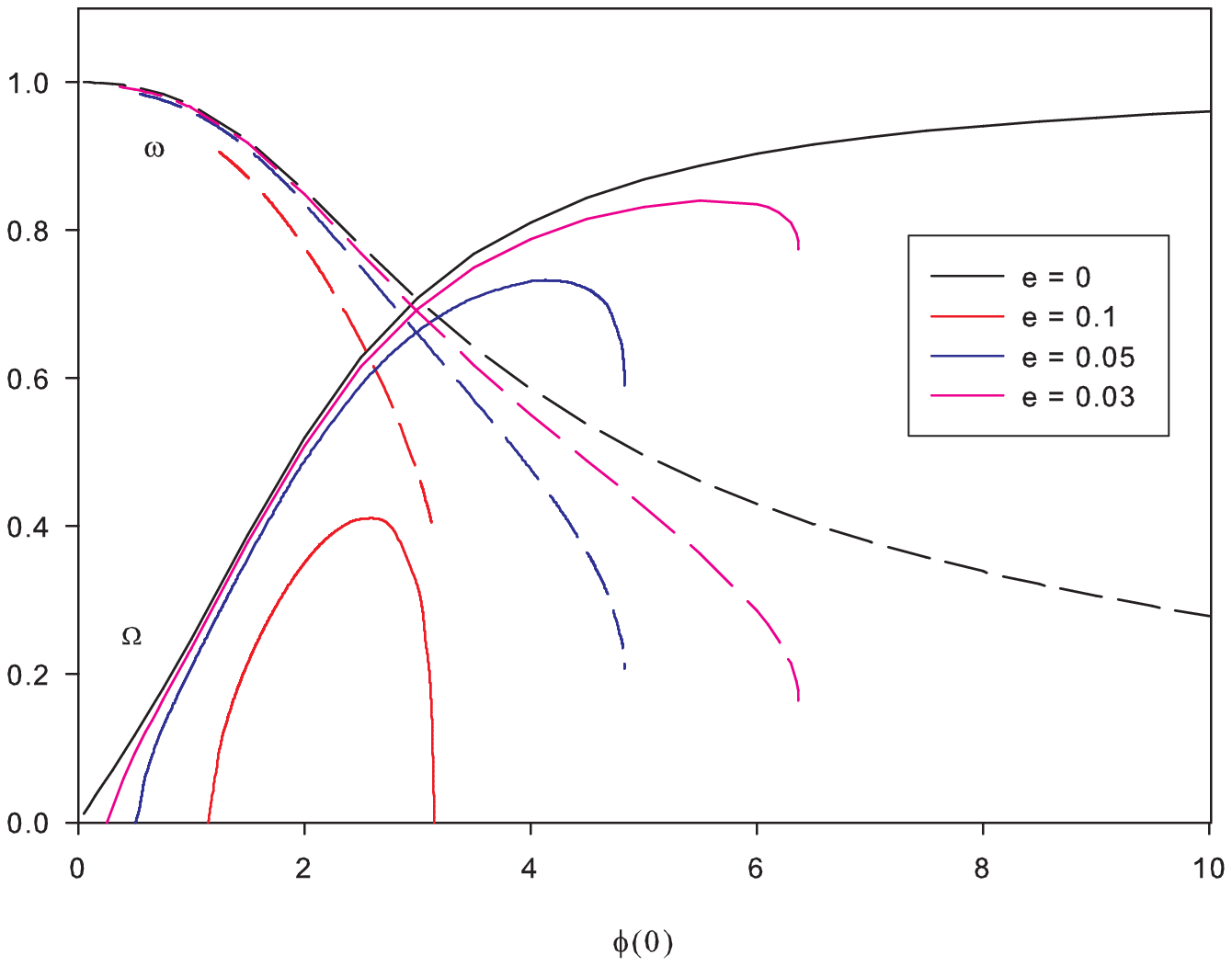}}
\subfigure[][$Q$ as function of $\phi(0)$]{\label{Q_phi0}\includegraphics[width=8cm]{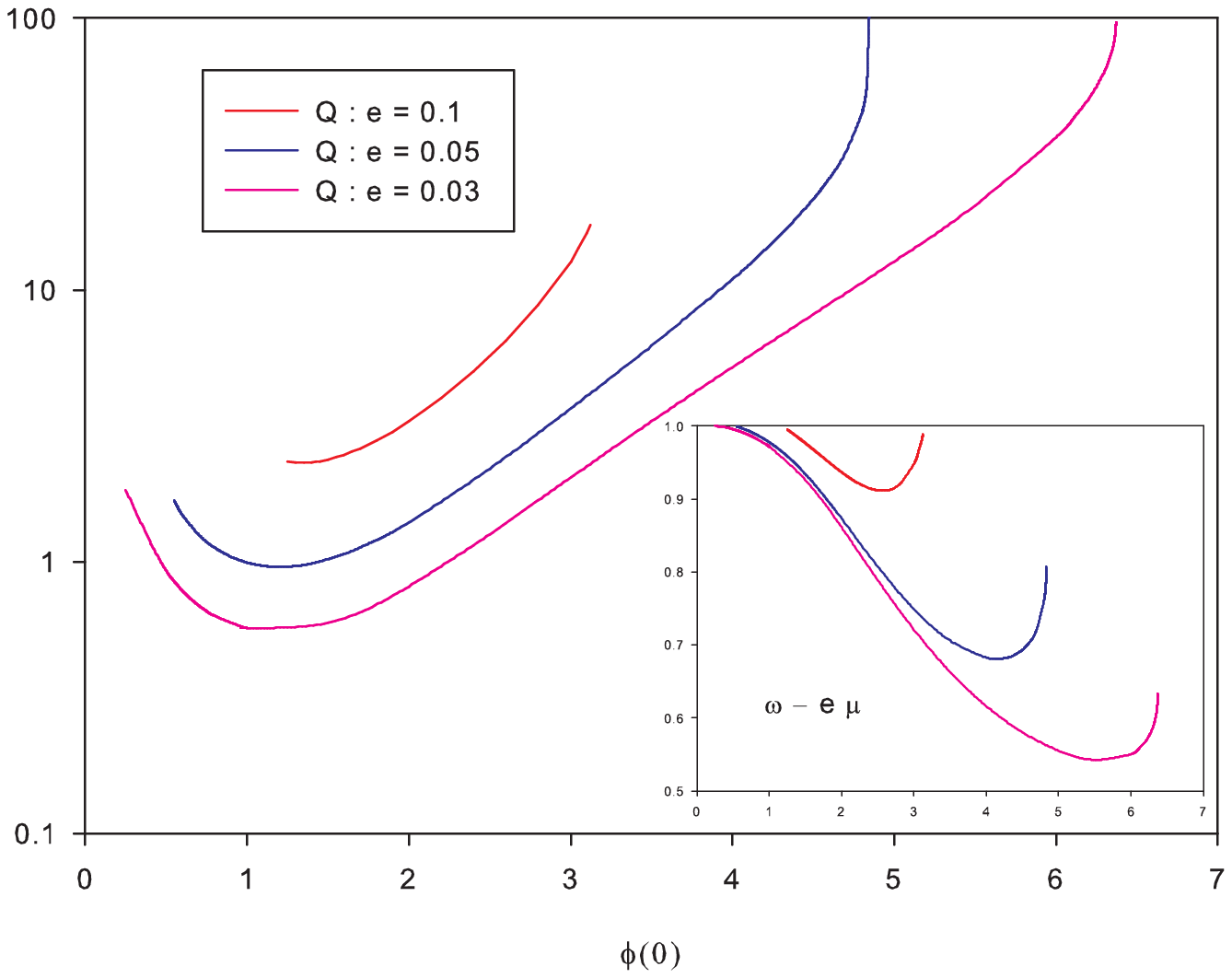}}
\end{center}
\caption{(Left:) We give the value of $\Omega:=\sqrt{1-(e\mu -\omega)^2}$ (solid) and $\omega$ (dashed) 
as functions of $\phi(0)$ for uncharged
and charged $Q$-balls. From top to bottom the curves correspond to $e=0.0$, $e=0.03$, $e=0.05$ 
and $e=0.1$, respectively. (Right:) We give the corresponding values of the charge $Q$
as function of $\phi(0)$ for $e\neq 0$. From bottom to top the values of $e$ are: $e=0.03$, $e=0.05$ 
and $e=0.1$, respectively. The insert shows the value of $\omega - e\mu$ as function of $\phi(0)$.  \label{charged_critical}}
\end{figure*}

For $e\neq 0$, we now observe that a different phenomenon exists such that the solutions do not exist on
the full interval $\phi(0)\in [0:\infty[$, but that $Q$-balls for $e > 0$ exist only on a finite
interval of $\phi(0)\in [\phi(0)_{\rm min}:\phi(0)_{\rm max}]$, where both $\phi(0)_{\rm min}$ and 
$\phi(0)_{\rm max}$ are finite and depend on $e$. Hence, charged $Q$-balls have neither a {\it thin wall limit}
nor a {\it thick wall limit} considering the scalar field only.   

This is shown in Fig.\ref{charged_critical}, where we give 
$\omega$ and the quantity $\Omega:=\sqrt{1-(e\mu-\omega)^2}$ in dependence on $\phi(0)$ (left)
as well as $Q$ and $\omega -e\mu$ as function of $\phi(0)$ (right). At both $\phi(0)=\phi(0)_{\rm min}$  and
$\phi(0)=\phi(0)_{\rm max}$ we find that
the solutions cease to exist because $\Omega \rightarrow 0$ and at the same 
time the particle number $N$ (and with it $Q$)
tend to infinity. This is related to the argument with the potential given above (see (\ref{bounds})), 
but here the lower bound on $(e\mu-\omega)$ is not really a constraint. In fact, the solutions cease to exist
because the quantity $\omega - e\mu$ tends to unity at both $\phi(0)_{\rm min}$ and $\phi(0)_{\rm max}$ (see the
subfigure of Fig.\ref{Q_phi0}). Our results seem to indicate that in the limit $\phi(0)\rightarrow \phi(0)_{\rm min}$ the electric field $\propto
\partial_r A$ tends to zero, while in the limit $\phi(0)\rightarrow \phi(0)_{\rm max}$ the electric field
spreads over all space. The physical interpretation is the following: remembering that the integral of the
scalar field over $r\in [0:\infty[$ corresponds to the particle number $N$, our results suggest that
a minimal number of $N$ needs to be present to allow for charged $Q$-balls. 
Furthermore, if $N$ becomes too large (at $\phi(0)_{\rm max}$) the electric repulsion among the individual
particles becomes too strong and the solution spreads over all space.

\begin{figure}[h]
\begin{center}
{\includegraphics[width=8cm]{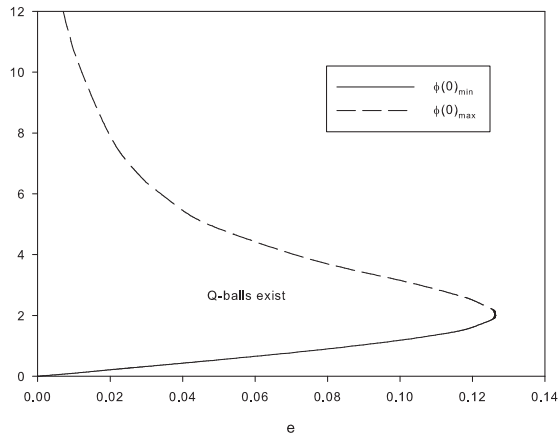}}
\end{center}
\caption{We give the minimal and maximal value of $\phi(0)$ of $Q$-balls in dependence on the charge $e$. 
Charged $Q$-balls exist only in the region in between these curves. 
In particular, for $e \gtrsim 0.1262$ no charged $Q$-balls exist at all. 
\label{fig_domain}
}
\end{figure}

We find that the minimal value of $\phi(0)$
increases with $e$, while the maximal value of $\phi(0)$ decreases with $e$.
This is shown in Fig.\ref{fig_domain}, where we give $\phi(0)_{\rm min}$ and $\phi(0)_{\rm max}$ in
dependence on $e$. $\phi(0)_{\rm min}$ increases from zero at $e=0$, while $\phi(0)_{\rm max}$ decreases
from infinity. At a sufficiently large value of 
$e \approx 0.1262$ we find that $\phi(0)_{\rm min}=\phi(0)_{\rm max}$. This means that for $e \gtrsim 0.1262$ {\bf no}
$Q$-balls can be constructed. This is related to the electric repulsion that becomes too strong to support
stable configurations.

\subsection{Charged boson stars}
\begin{figure}[h]
\begin{center}
{\includegraphics[width=8cm]{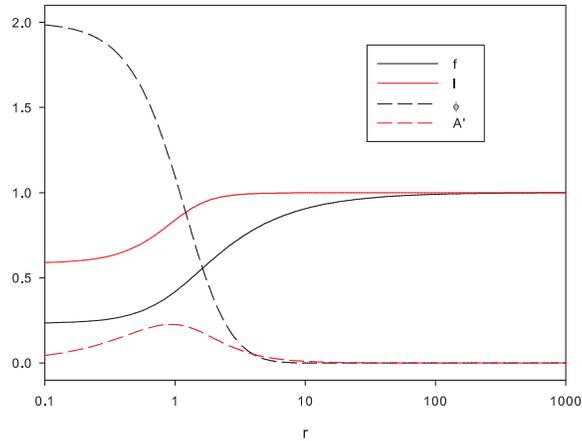}}
\end{center}
\caption{The metric functions $f$ and $l$, the scalar field function $\phi$ and the derivative of the gauge
field function $A'$, which is equivalent the radial electric field, are shown for a typical charged boson star solution.
\label{fig0}
}
\end{figure}

Here, the matter fields are coupled to the metric field, i.e. $\alpha\neq 0$. 
In Fig.\ref{fig0} we show the profiles of the functions of a typical charged boson star solution.

\begin{figure}[h]
\begin{center}
{\includegraphics[width=12cm]{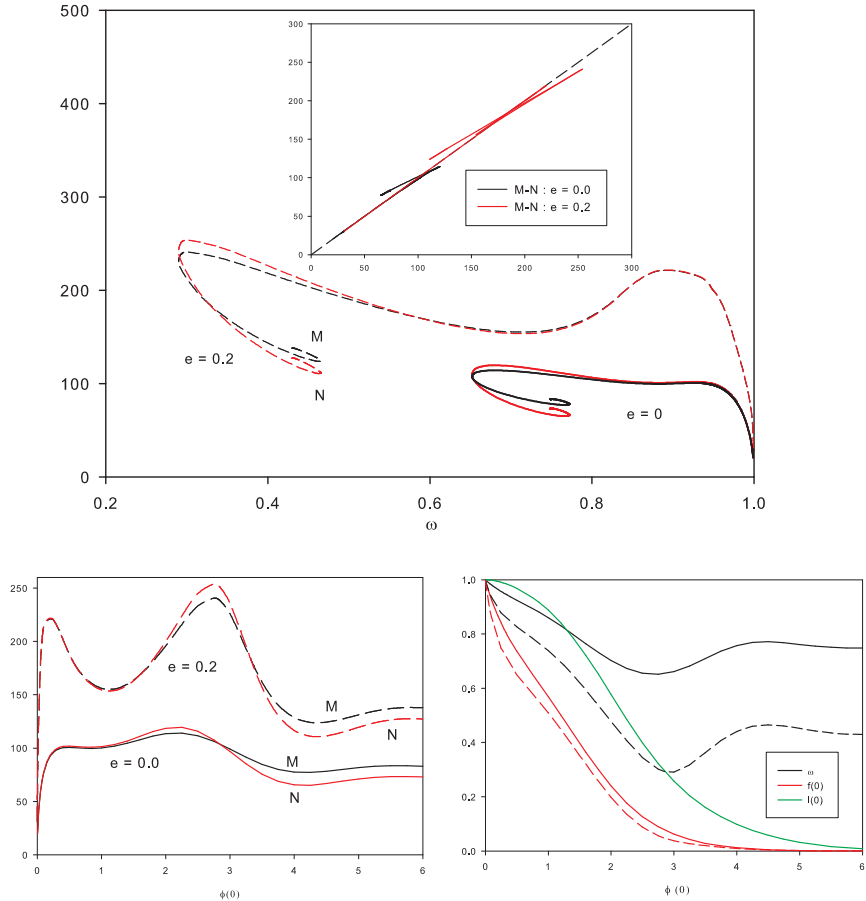}}
\end{center}
\caption{We show the mass $M$ and the particle number $N$ of boson stars in dependence on $\omega$ for $e=0$ and
$e=0.2$, respectively (upper figure). We also present $M$ and $N$ as function of $\phi(0)$ (lower left) 
as well as $\omega$, $f(0)$ and $l(0)$ as function of $\phi(0)$ (lower right). 
Here $\alpha=0.1$.}
\label{fig_uncharged_charged}
\end{figure}

First, we have fixed $\alpha=0.1$ and $e$ and varied $\phi(0)$ in order to understand the dependence
of the physical quantities on $\omega$.  
In the limit $e=0$, the solutions coincide with the uncharged boson stars constructed in \cite{hartmann_riedel}.
We were able to construct a branch of 
charged boson stars for $e >0$. The numerical results suggest that 
the branch is not limited by any critical value of $\phi(0)$ as in the $Q$-ball case.
In the limit $\phi(0)\to 0$, i.e. the 
{\it thick wall limit} the gauge field function $A(r)$ becomes a constant function, where
the constant depends on the value of $e$.  
This is shown in Fig.\ref{fig_uncharged_charged} (upper figure), where we give the mass $M$ and the charge $Q$ 
as function of $\omega$ for $e=0$ and $e=0.2$, respectively. Both the uncharged and the charged boson stars show a typical 
spiraling behaviour
at the minimal value of $\omega$, while $M$ and $Q$ both tend to zero at the maximal value of $\omega$. 
We find that the maximal value of $\omega$ is independent of $e$ (like it is independent of $\alpha$). However,
the minimal value of $\omega$ decreases for increasing $e$. This can be understood when noticing that the gravitational
field acts as an attractive force, while the electric field is repulsive in nature. Hence the electric repulsion
counterbalances gravity. As such boson stars with larger masses and particle numbers are possible as compared to
the uncharged case. The mass of the boson stars in dependence on the particle number 
shows a spike-like behaviour typical for boson star solutions. On the main branch of solutions boson stars
are stable ($M < N$), while on the second branch (corresponding to the spirals in the $\omega$-$M$-plot) 
the solutions develop an instability to decay into $N$ individual scalar boson (since $M > N$).  

We also observe that the mass and charge possess several local minima and maxima in dependence on $\phi(0)$. This is
shown in Fig.\ref{fig_uncharged_charged} (bottom left). Now, the question is what happens at the minimal value
of $\omega$ in this case. This is shown in Fig.\ref{fig_uncharged_charged} (bottom right), where we give
$\omega$, $f(0)$ and $l(0)$ as function of $\phi(0)$. Clearly, $\omega$ tends to a finite value for $\phi(0)\rightarrow \infty$. 
In contrast
to the case of $Q$-balls, however, this {\it thin wall limit} is not regular since at the same time $f(0)$ and $l(0)$ tend
both to zero.

Fixing $e$ and increasing $\phi(0)$, our results hence indicate that charged boson stars
exist up to very large values of $\phi(0)$. This contrasts with the case of charged $Q$-balls.
This behaviour can be easily understood when realizing that increasing the central
density increases the gravitational attraction which compensates the electrostatic repulsion.
In particular, the values of the metric functions at $r=0$, i.e. $f(0)$ and $l(0)$ tend to zero
for $\phi(0)$ becoming large. At the same time, however, the mass
and particle number remain finite and the parameter $\Omega$ does not approach zero.

Although we only give the data for $\alpha = 0.1$ and $e=0.2$ here, we have checked that the
underlying properties  hold for generic values of these parameters.

In \cite{Jetzer:1993nk} it was argued that the vacuum becomes unstable to pair production in the presence of a boson star
when the charge of a boson star fulfills $Q \geq 1/\alpha_{\rm fs}$.  Using all the rescalings used in this paper
we find that this
bound translates into a bound for the (dimensionless) particle number $N$ computed in our numerical analysis:
\begin{equation}
 N \geq \frac{4\pi}{e^3} \frac{\eta_{\rm susy}}{m} \ .
\end{equation}
Now, this bound of course depends strongly on the value of $\eta_{\rm susy}$. For the upper bound 
$\eta_{\rm susy}=10^8$ TeV and choosing $e=0.2$ we find that for values of $N \gtrsim 10^{12}$ pair production would occur, while for the
lower bound $\eta_{\rm susy}=1$ TeV and again $e=0.2$ we find $N \gtrsim 10^{4}$. As can be clearly seen in Fig.\ref{fig_uncharged_charged} 
the maximal value of $N$ is always on the order of $10^2$ and we would hence expect the vacuum to be stable
with respect to pair production in the presence of our charged boson stars.

\section{Motion of a charged test particle in the space-time of a boson star}
In the following, we want to study the motion of test particles in the space-time of a charged boson star.
The Hamilton-Jacobi equation describing such a motion reads
\begin{equation}
- 2\frac{\partial S}{\partial\tau} = g^{ \mu\nu}\left(\frac{\partial S}
{\partial x^\mu} - qA_\mu \right)\left( \frac{\partial S}{\partial x^\nu} - qA_\nu  \right) \label{HJeq}  \ ,
\end{equation}
where $q$ denotes the charge of the test particle and $S$ is the action.  

These have a solution of the form~\cite{Chandrasekhar83}
\begin{equation}
S=\frac{1}{2} \delta\tau - E t + L \varphi + S(r) + S(\theta) \label{HJ_sol} \ . 
\end{equation} 
Here $\tau$ is an affine parameter which corresponds to proper time for
massive test particles. The parameter $\delta$ is equal to $0$ for a massless 
and equal to $1$ for a massive test particle, respectively. Since no massless, charged particles 
exist in nature, we will here concentrate on the massive particles and will set $\delta=1$ in the following.
 
The constants $E$ and $L$ are the conserved energy and the angular momentum of a test particle related to
the Killing vector $\partial/\partial t$ and $\partial/\partial \varphi$, respectively.
Using~\eqref{HJ_sol} and the components of the metric tensor (\ref{metric}) we can rewrite the equation~\eqref{HJeq} as follows
\begin{equation}
- \delta r^2\frac{l(r)}{f(r)} + \frac{l(r)}{f^2(r)} r^2 (E + q A(r))^2 - r^2 \left(\frac{\partial S(r)}{\partial r} \right)^2 = \left(\frac{\partial S(\theta)}{\partial \theta} \right)^2 + \frac{L^2}{\sin^2\theta} \ , \label{HJeq2}
\end{equation}
where we collected radial coordinate dependent terms on the left 
and polar coordinate dependent terms on the right hand side.
Since the right and left sides of~\eqref{HJeq2} are equal and each of them depends only on one 
coordinate they have to be equal to a constant $K$ (a separation constant). 
Thus, the expressions for $\frac{\partial S(r)}{\partial r}$ and $\frac{\partial S(\theta)}{\partial \theta}$ yield
\begin{eqnarray}
&& \left(\frac{\partial S(r)}{\partial r} \right)^2 = \frac{l(r)}{f^2(r)} (E + q A(r))^2 - \delta \frac{l(r)}{f(r)} - \frac{K}{r^2} \label{HJ_R} \ , \\
&& \left(\frac{\partial S(\theta)}{\partial \theta} \right)^2 = K - \frac{L^2}{\sin^2\theta}  \label{HJ_Theta} \ .
\end{eqnarray}
Now introducing 
\begin{equation}
R \equiv \left(\frac{\partial S(r)}{\partial r} \right)^2  \ \ \ , \ \ \ 
\Theta \equiv \left(\frac{\partial S(\theta)}{\partial \theta} \right)^2 \label{def_Theta_R} \,
\end{equation}
we rewrite the solution~\eqref{HJ_sol} in the following way
\begin{equation}
S=\frac{1}{2} \delta\tau - E t + L \varphi + \int_{r} \sqrt{R} dr + \int_{\theta} \sqrt{\Theta} d\theta \label{HJ_sol2} \ . 
\end{equation} 

Differentiating~\eqref{HJ_sol2} with respect to $K$, $\delta$, $E$ and $L$, 
using~\eqref{HJ_R} and~\eqref{HJ_Theta} and setting the result equal to zero (since the variation of 
the action with respect to the constants of motion should vanish), we get 
the following four expressions
\begin{eqnarray}
&& \int_r \frac{1}{r^2} \frac{dr}{\sqrt{R}} = \int_\theta \frac{d\theta}{\sqrt{\Theta}} \label{HJ_K} \ , \\
&& \tau = \int_r \frac{l(r)}{f(r)} \frac{dr}{\sqrt{R}} \label{HJ_delta} \ ,  \\
&& t = \int_r \frac{l(r)}{f^2(r)} (E + q A(r)) \frac{dr}{\sqrt{R}} \label{HJ_E} \ , \\
&& \varphi = \int_\theta \frac{L}{\sin^2\theta} \frac{d\theta}{\sqrt{\Theta}} \label{HJ_L} \ .
\end{eqnarray}
From~\eqref{HJ_delta} we infer the differential equation for the radial coordinate and the polar coordinate, respectively
\begin{equation}
\frac{dr}{d\tau} = \frac{f(r)}{l(r)} \sqrt{R} \ \ \ , \ \ \ 
\frac{d\theta}{d\tau} = \frac{f(r)}{r^2 l(r)} \sqrt{\Theta} \label{HJ_pol_rad} \ .
\end{equation}
To omit the dependence on the radial coordinate in the equation for the polar coordinate $\theta$
we introduce a new affine parameter 
$\lambda$~\cite{Mino03}
\begin{equation}
d\lambda = \frac{f(r)}{r^2 l(r)} d\tau  \label{HJ_lambda} \ .
\end{equation} 
With this the equations~\eqref{HJ_pol_rad}  simplify to
\begin{equation}
\frac{d\theta}{d\lambda} =  \sqrt{\Theta} \ \ \ , \ \ \ 
\frac{dr}{d\lambda} = r^2 \sqrt{R} \label{HJ_rad_pol2} \ .
\end{equation}
The differential equations for the remaining two coordinates $\varphi$ and $t$ are now easy to derive. 
For the azimuthal coordinate $\varphi$ and the time coordinate $t$, respectively, we get
\begin{equation}
\frac{d\varphi}{d\lambda} = \frac{L}{\sin^2\theta}   \ \ \ , \ \ \ 
\frac{d t}{d\lambda} = \frac{r^2 l(r)}{f^2(r)} (E + q A(r))  \label{HJ_azi_time} \ .
\end{equation}
Equations~\eqref{HJ_rad_pol2} and~\eqref{HJ_azi_time} describe 
the motion of a charged test particle in the boson star space-time given by the metric~\eqref{metric}.

\subsection{Components of the equation}

To visualize the motion we need to integrate the differential 
equations~\eqref{HJ_rad_pol2} and~\eqref{HJ_azi_time}. This reads:
\begin{eqnarray}
&& \lambda - \lambda_0 = \int^r_{r_0} \frac{dr}{r^2\sqrt{R}} \label{HJ_rad3} \ , \\
&& \lambda - \lambda_0 = \int^\theta_{\theta_0} \frac{d\theta}{ \sqrt{\Theta} } \label{HJ_pol3} \ , \\
&& \varphi - \varphi_0 = \int^\theta_{\theta_0} \frac{L}{\sin^2\theta} \frac{d\theta}{ \sqrt{\Theta} } \label{HJ_azi2} \ , \\
&& t - t_0 = \int^r_{r_0} \frac{l(r)}{f^2(r)} (E + q A(r)) \frac{dr}{\sqrt{R}}  \label{HJ_time2} \ ,
\end{eqnarray}
where index zero denotes an initial value.
Since the coordinate $r$ and functions $f(r)$, $l(r)$ and $A(r)$ 
(and correspondingly $R$) are given numerically, 
only numerical solutions to the radial and time equations~\eqref{HJ_rad3} and~\eqref{HJ_time2} are possible.
The integrals in the equations~\eqref{HJ_pol3} and~\eqref{HJ_azi2}, on the other hand, 
can be solved analytically.

\subsubsection{Polar motion}~\label{subsubsec:polarmotion}
Let us first consider equation~\eqref{HJ_pol3} for the polar coordinate $\theta\in[0,\pi]$. 
Introducing a new variable $\xi=\cos\theta$ we can rewrite the equation as follows
\begin{equation}
\lambda - \lambda_0 = \int^\theta_{\theta_0} \frac{d\theta}{ \sqrt{\Theta} } = - \int^\xi_{\xi_0} \frac{d\xi}{ \sqrt{ K-L^2 - K\xi^2 } } \label{HJ_pol4} \ ,
\end{equation}
where $\Theta = K - \frac{L^2}{\sin^2\theta}$.
For physical motion the values of $\theta$ (as well as $\xi$) must be real, moreover, 
the condition $\xi\in[-1,1]$ must be fulfilled.
Introducing
\begin{equation}
\Theta_\xi = K-L^2 - K\xi^2 \label{def_Theta_xi} \,
\end{equation}
in~\eqref{HJ_pol4} and studying its zeros given by
\begin{equation}
\xi_{1,2} = \pm  \frac{\sqrt{K(K-L^2)}}{-K} \label{def_xizeros} \,
\end{equation}
we can get limits on the values of $K$. In the following we will consider the different possible values:
\begin{itemize}
\item {\bf $K > 0:$} Since the square root in~\eqref{def_xizeros} must be positive 
it immediately follows that $K\geq L^2$. 
Rewriting~\eqref{def_xizeros} in the form 
\begin{equation}
\xi_{1,2} = \mp  \sqrt{\frac{ K-L^2 }{K}} \equiv \mp  \sqrt{ 1- \frac{ L^2 }{K}} \label{def_xizeros2} \ .
\end{equation}
we find that $-1 \leq \xi \leq 1$ which leads to 
physical values of $\theta$. Moreover, $\theta$ will be bounded from above and below such that
the motion will be between a minimal and maximal value of $\theta$ which can be between $0$ and $\pi$. 
If $K=L^2$ then $\xi_{1,2}=0$. This corresponds to $\theta=\pi/2$, i.e. corresponds
to motion in the equatorial plane.
\item {\bf $K<0:$} Introducing $K_1 \equiv-K$ where $K_1>0$ and rewriting~\eqref{def_xizeros} as
\begin{equation}
\xi_{1,2} = \pm  \sqrt{\frac{ K_1+L^2 }{K_1}} \equiv \mp  \sqrt{ 1 + \frac{ L^2 }{K_1}} \label{def_xizeros3} \ .
\end{equation}
we find that the values of $\xi_{1,2}$ in~\eqref{def_xizeros3} belong to unphysical branches with 
$\xi>1$ and $\xi<-1$ for non-vanishing $L^2$. 
If $L^2=0$ then $\xi_{1,2}=\pm 1$ and the result of integration in~\eqref{HJ_pol4} 
contains complex values which is unphysical ($\lambda$ must be real). 
\item {\bf $K=0:$} This case is not possible for non-vanishing $L^2$ since the values 
of $\theta$ are complex (because of $\sqrt{-L^2}$ in~\eqref{HJ_pol4}) and hence unphysical.
\end{itemize}
To summarize we note that physical values of the coordinate $\theta$ in~\eqref{HJ_pol4} 
are possible provided that 
\begin{equation}
K > 0 \qquad \text{and} \qquad K\geq L^2 \label{cond_fin} \ . %\\
%&& K < 0 \qquad \text{and} \qquad L^2=0 \qquad \text{which implies} \qquad \theta=0,\pi \label{cond_fin}  \ .
\end{equation}
Integration of~\eqref{HJ_pol4} for $K > 0$ and $K\geq L^2$ thus yields:
\begin{equation}
\lambda - \lambda_0 = - \left( -\frac{1}{\sqrt{K}} \arcsin{\left( \frac{-K\xi}{\sqrt{K(K-L^2)}} \right)} \right) 
\left\vert^{\xi}_{\xi_{0} } \label{HJ_pol5} \       \right. \ .
\end{equation}
Inversion of~\eqref{HJ_pol5} where $\xi=\cos\theta$ gives the final solution for the coordinate $\theta$:
\begin{equation}
\theta = \arccos\left(  \frac{\sqrt{K(K-L^2)}}{K} \sin{ \left( \sqrt{K}(\lambda^\prime - \lambda) \right) } \right) \label{HJ_pol6} \ ,
\end{equation}
where $\lambda^\prime = \lambda_0 + \frac{1}{\sqrt{K}} \arcsin{\left( \frac{K\xi_0}{\sqrt{K(K-L^2)}} \right)}$ is a constant.

\subsubsection{Azimuthal motion}
We have the following integral for the azimuthal coordinate $\varphi$
\begin{equation}
\varphi - \varphi_0 = \int^\theta_{\theta_0} \frac{L}{\sin^2\theta} \frac{d\theta}{ \sqrt{\Theta} } =  - \int^\xi_{\xi_0} \frac{L}{1-\xi^2 } \frac{d\xi}{ \sqrt{ K-L^2 - K\xi^2 } }    \label{HJ_azi3} \ .
\end{equation}
The statements made on the values of the separation constant $K$ in  Sec.~\ref{subsubsec:polarmotion} 
are also true for the equation~\eqref{HJ_azi3}. The integral in~\eqref{HJ_azi3} can be solved by 
elementary functions as follows
\begin{equation}
\varphi  = \varphi_0 + 
\frac{1}{2} \frac{L}{|L|} \arctan{ \frac{2\xi\sqrt{L^2(K-L^2-K\xi^2)}}{ \xi^2(L^2+K) - K + L^2 } } \Bigl|^{\xi}_{\xi_{0} } \label{HJ_azi4} \ .
\end{equation}
Substituting $\xi=\cos\theta$ with $\theta$ given by~\eqref{HJ_pol6} into~\eqref{HJ_azi4} 
we get $\varphi=\varphi(\lambda)$.
 
Using the components of the equations for the $\theta$- and the $\varphi$-coordinate we can give the
$\theta(\varphi)$ motion explicitly. This reads
\begin{equation}
 \frac{d\theta}{d\varphi} = \sin\theta \sqrt{\frac{K}{L^2}\sin^2\theta -1} \ ,
\end{equation}
which can be solved by
\begin{equation}
\label{theta_varphi}
 \cot^2\theta = \left(\frac{K}{L^2} - 1 \right) \sin^2\varphi \ .
\end{equation}
The turning points of the motion in $\theta$-direction are given by (using (\ref{HJ_pol_rad}))
\begin{equation}
 \frac{d\theta}{d\tau}=0 \ \ \Longrightarrow \ \  \sin^2\theta = \frac{L^2}{K} \ \  \ .
\end{equation}
Hence, the $\theta$-coordinate fulfills
\begin{equation}
 \arcsin\left(\frac{L}{\sqrt{K}}\right) \leq \theta \leq \pi - \arcsin\left(\frac{L}{\sqrt{K}}\right) \ .
\end{equation}
Inserting this result into (\ref{theta_varphi}) we find that the turning points in $\varphi$-direction correspond to
\begin{equation}
 \sin^2 \varphi = 1 \ \ \Longrightarrow \ \ \varphi = \frac{\pi}{2} + n\pi \ \ \ , \ \ n\in \mathbb{Z} \ . 
\end{equation}
The angular motion of test particles is hence planar, where the inclination of the
plane with respect to the equatorial plane is determined by the ratio $L^2/K$. For $L^2=K$ the
particle moves in the equatorial plane $\theta=\pi/2$. Hence, the stronger $K$ differs from $L^2$
the stronger the plane in which the particle moves is inclined with respect to the equatorial plane.
Note that this is in stark contrast to the motion
of charged test particles in the space-time of charged black holes \cite{GruKa}. In this latter case, the charged
particles move in general on non-planar orbits.

\subsection{Numerical results}
In the following we will describe our results for the motion of charged, massive test particles in
charged boson star space-times. We solve the radial differential equation~\eqref{HJ_rad3} numerically using MATLAB with the recursive adaptive Lobatto quadrature with an 
absolute error tolerance of $10^{-6}$. For the polar $\theta$- and azimuthal $\varphi$-coordinates we use the results~\eqref{HJ_pol6} and~\eqref{HJ_azi4}. For visualization of the trajectories we use a spherical coordinate system with $x=r\sin\theta\cos\varphi$, $y=r\sin\theta\sin\varphi$ and $z=r\cos\theta$.

We have
chosen the space-time of a boson star with $e=0.1$, $\alpha=0.1$ and some particular choices of the parameter $\phi(0)$.
The physical values of the corresponding solutions are given in Table~\ref{table_boson}. Whenever we give the value
of $\phi(0)$ in the following, the corresponding values of mass and charge can be read off from this table.

\begin{table}
\begin{center}
  \begin{tabular}{| c | c | c | c | c |}
\hline
$\phi(0)$ &  $\omega$ & $M$ & $Q$ & comments \\                
\hline\hline
0.405   &  0.9243 &  112.8185 &  11.3439     & {\it local} maximum of $M$, $Q$ \\
2.250   &  0.6252 &  127.7272 & 13.2867      & {\it global} maximum of $M$, $Q$ \\
2.420   &  0.6083 &  126.8053 & 13.1553     & {\it minimal} value of $\omega$ \\
4.200   &  0.7088 &   84.0175 &  7.1502     & {\it local} minimum of $M$, $Q$ \\
\hline
\end{tabular}
\end{center}
  \caption{The values of $\omega$, the mass $M$ and the charge $Q$ for boson stars with 
$\alpha=0.1$, $e=0.1$ and given $\phi(0)$. }
\label{table_boson}
  \end{table}

\subsubsection{The effective potential}
In order to understand what type of orbits are possible, we 
define an effective potential
\begin{equation}
V^\pm_{\rm{eff}} = -q A(r) \pm \sqrt{\frac{f^2(r)}{l(r)}\left( \delta \frac{l(r)}{f(r)} +\frac{K}{r^2} \right)} 
\label{Veff} \ ,
\end{equation}
such that we can rewrite the equation describing radial motion as follows
\begin{equation}
\left(\frac{dr}{d\lambda}\right)^2 = r^4 R \equiv r^4 \frac{l(r)}{f^2(r)} 
\left( E - V^+_{\rm{eff}} \right)\left( E - V^-_{\rm{eff}} \right) \label{rpot} \ .
\end{equation}
The values of $r$ at which $E =V^\pm_{\rm{eff}}(r)$ mark turning points of the motion. 
In order for $r$ to be real and positive, we have to require the positiveness of the right hand side
of~\eqref{rpot}. Values of $r$ for which the right hand side of ~\eqref{rpot} is negative are not allowed.
In the following, we will refer to these values of $r$ as {\it forbidden regions}. Note that
(\ref{rpot}) is completely invariant under $(E,q) \leftrightarrow (-E,-q)$. It hence contains
a charge conjugation (C) - time reversal (T) symmetry and furthermore a parity (P) symmetry since
only the angular momentum of the particle enters. Thus, the equation (\ref{rpot})
is invariant under a CPT transformation and particles with negative energy and charge $q$ can be interpreted
as anti-particles with positive energy and charge $-q$. 
Now, we can distinguish different type of orbits~:
\begin{itemize}
 \item {\bf Bound orbits (BOs)}: these orbits have two turning points corresponding to
two values of $r$ at which $E =V^\pm_{\rm{eff}}(r)$. The motion of the particle varies between a minimal
radius and a maximal radius, where both have finite values. This is often also referred to as {\it planetary motion}.
\item {\bf Escape orbits (EOs):} these orbits have only one turning point corresponding to one
value of $r$ at which $E =V^\pm_{\rm{eff}}(r)$. The motion of the particle varies between a finite
minimal radius and $r=\infty$. Hence, the particle approaches the boson star from infinity, scatters of
the boson star and moves back to infinity.
\end{itemize}

\begin{figure}[h]
\begin{center}
\subfigure[][$\phi(0)=2.25$]{\label{potential_massive1}\includegraphics[width=6.0cm]{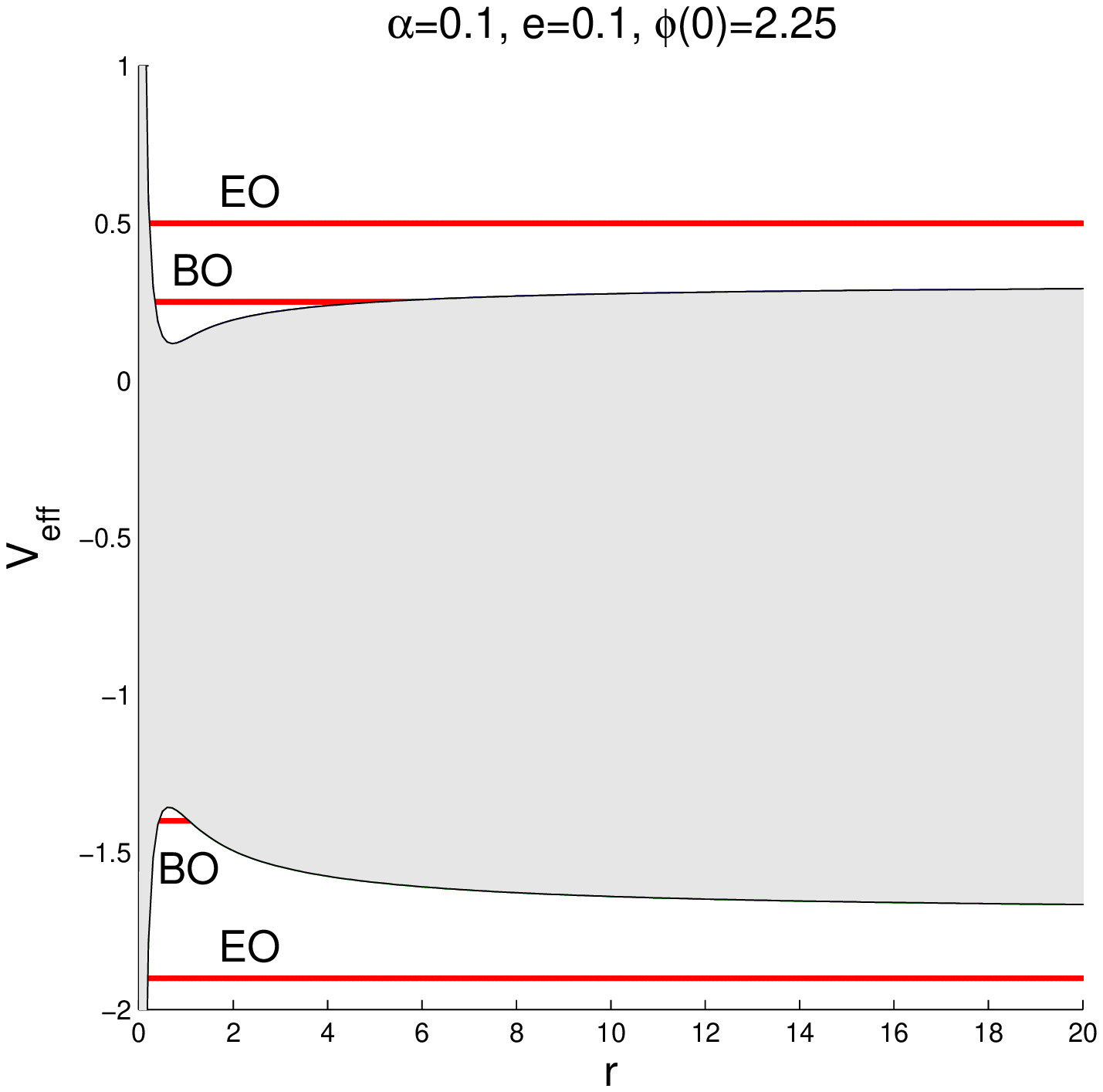}}
\subfigure[][$\phi(0)=2.42$]{\label{potential_massive2}\includegraphics[width=6.0cm]{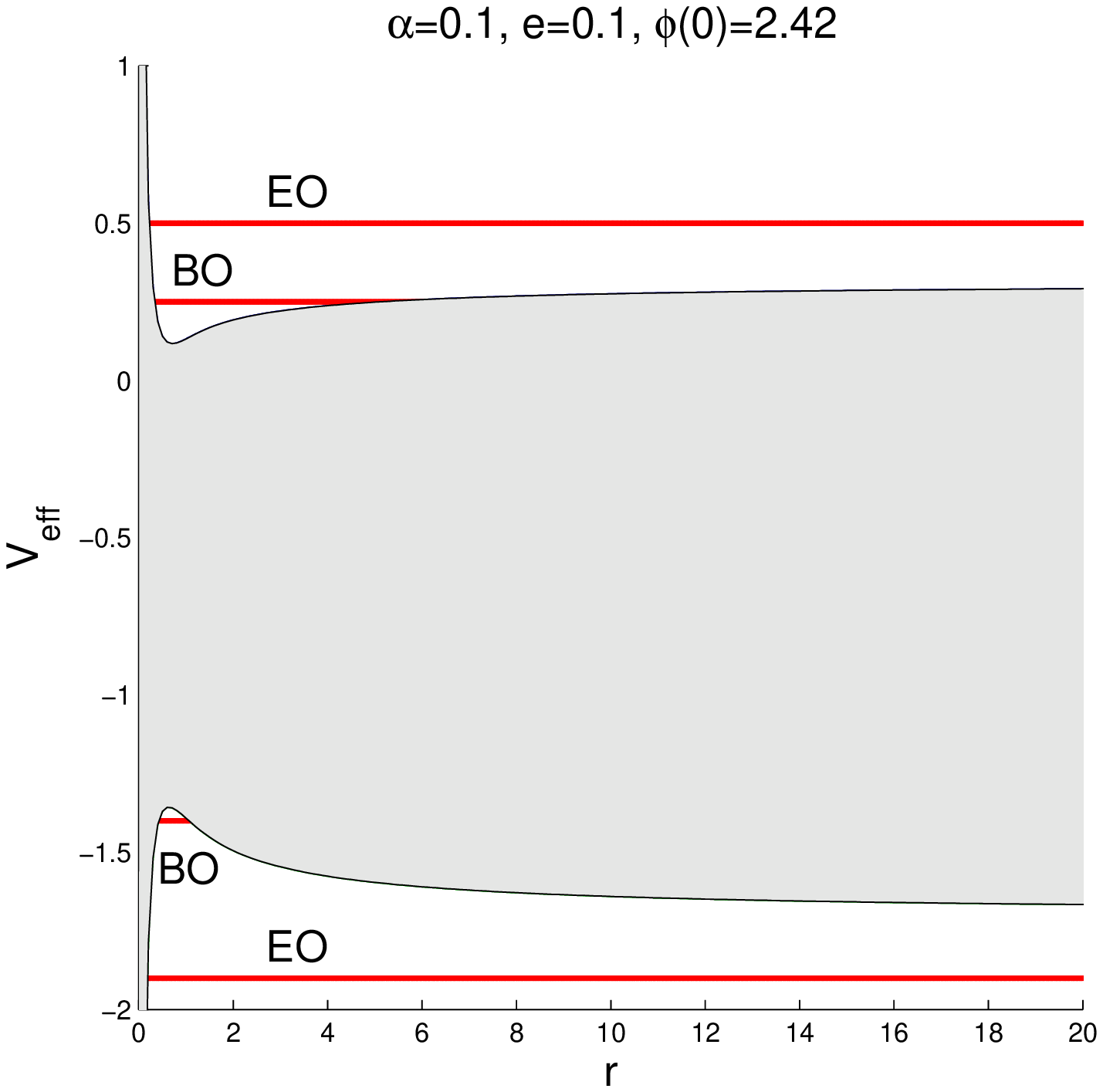}}
\subfigure[][$\phi(0)=0.405$]{\label{potential_massive3}\includegraphics[width=6.0cm]{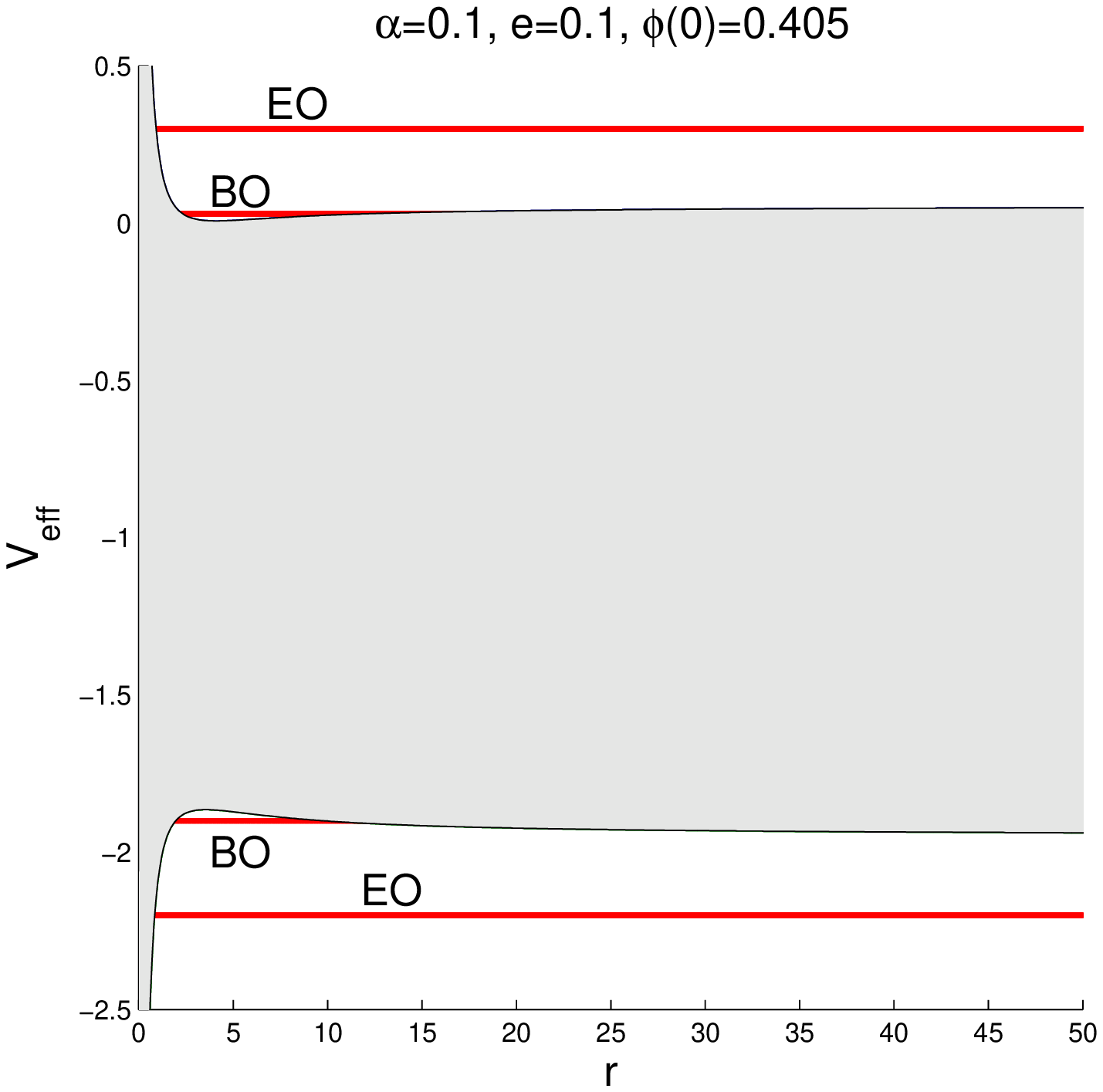}}
\subfigure[][$\phi(0)=4.2$]{\label{potential_massive4}\includegraphics[width=6.0cm]{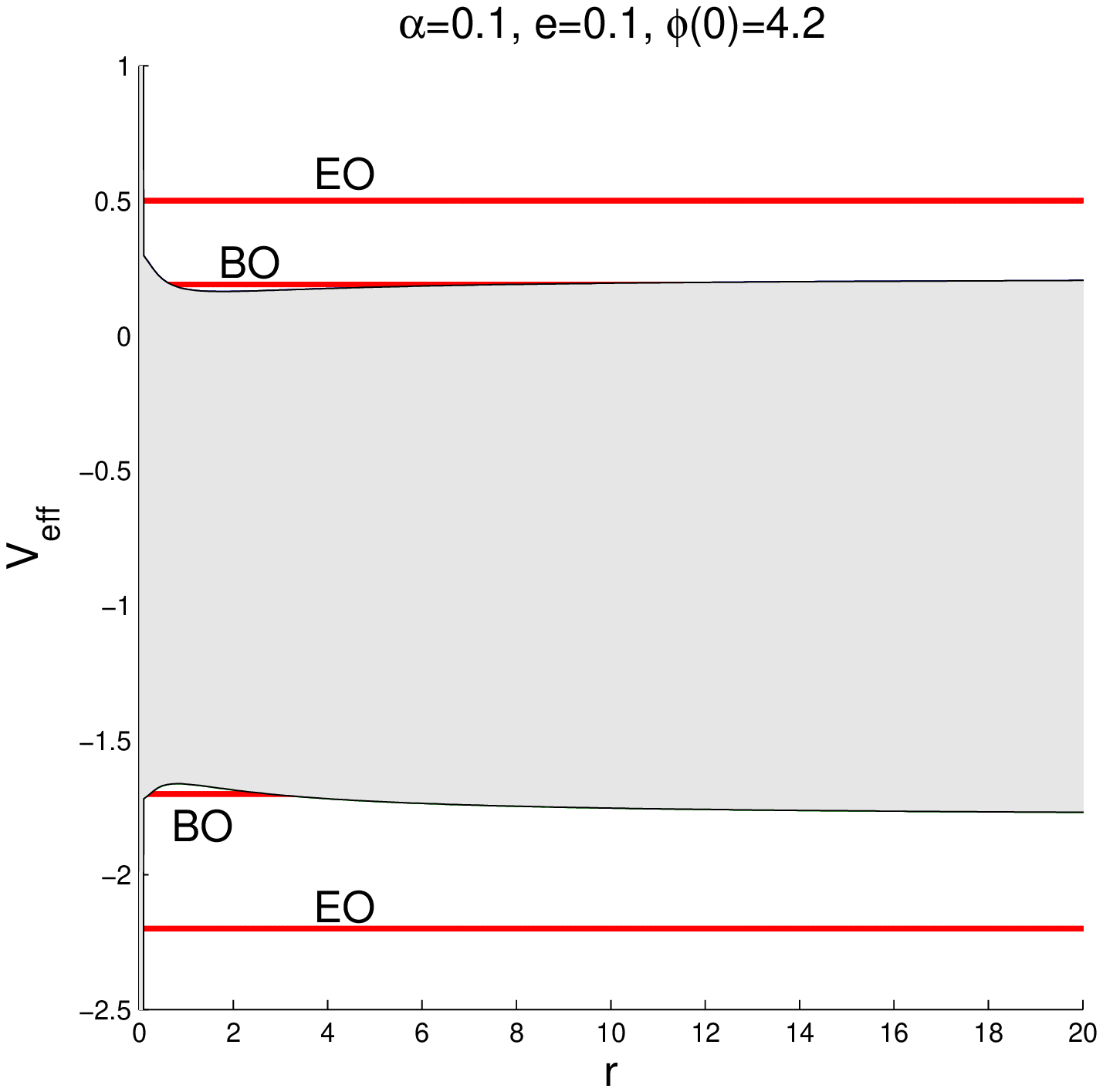}}
\end{center}
\caption{We show the effective potential (\ref{Veff}) for massive, charged test particles 
with charge $q=0.1$, angular momentum $L=0.5$ and separation constant $K=1$ (corresponding
to $\theta\in [\frac{\pi}{6}:\frac{5\pi}{6}]$) in the space-time of
a boson star with $\alpha=0.1$, $e=0.1$ and different values of $\phi(0)$. 
The grey region is forbidden. 
  \label{effective}}
\end{figure}

In Fig.\ref{effective} we show the effective potential~\eqref{Veff} for a charged test particle with $q=0.1$,
angular momentum $L=0.5$ and separation constant $K=1$. This corresponds to motion
in a plane with $\theta\in [\frac{\pi}{6}:\frac{5\pi}{6}]$. The particles move in a boson star space-time with
$\alpha=0.1$, $e=0.1$ and different values of the central value of the scalar field $\phi(0)$.
We find that depending on the value of $E$ either BOs or EOs exist and 
that two regions for $E$ are possible:
$E > 0$ and $E < 0$. Now, remembering that our particles are charged, we can interpret 
particles with $E < 0$ and charge $q$ as anti-particles with energy $-E$ and charge $-q$. 

\begin{figure}[h]
\begin{center}
\subfigure[][$\phi(0)=2.25$]{\label{potential_massive11}\includegraphics[width=6.0cm]{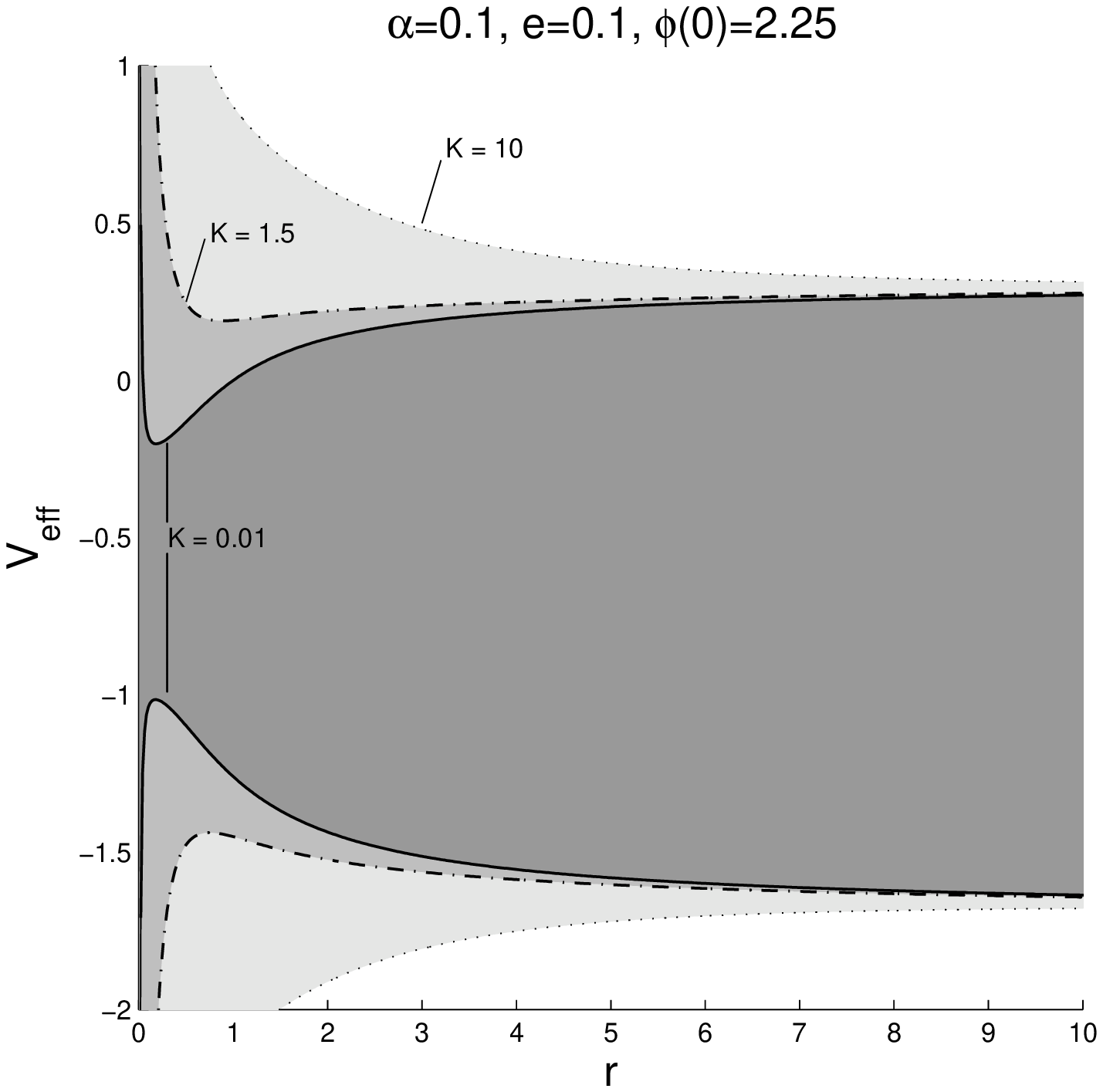}}
\subfigure[][$\phi(0)=0.405$]{\label{potential_massive33}\includegraphics[width=6.0cm]{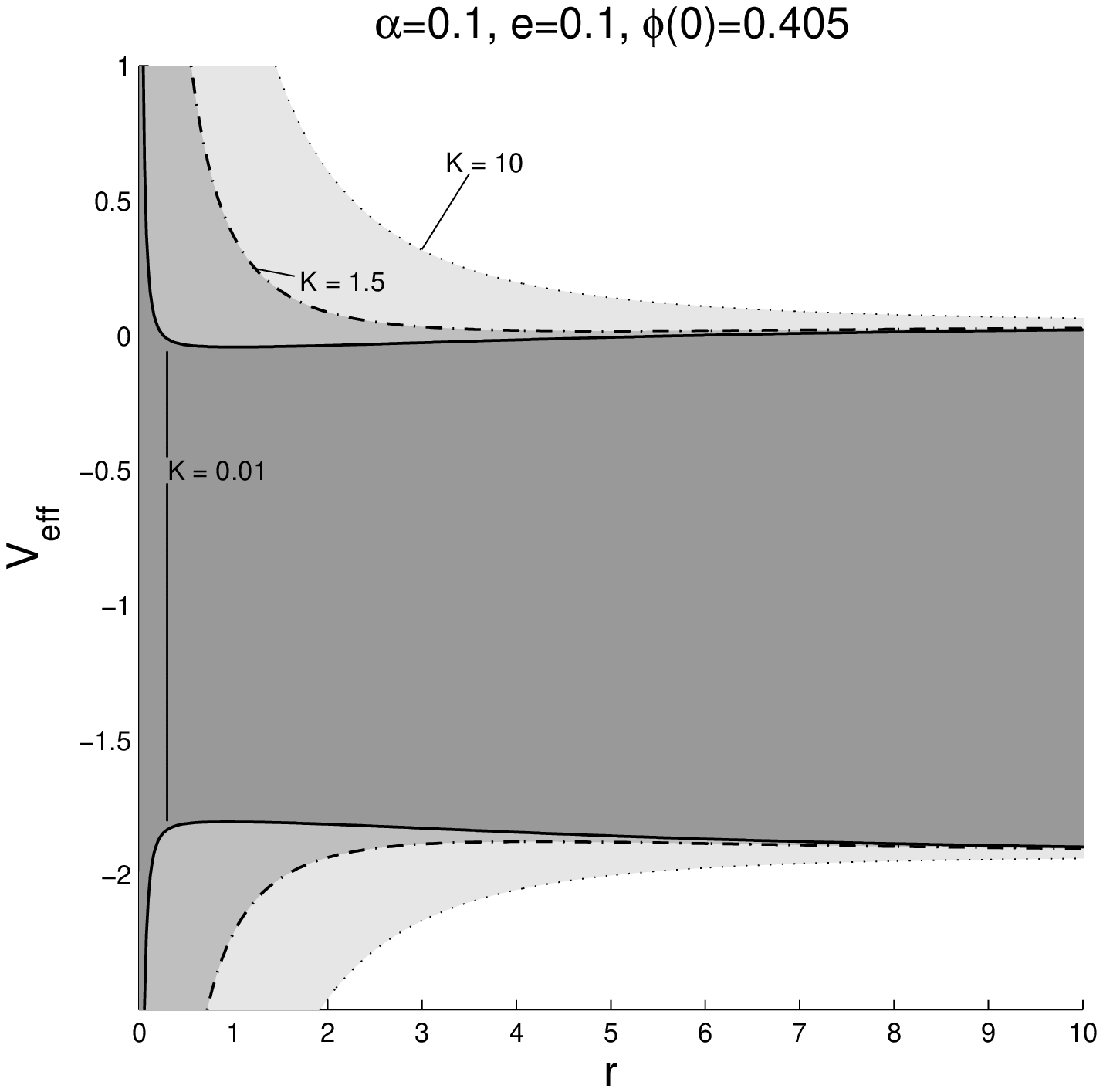}}
\subfigure[][$\phi(0)=4.2$]{\label{potential_massive44}\includegraphics[width=6.0cm]{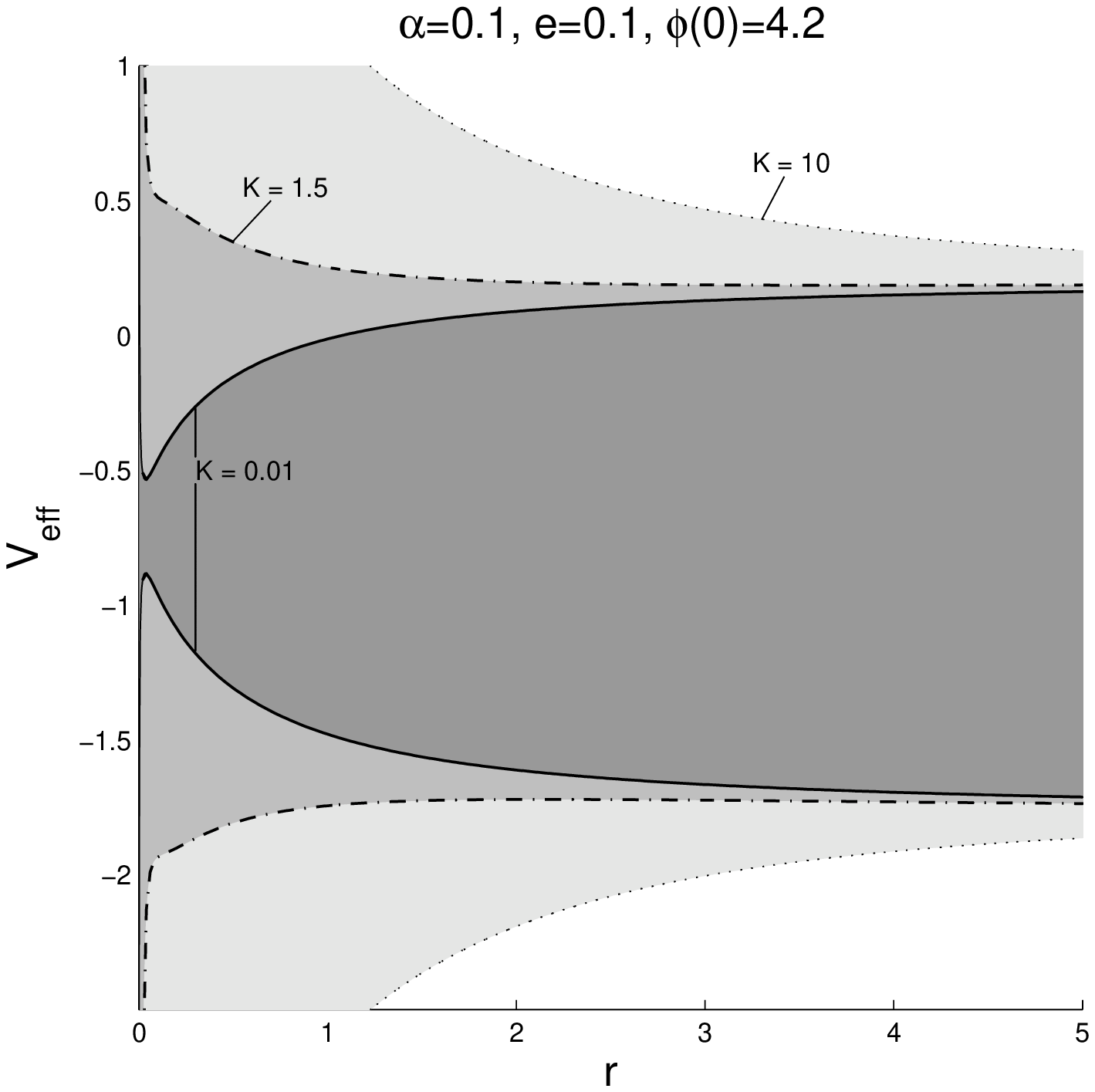}}
\end{center}
\caption{We show the effective potential~\eqref{Veff} for massive, charged test particles 
with charge $q=0.1$, angular momentum $L=0.5$ and varying values of the 
separation constant $K$ in the space-time of
a boson star with $\alpha=0.1$, $e=0.1$ and different values of $\phi(0)$. 
The grey region is forbidden for the given value of $K$. We choose 
$K=0.01$, $K=1.5$ and $K=10$, respectively.
  \label{effective2}}
\end{figure}

In Fig.~\ref{effective2} we show the effective potential~\eqref{Veff} for varying separation constant $K$ for 
$\phi(0)=2.25$ in Fig.~\ref{potential_massive11}, for $\phi(0)=0.405$ 
in Fig.~\ref{potential_massive33} and for $\phi(0)=4.2$ in Fig.~\ref{potential_massive44}. 
For small values of $K$ bound orbits exist, 
while for increasing $K$ only escape orbits are present. Since for a fixed value of $L^2$ the value of
$K$ determines how strong the plane of motion is inclined with respect to the equatorial plane this suggests that
bound orbits are only possible if the inclination is not too big, i.e. the range of $\theta$ does not vary too strongly
along the trajectory.

\begin{figure}[h]
\begin{center}
\subfigure[][$\phi(0)=4.2$, $q=1$]{\label{potential_massive11_q1}\includegraphics[width=6.0cm]{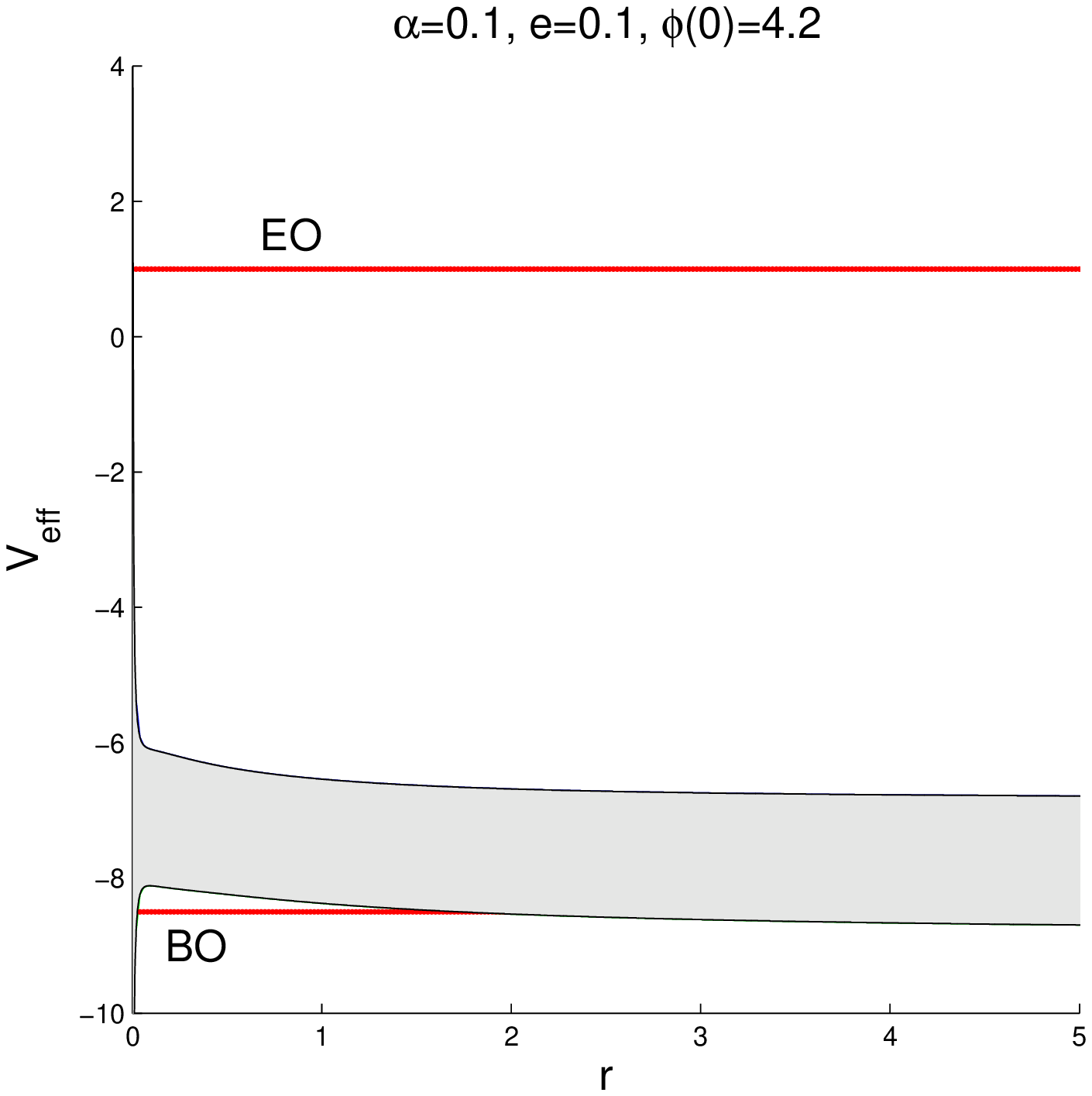}}
\subfigure[][$\phi(0)=4.2$, $q=3$]{\label{potential_massive11_q3}\includegraphics[width=6.0cm]{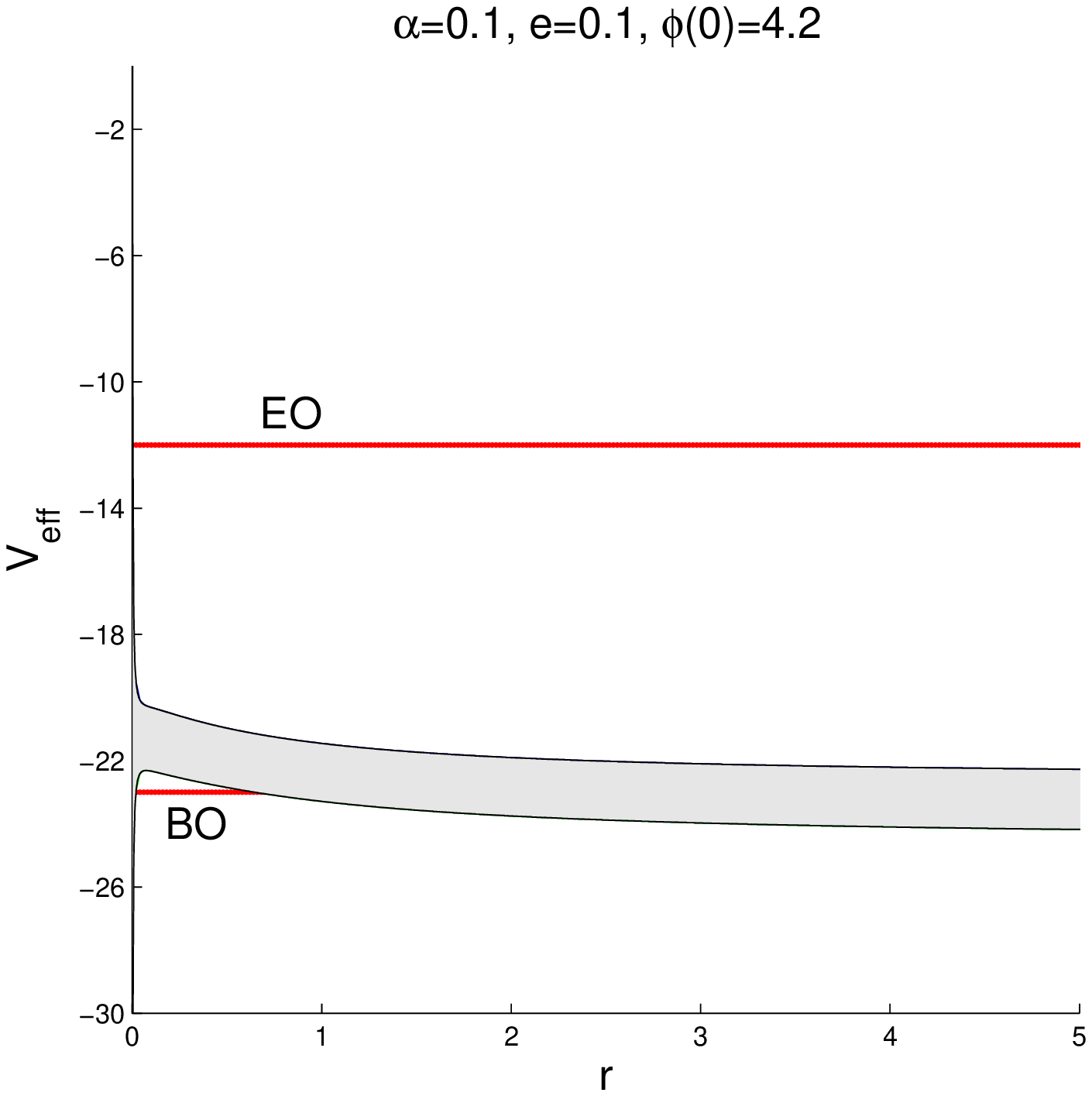}}
\end{center}
\caption{We show the effective potential~\eqref{Veff} for massive, charged test 
particles with angular momentum $L=0.5$ and  separation constant $K=1$ in the space-time of
a boson star with $\alpha=0.1$, $e=0.1$ and $\phi(0)=4.2$. In Fig.~\subref{potential_massive11_q1} 
the charge of the test particle is $q=1$ and in Fig.~\subref{potential_massive11_q3} the charge is $q=3$.
The grey region is forbidden for test particle motion. 
  \label{effective_qpos}}
\end{figure}

\begin{figure}[h]
\begin{center}
\subfigure[][$\phi(0)=2.25$, $q=-2$]{\label{potential_massive11_q-2}\includegraphics[width=6.0cm]{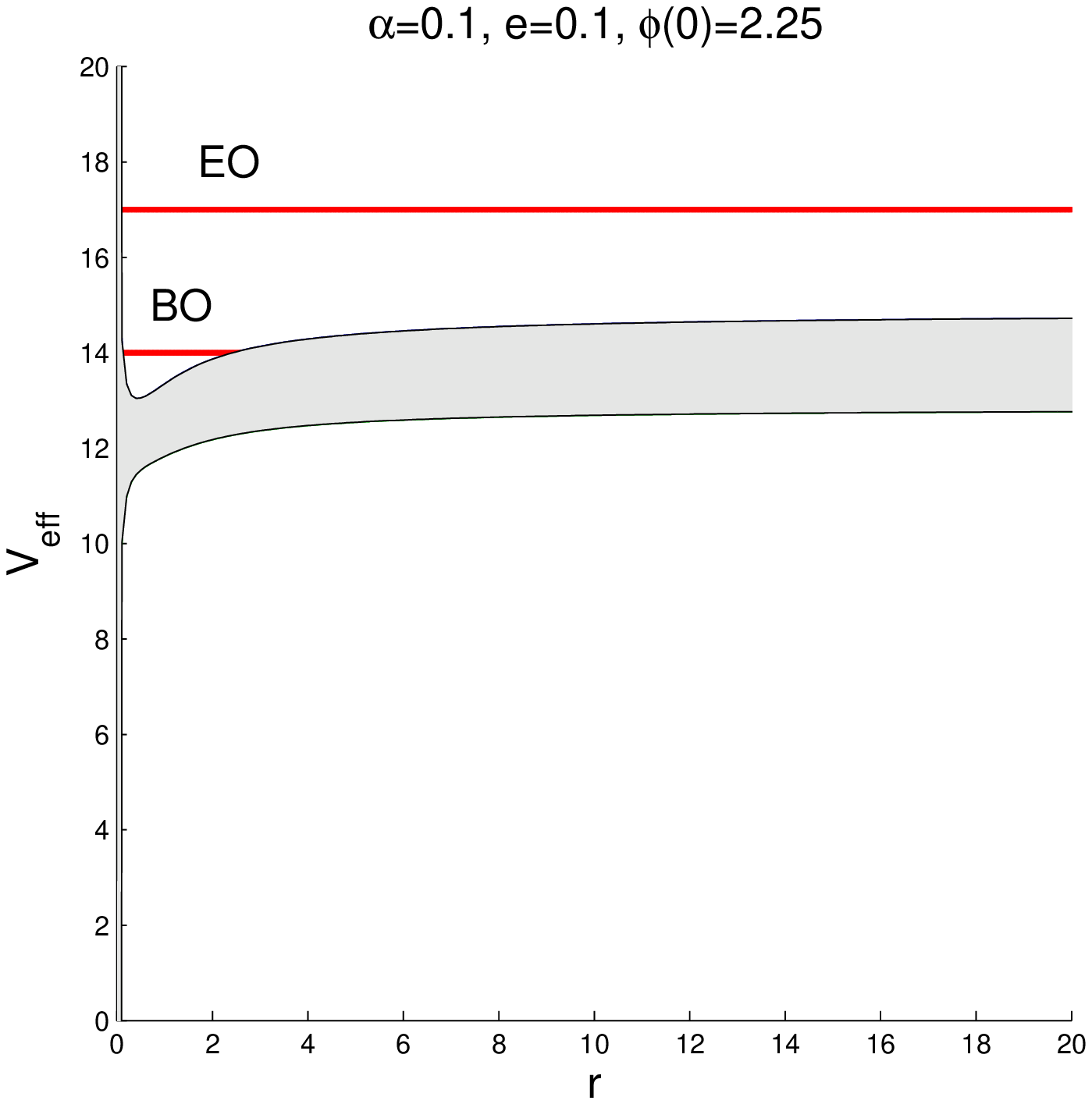}}
\subfigure[][$\phi(0)=2.25$, $q=-4$]{\label{potential_massive11_q-4}\includegraphics[width=6.0cm]{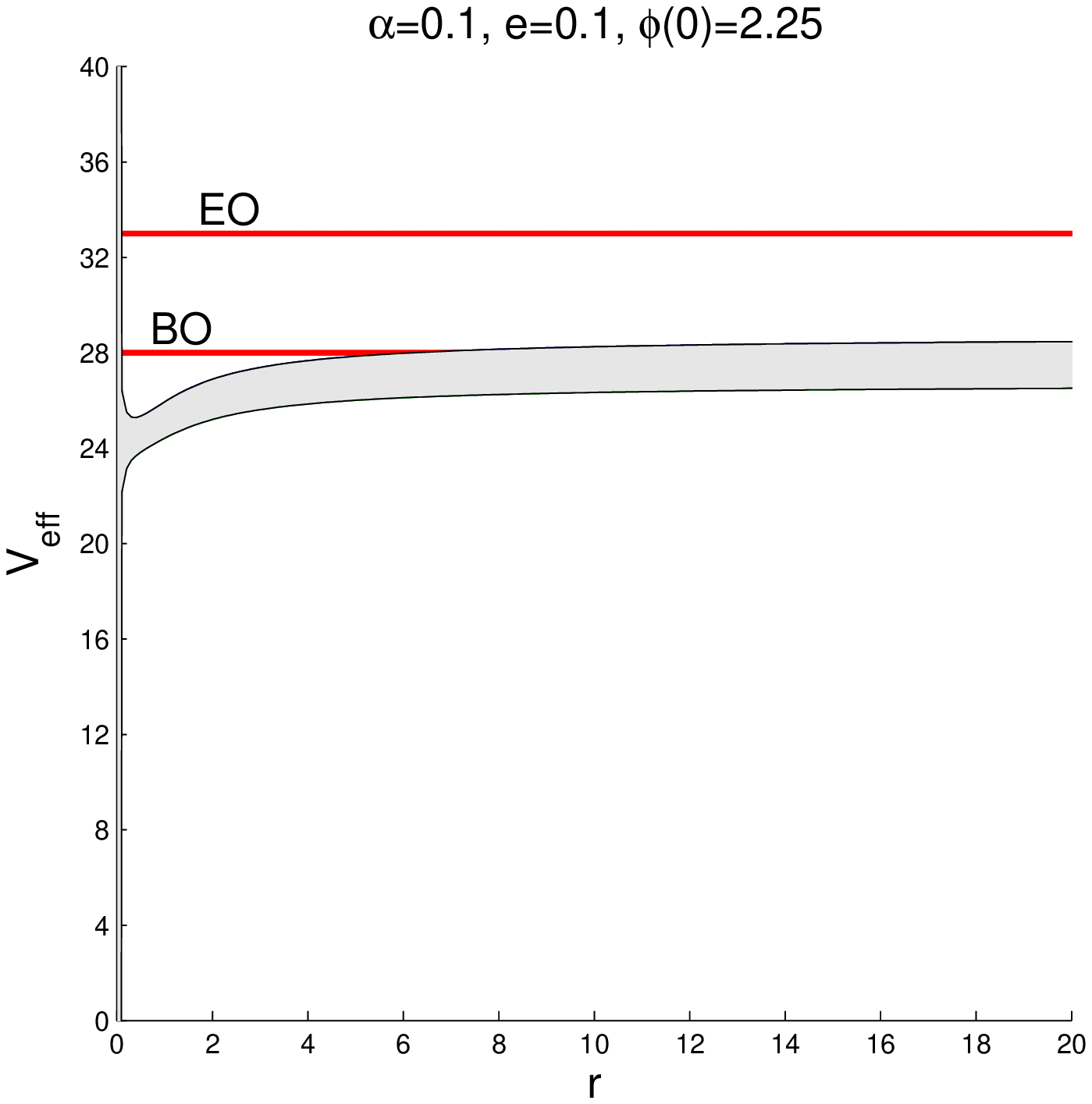}}
\end{center}
\caption{We show the effective potential~\eqref{Veff} for massive, charged test particles with angular 
momentum $L=0.5$ and separation constant $K=1$ in the space-time of
a boson star with $\alpha=0.1$, $e=0.1$ and $\phi(0)=2.25$. 
In the plot~\subref{potential_massive11_q-2} the charge of a 
test particle $q=-2$ and in the plot~\subref{potential_massive11_q-4} the charge $q=-4$. The grey region is forbidden.  
  \label{effective_q}}
\end{figure}

The charge $q$ of a test particle also influences the effective potentials. For large positive charges $q$ 
the effective potential will be shifted downwards along the $V_{\rm{eff}}$-axis, which means that for positive
$E$ bound orbits are no longer possible. This is related to the electromagnetic repulsion between
the positively charged boson star and the positively charged test particle. If $q$ is too big, the electromagnetic
repulsion is too strong and the particle gets simply scattered by the boson star. On the other hand,
bound orbits exist for negative values of $E$. With the anti-particle interpretation this is
also obviously explained. Since a particle with negative energy and large positive $q$ is equivalent to 
an anti-particle with positive energy and large negative charge $-q$ the electromagnetic interaction between the
positively charged boson star and the negatively charged test particle is attractive and bound orbits become
possible. We present potentials for large positive charges $q$ in Fig.~\subref{potential_massive11_q1}
for $q=1$ and in Fig.~\subref{potential_massive11_q3} for $q=3$. The boson star has $\phi(0)=4.2$ in this case. 
For decreasing negative charges the potential moves upwards along the $V_{\rm{eff}}$-axis,  
which leads to two possible regions for bound orbits with positive energies. 
In this case for negative energies only escape orbits can be found. Again, the reason is the
increased electromagnetic attraction in the case of decreasing negative $q$ and the increased
repulsion between the particle with negative energy and negative charge $q$ which can be interpreted as 
an anti-particle with positive energy and positive charge $-q$.
Examples of potentials for negative charges $q$ are shown in Fig.~\ref{potential_massive11_q-2} for $q=-2$ 
and in Fig.~\subref{potential_massive11_q-4} for $q=-4$. 

\subsubsection{Examples of orbits}

In the following, we will present some typical orbits possible in the space-time of boson stars.

\begin{figure*}[h!]
\begin{center}
\subfigure[][BO for $E=0.25$]{\label{orb1}\includegraphics[width=6cm]{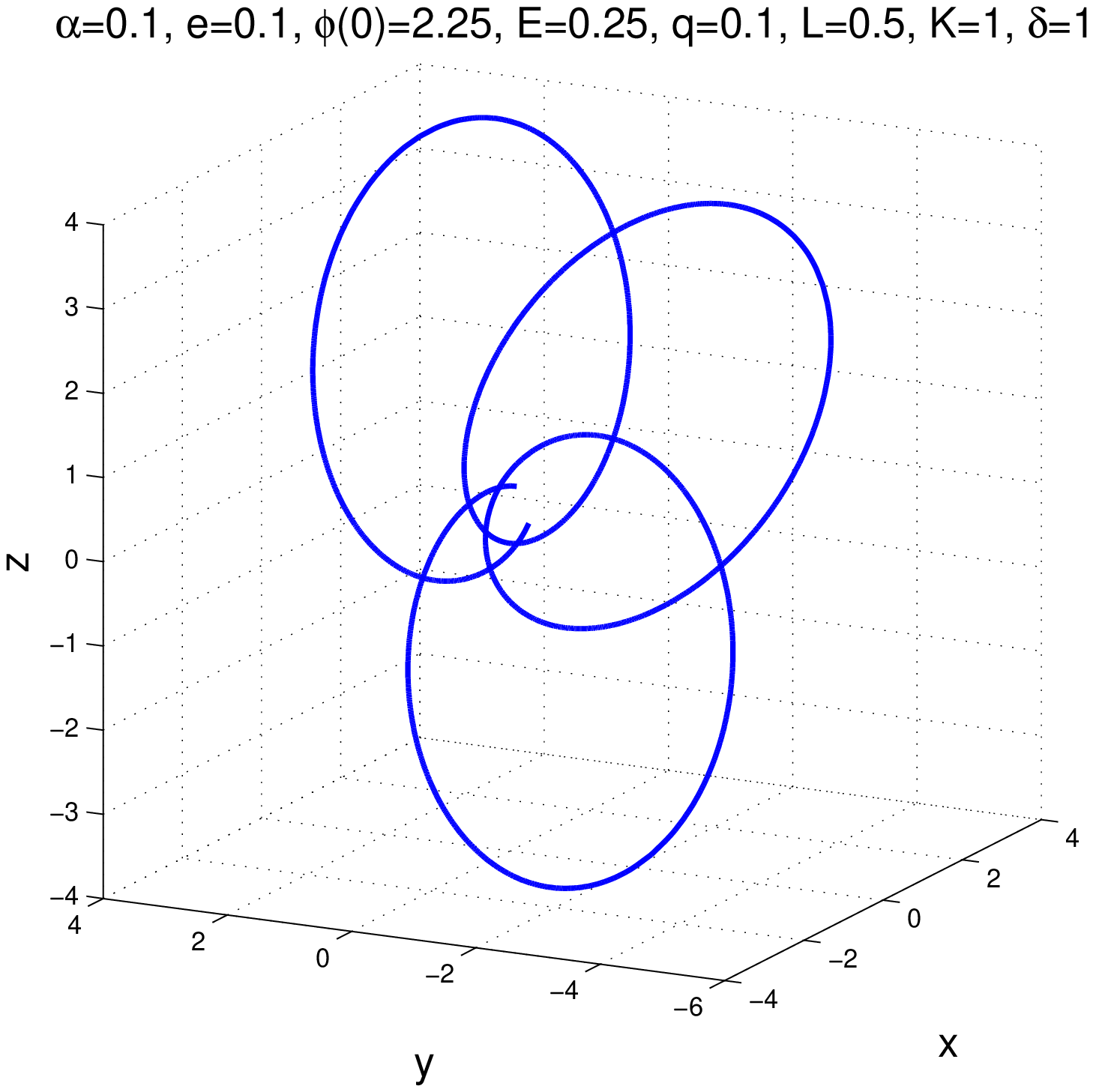}}
\subfigure[][$r$-$\theta$ plot for figure~\subref{orb1}]{\label{orb1-rth}\includegraphics[width=7cm]{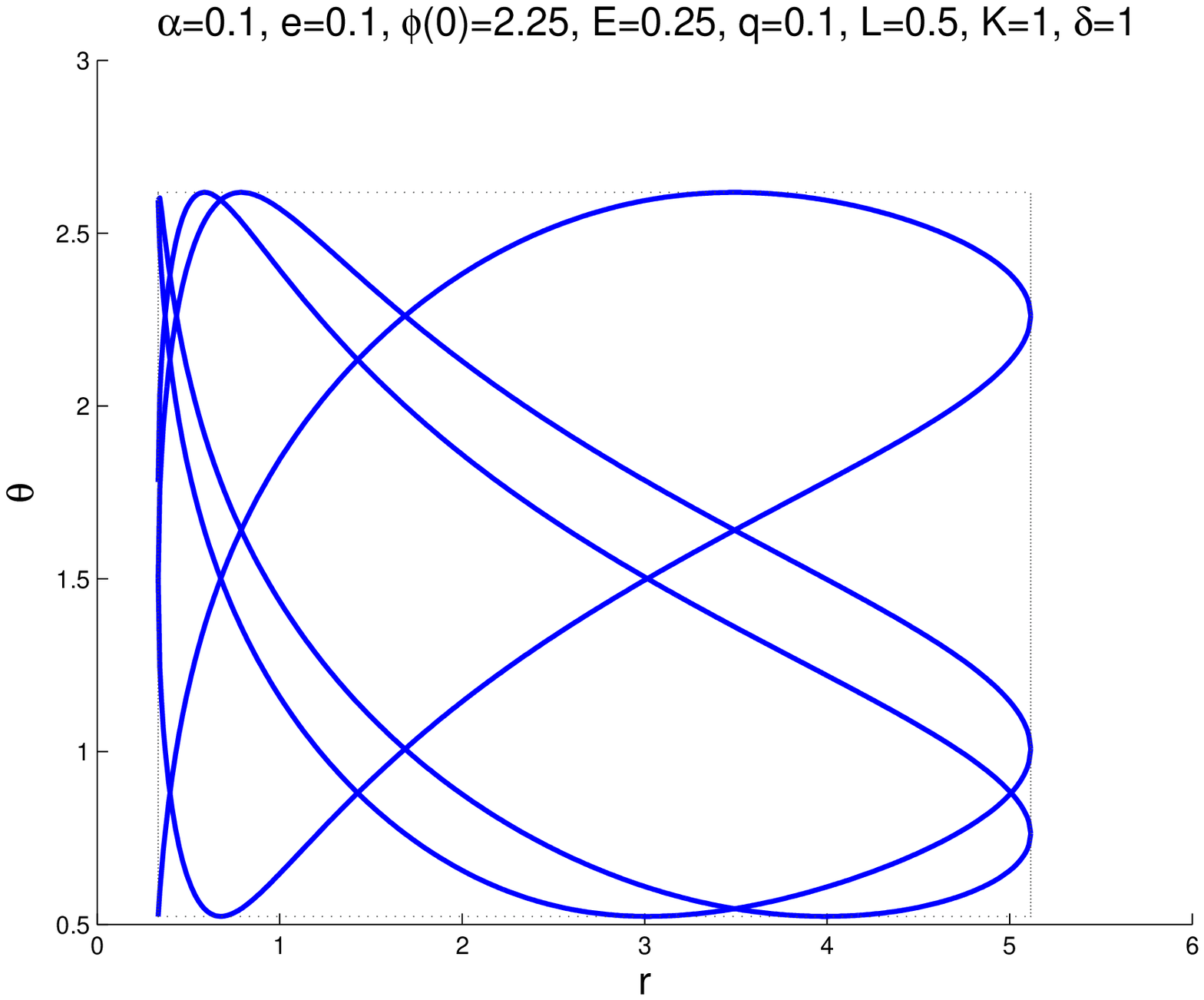}}
\subfigure[][BO for $E=0.13$]{\label{orb12}\includegraphics[width=6cm]{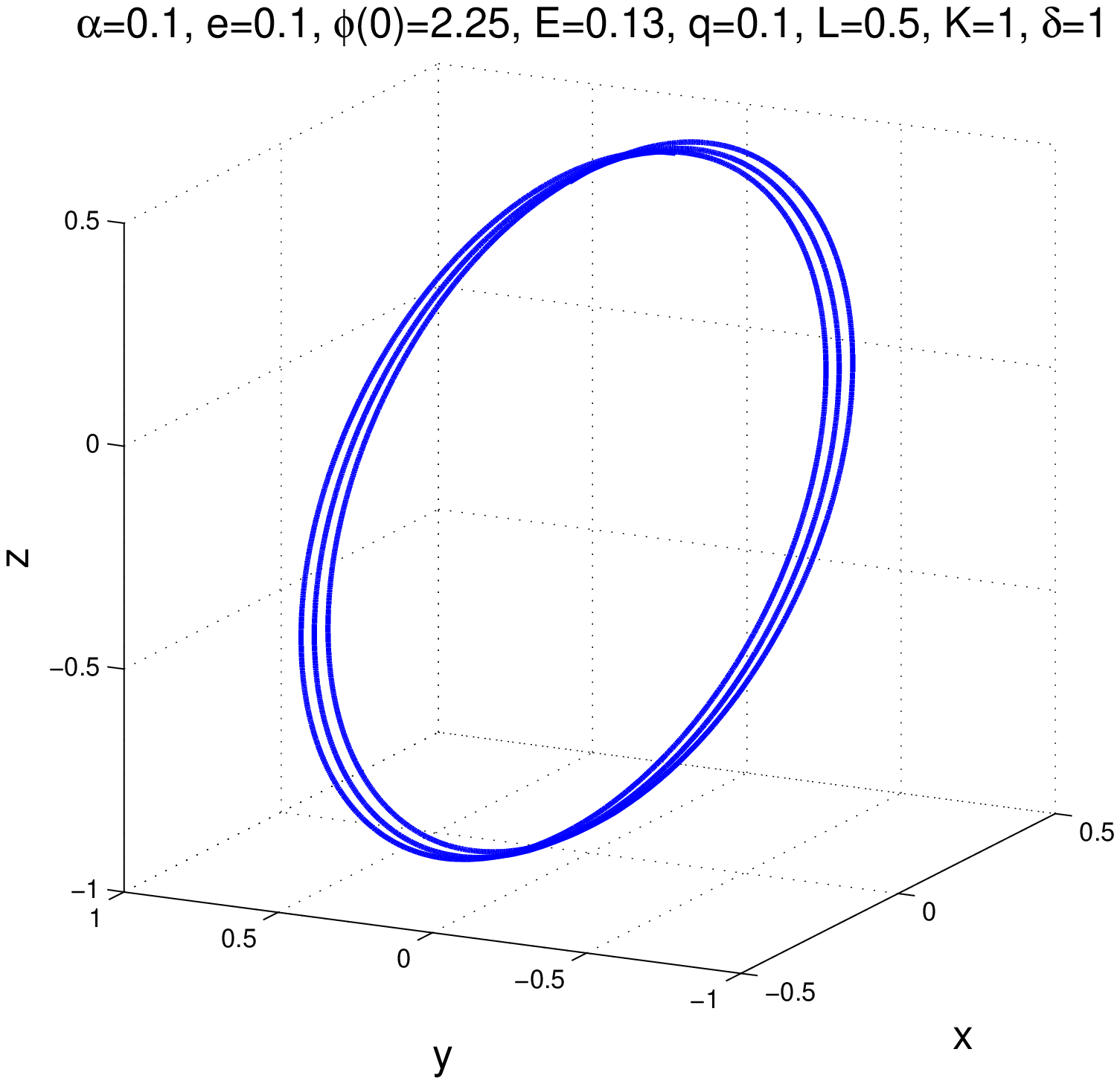}}
\subfigure[][$r$-$\theta$ plot for figure~\subref{orb12}]{\label{orb12-rth}\includegraphics[width=7cm]{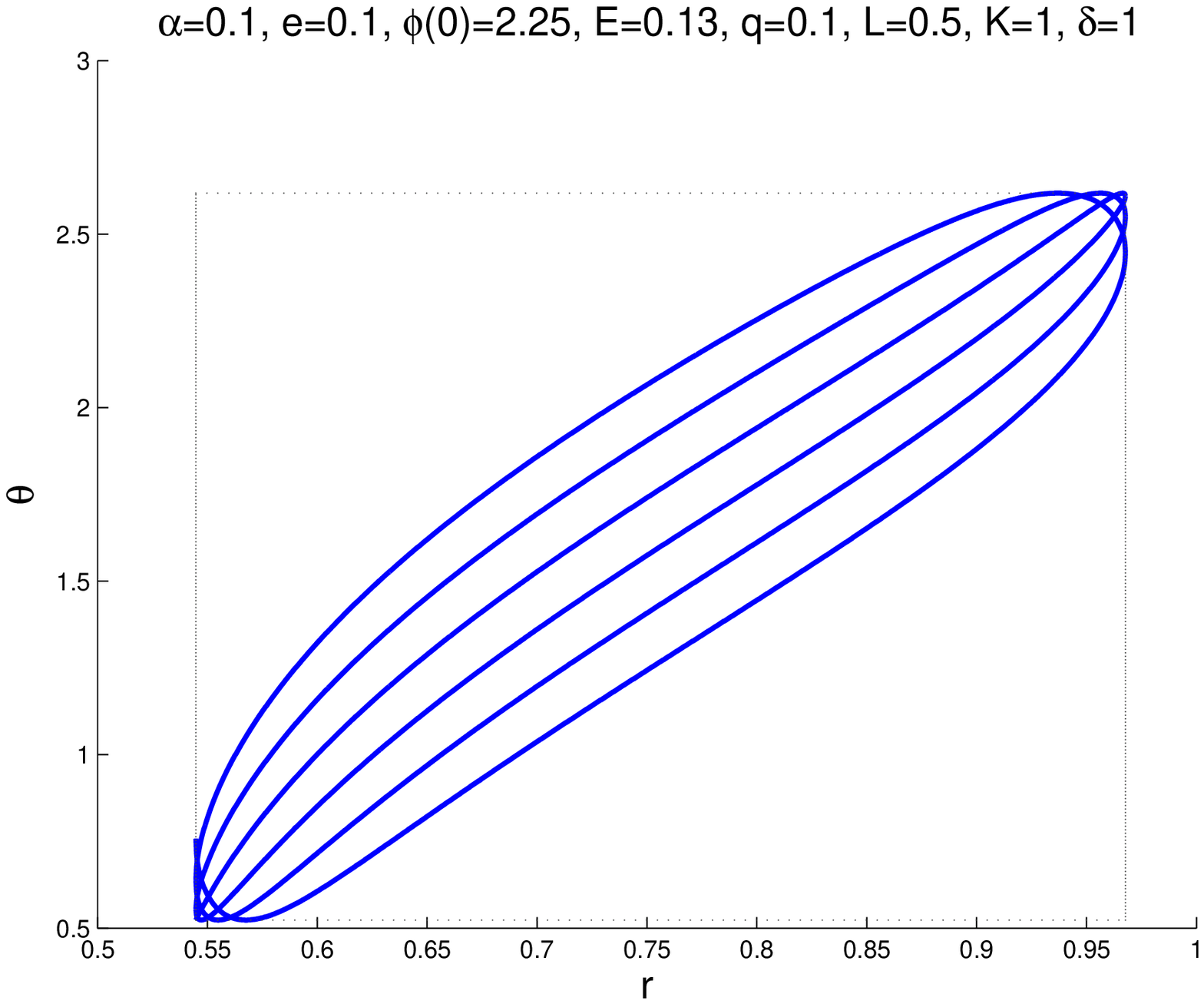}}
\subfigure[][EO for $E=0.33$]{\label{orb13}\includegraphics[width=6cm]{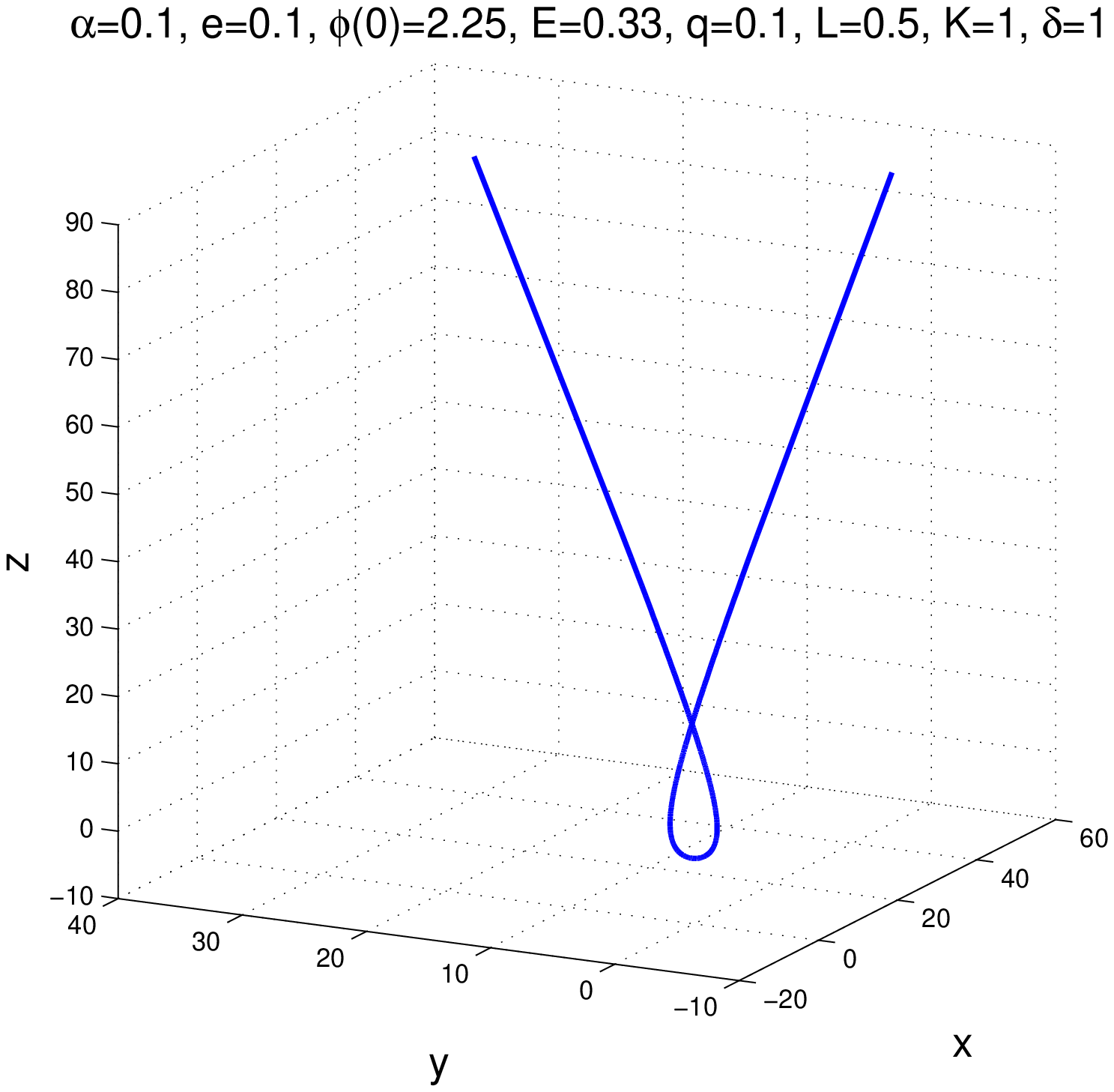}}
\subfigure[][EO for $E=0.5$]{\label{orb11}\includegraphics[width=6cm]{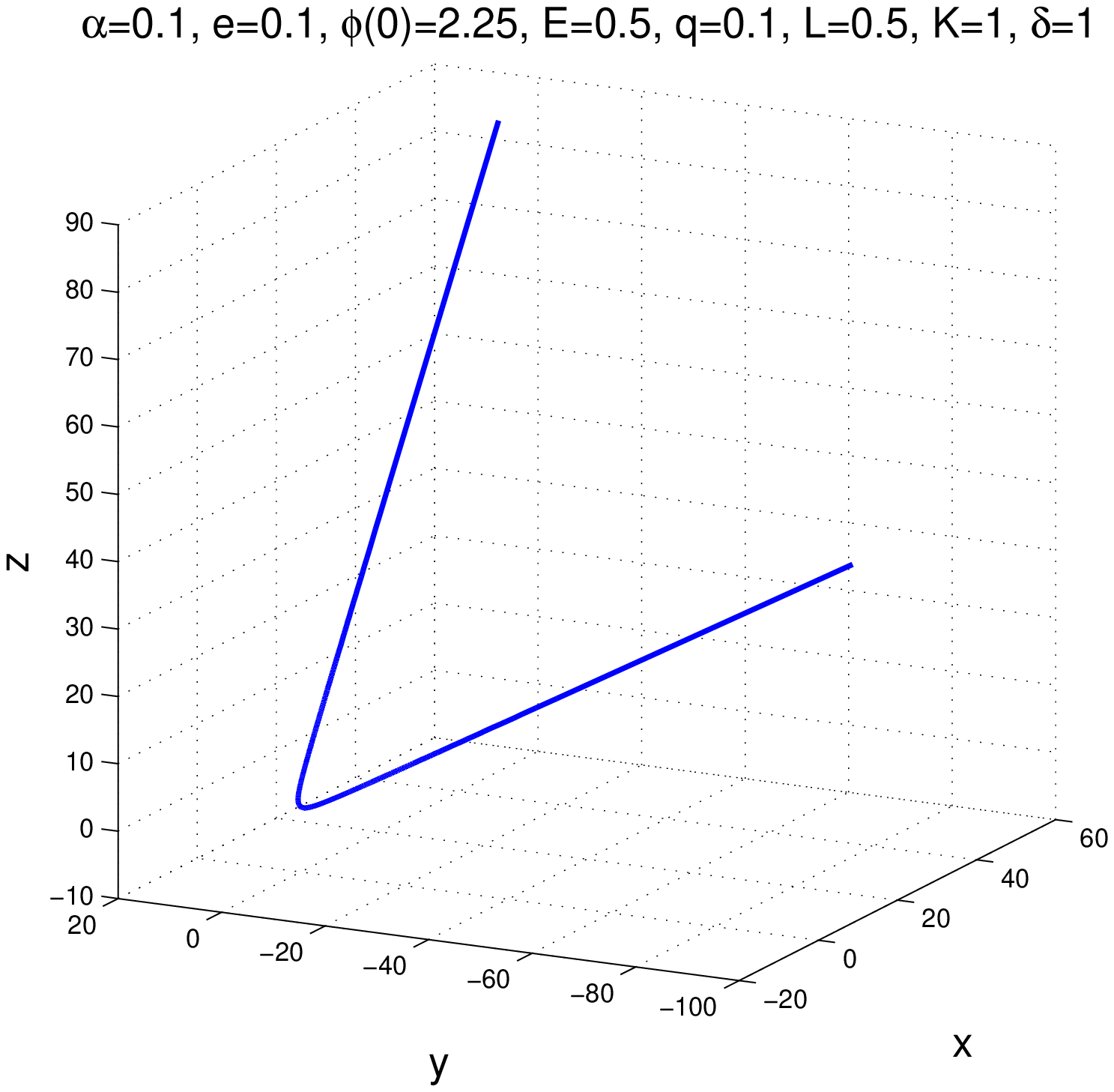}}
\end{center}
\caption{Motion of a charged, massive test particle with $q=0.1$, $L=0.5$ and $K=1.0$ in the space-time
of a boson star with $\alpha=0.1$, $e=0.1$ and $\phi(0)=2.25$. \label{fig:orb1}}
\end{figure*}

In the Fig.~\ref{fig:orb1}
we give an example of orbits for $\phi(0)=2.25$ which correspond to the effective potential in 
Fig.~\ref{potential_massive1}. In Fig.~\ref{orb1} and Fig.~\ref{orb12} we show bound orbits. 

\begin{figure*}[h!]
\begin{center}
\subfigure[][BO for $E=14$, $q=-2$]{\label{orb11_q-2}\includegraphics[width=6cm]{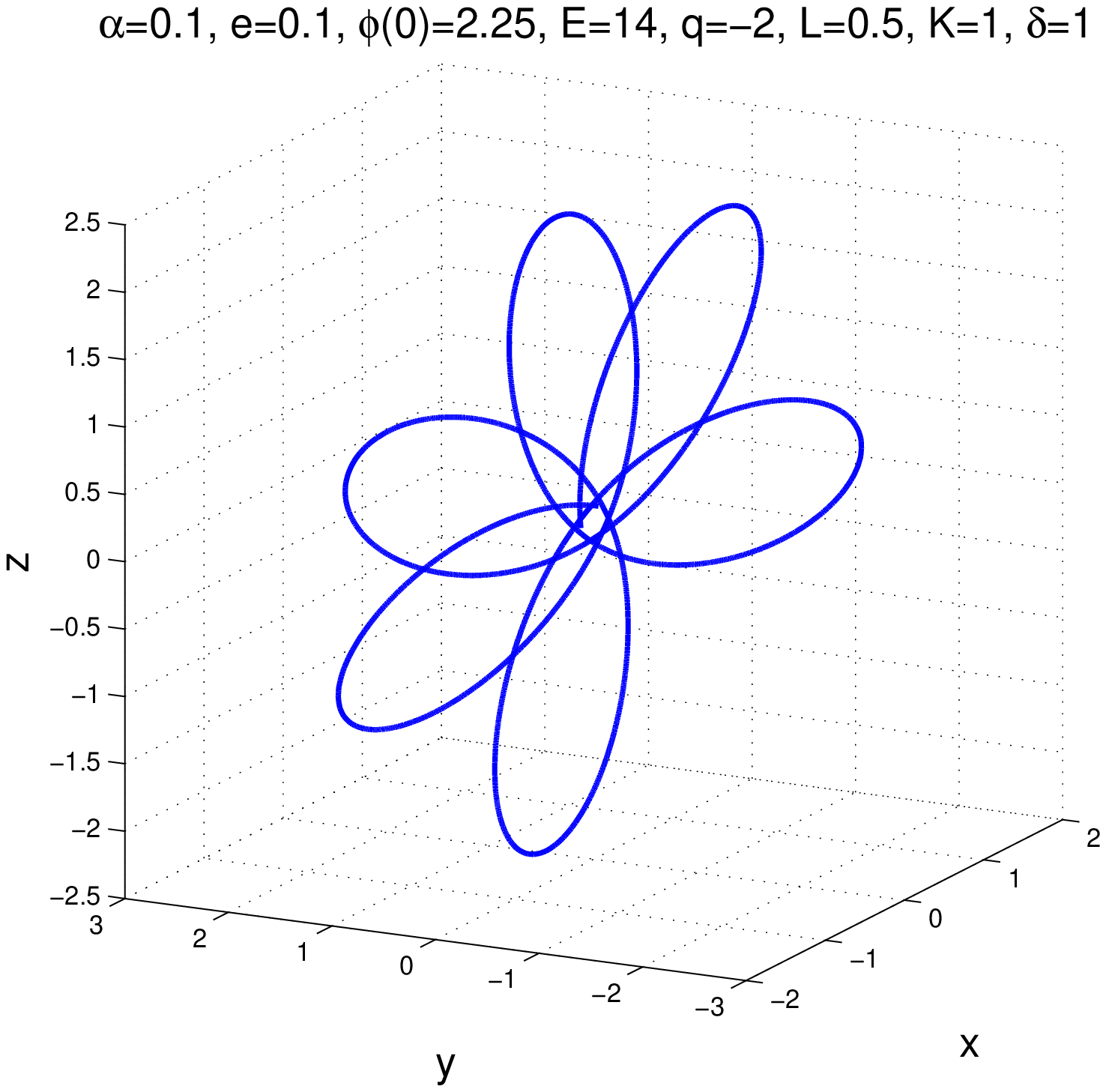}}
\subfigure[][BO for $E=28$, $q=-4$]{\label{orb11_q-4}\includegraphics[width=6cm]{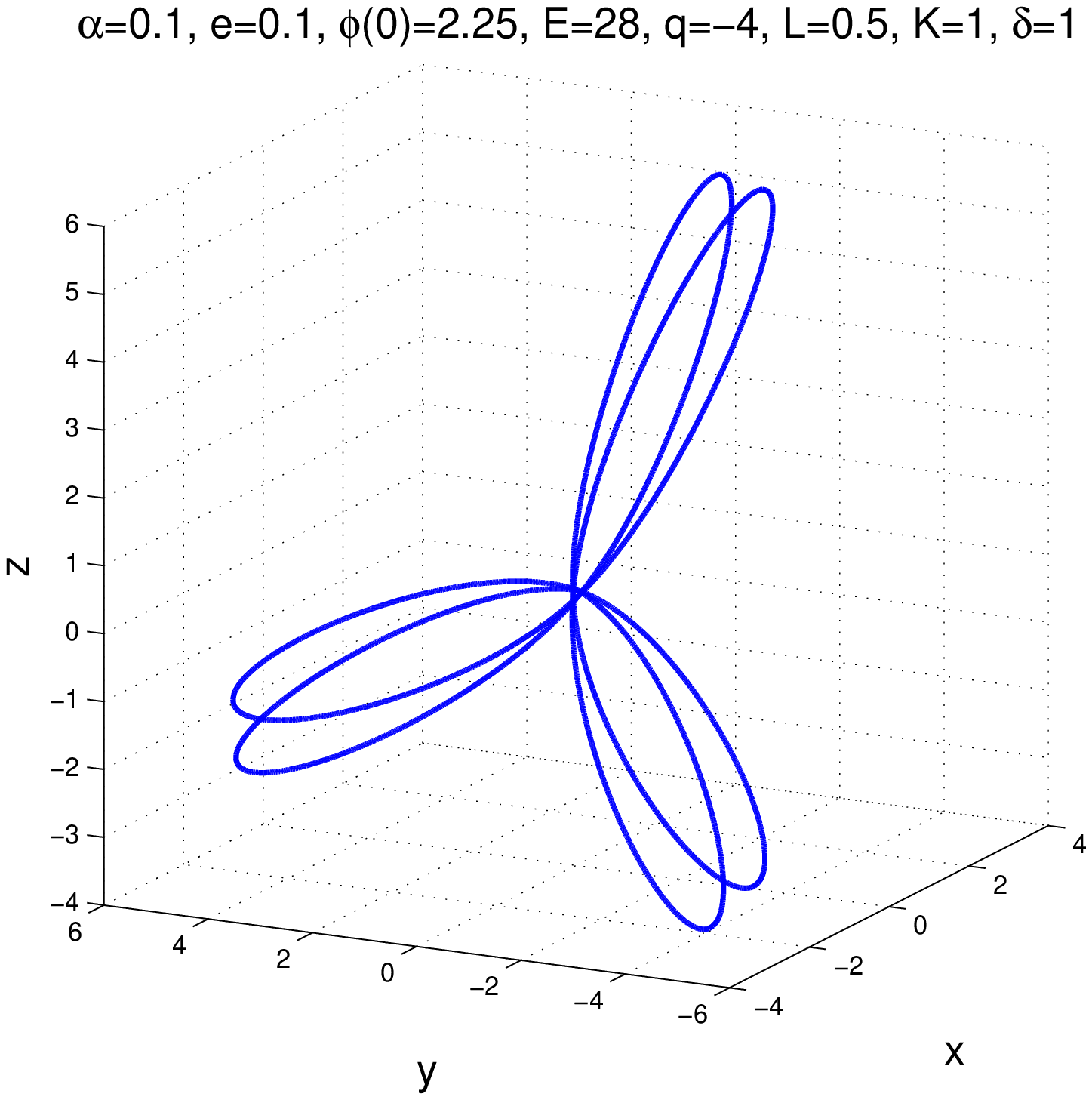}}
\end{center}
\caption{Motion of a charged, massive test particle with $L=0.5$ and $K=1.0$ in the space-time
of a boson star with $\alpha=0.1$, $e=0.1$ and $\phi(0)=2.25$.  \label{fig:orb11_q}}
\end{figure*}

In Fig.~\ref{orb12} we give the motion of test particles with smaller energies. 
In Figs.~\ref{orb1-rth} and~\ref{orb12-rth} 
the corresponding $r$-$\theta$ motion is shown. 
Two escape orbits for larger values of $E$ are presented in the Figs.~\ref{orb11} and~\ref{orb13}. 
We comparing the orbits we find that for larger values of $E$ the orbit has a simple parabolic form, 
whereas the particle with lower energy moves around the boson star, slightly changes its 
direction before going to infinity again. 
In Fig.~\ref{fig:orb11_q} we show bound orbits for small negative charges of the test particle. 
A bound orbit for $q=-2$ is given in Fig.~\ref{orb11_q-2} and corresponds to the potential in 
Fig.~\ref{potential_massive11_q-2}. For $q=-4$ a corresponding bound orbit is shown in 
Fig.~\ref{orb11_q-4} for the effective potential in Fig.~\ref{potential_massive11_q-4}.

All orbits are confined to a plane which is inclined with respect to the equatorial plane. 
This feature is peculiar to charged boson star space-time and can be observed in all cases 
considered below. Even a large absolute value of a charge of a test particle does not tear 
the orbit out of a plane as it is known from the charged particle motion in 
the Reissner-Nordstr\"om space-time~\cite{GruKa}. 

\begin{figure*}[h!]
\begin{center}
\subfigure[][BO for $E=0.25$]{\label{orb2}\includegraphics[width=6cm]{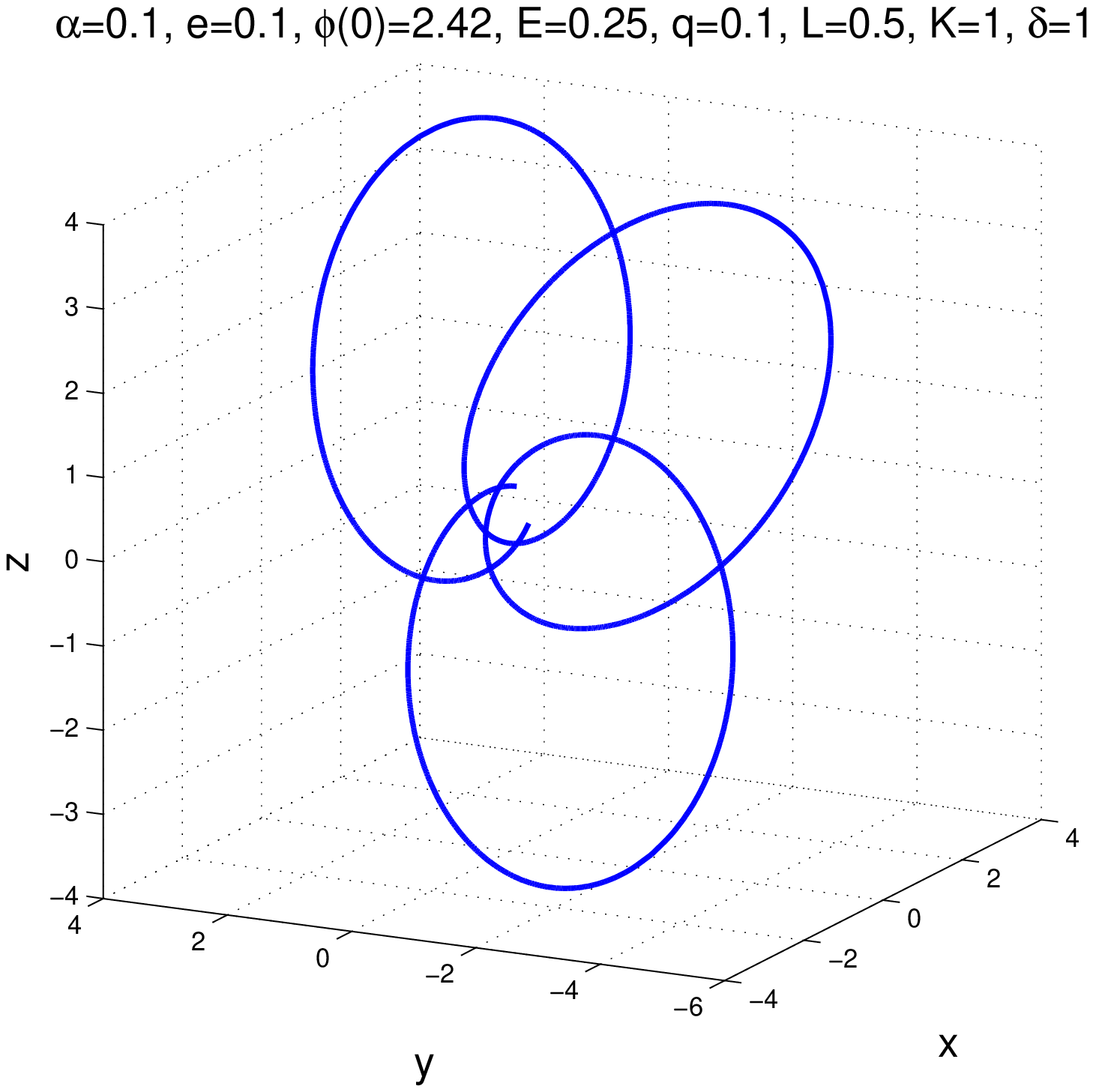}}
\subfigure[][BO for $E=0.13$]{\label{orb21}\includegraphics[width=6cm]{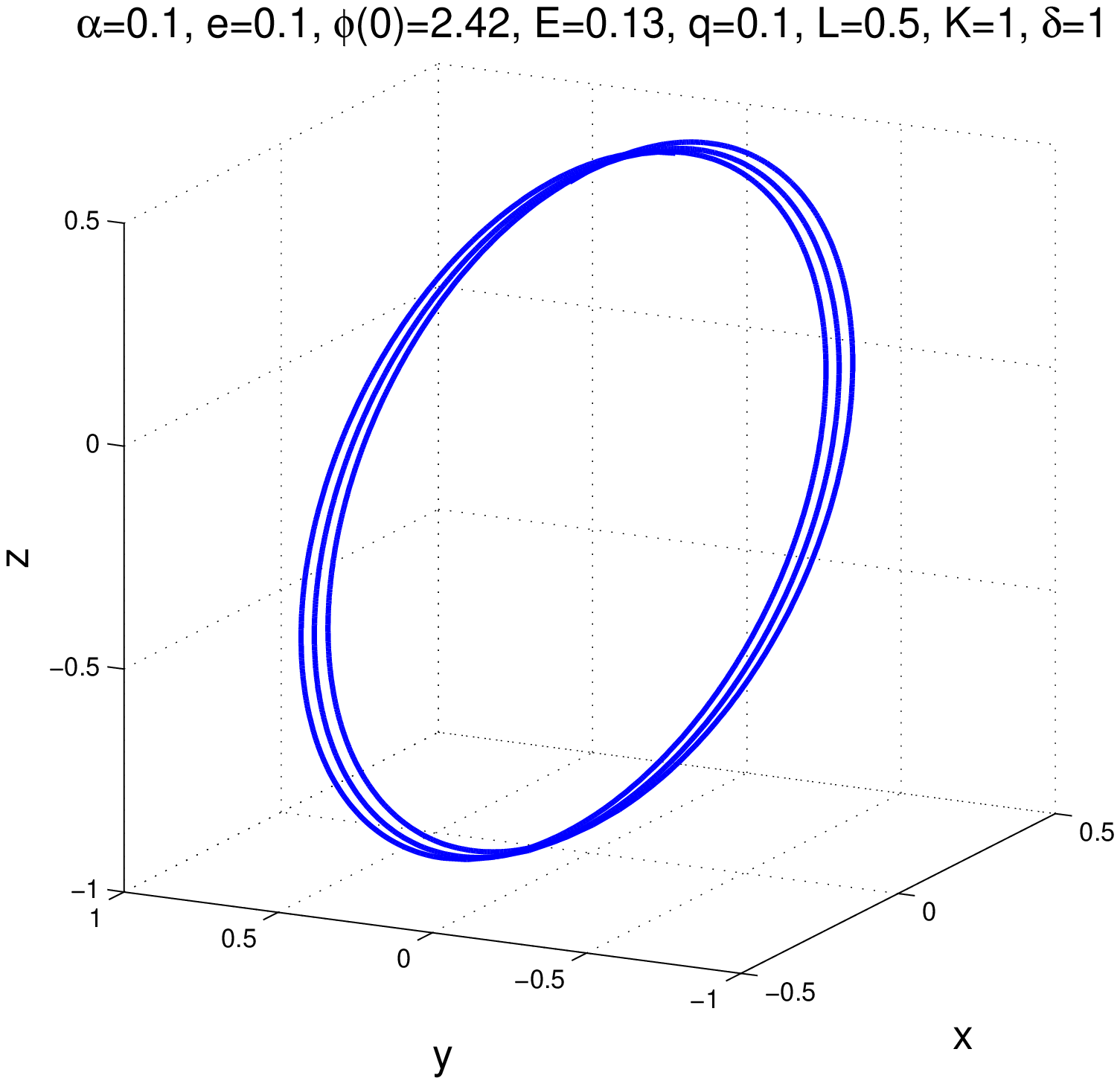}}
\subfigure[][EO for $E=0.5$]{\label{orb22}\includegraphics[width=6cm]{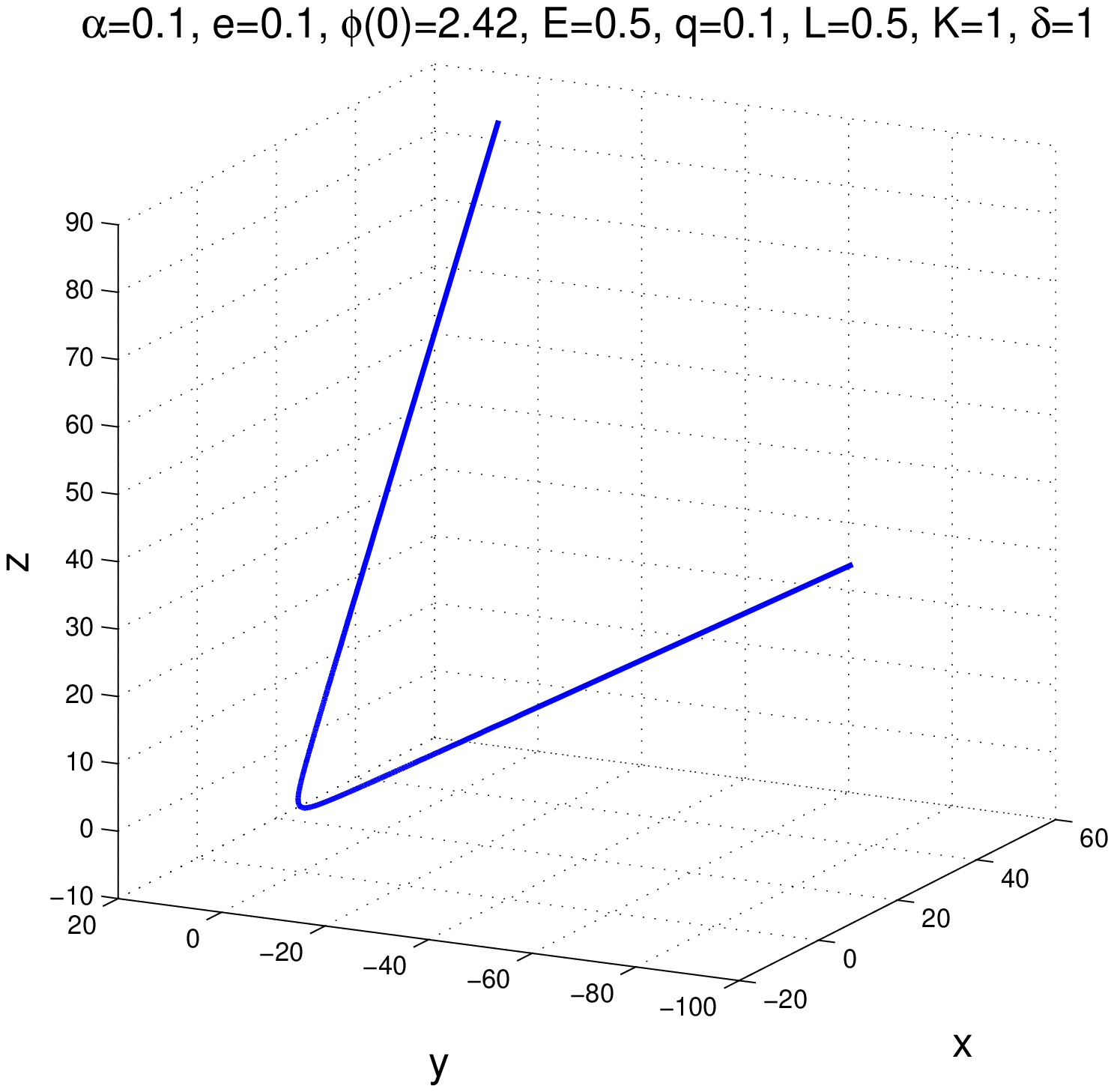}}
\subfigure[][EO for $E=0.33$]{\label{orb23}\includegraphics[width=6cm]{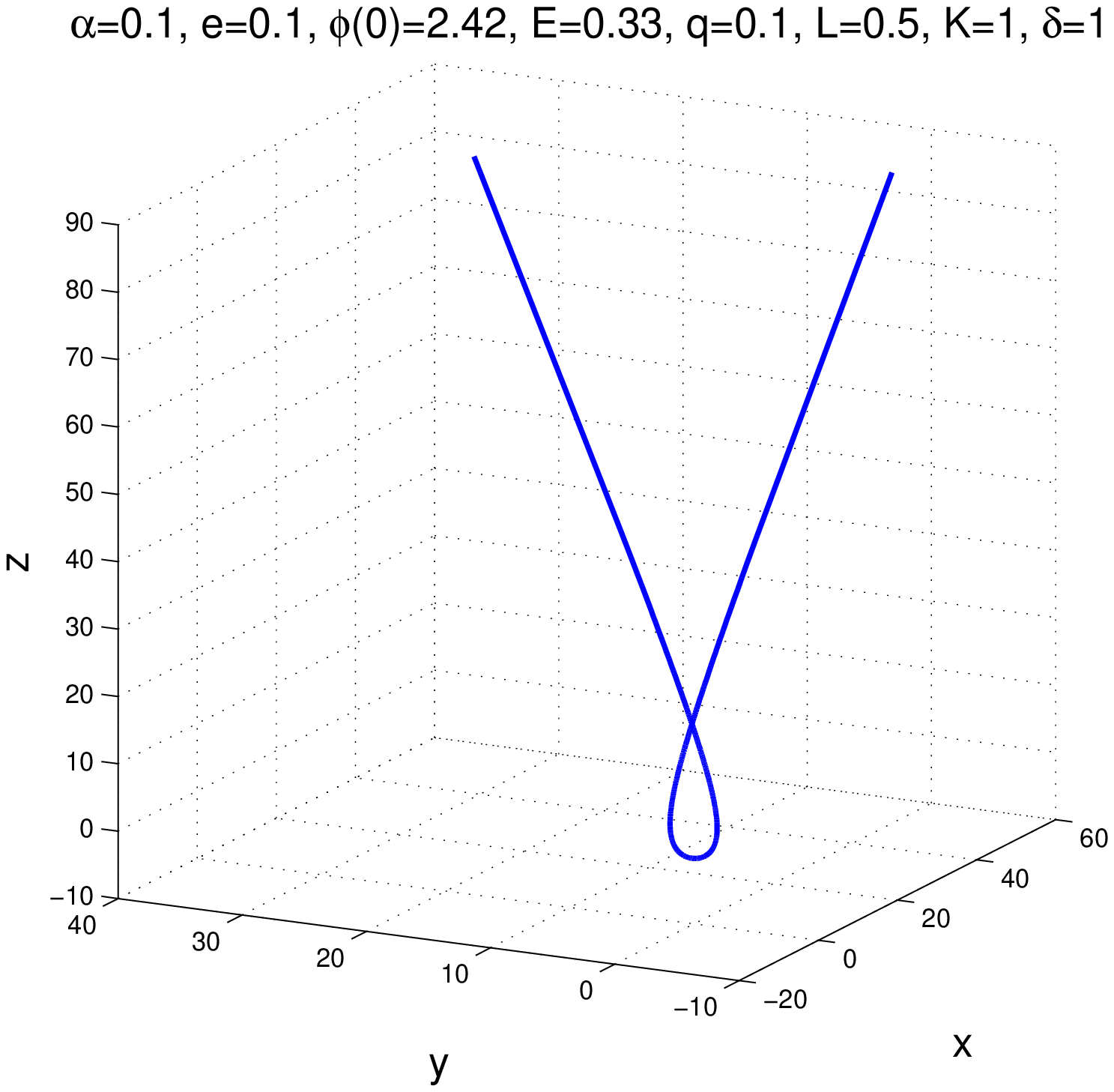}}
\end{center}
\caption{Motion of a charged, massive test particle with $q=0.1$, $L=0.5$ and $K=1.0$ in the space-time
of a boson star with $\alpha=0.1$, $e=0.1$ and $\phi(0)=2.42$.   \label{fig:orb2}}
\end{figure*}

A small increase of $\phi(0)$ in the potentials for $\phi(0)=2.25$ and $\phi(0)=2.42$ in 
Figs.~\ref{potential_massive1} and~\ref{potential_massive2} does not influence the potentials visibly.
Hence, the case of maximal mass (and charge) does not differ strongly from that of minimal internal frequency
$\omega$ at $\phi(0)=2.42$. This is apparent when comparing Fig.~\ref{fig:orb2}
with Fig.~\ref{fig:orb1}.

 but have closely lying values of 
turning points as can be seen from the Table~\ref{rminmax}.

\begin{figure*}[h!]
\begin{center}
\subfigure[][BO for $E=0.03$]{\label{orb3}\includegraphics[width=6cm]{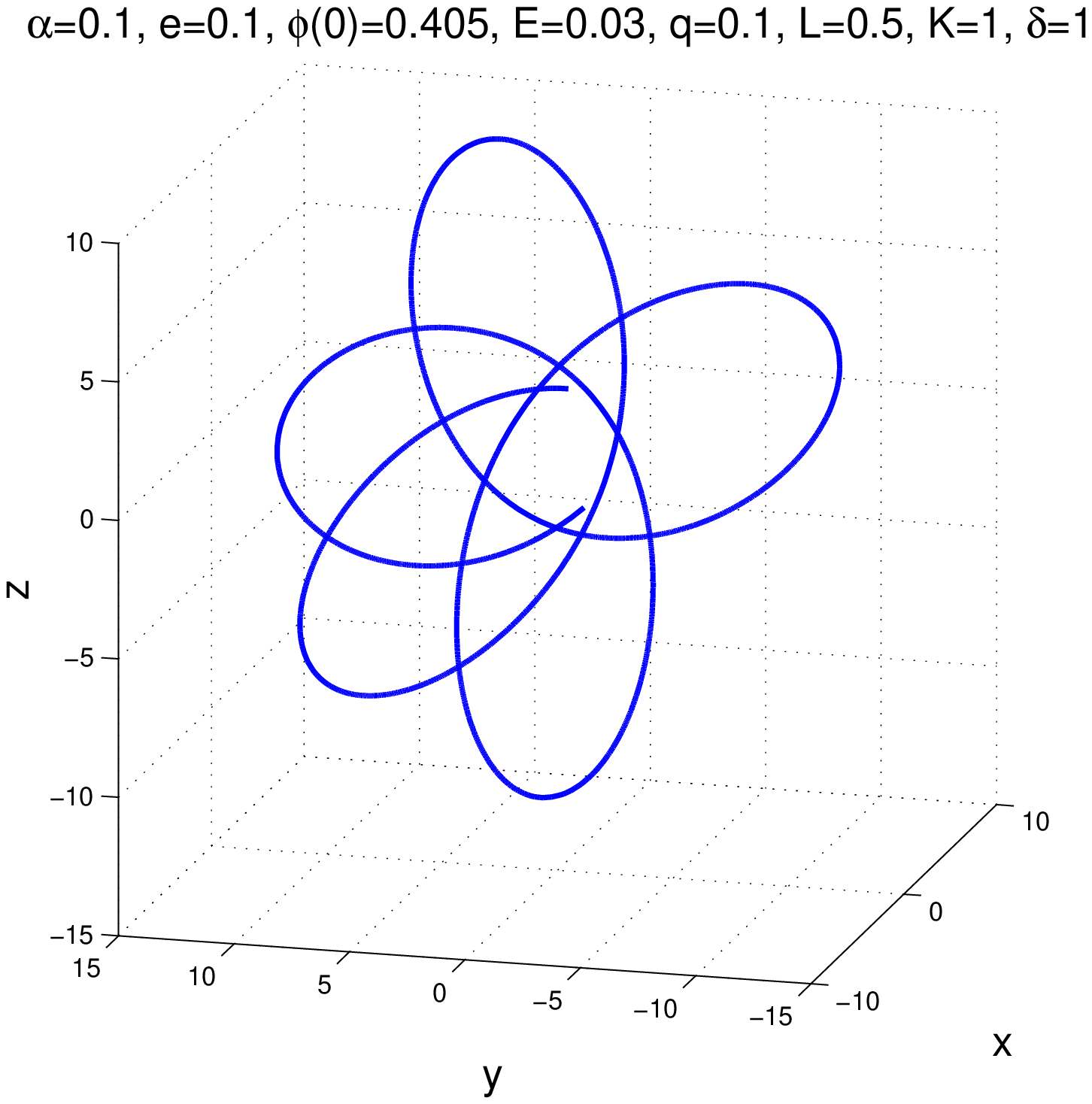}}
\subfigure[][$r$-$\theta$ plot for figure~\subref{orb3}]{\label{orb3-rth}\includegraphics[width=7cm]{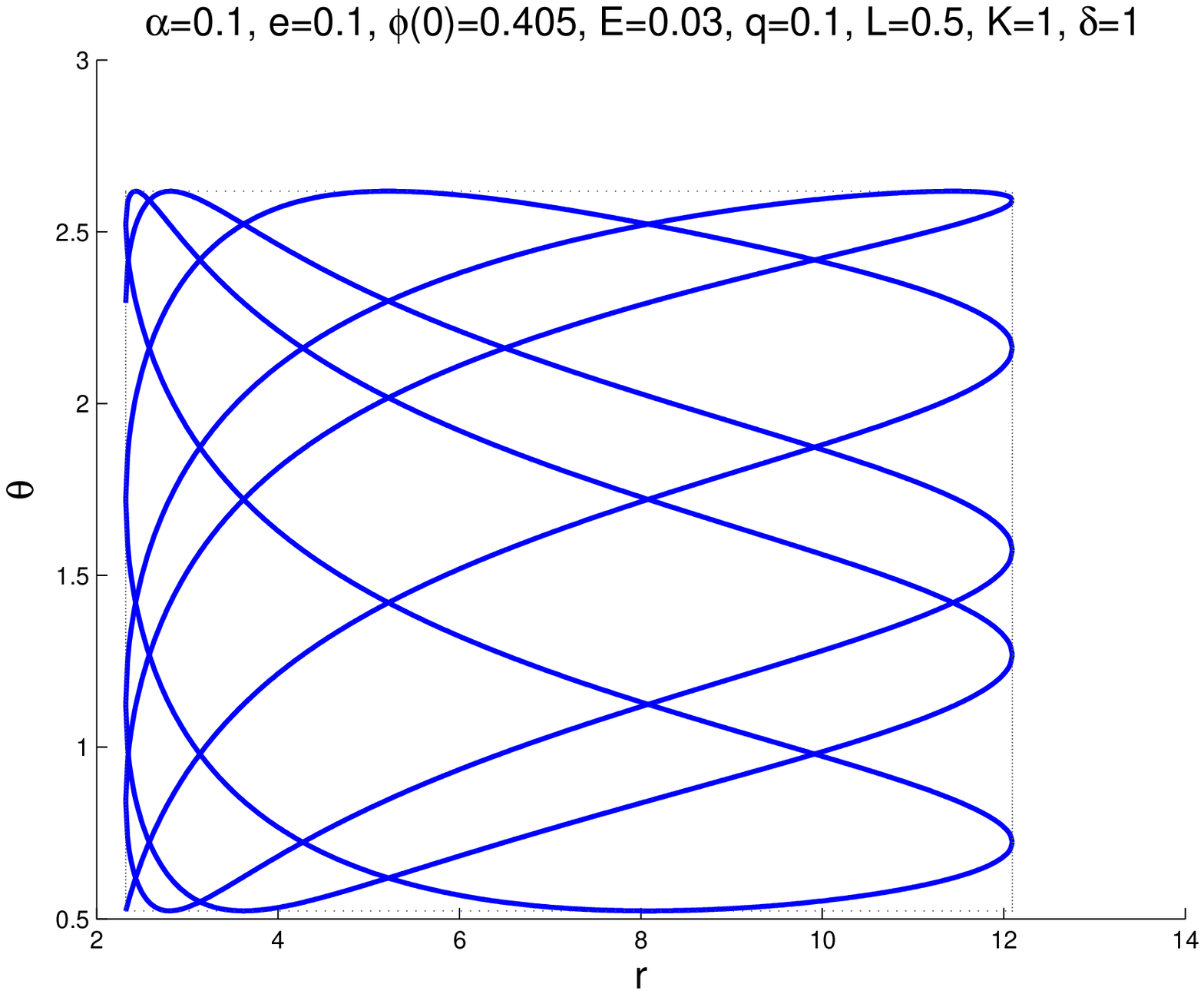}}
\subfigure[][BO for $E=0.008$]{\label{orb31}\includegraphics[width=6cm]{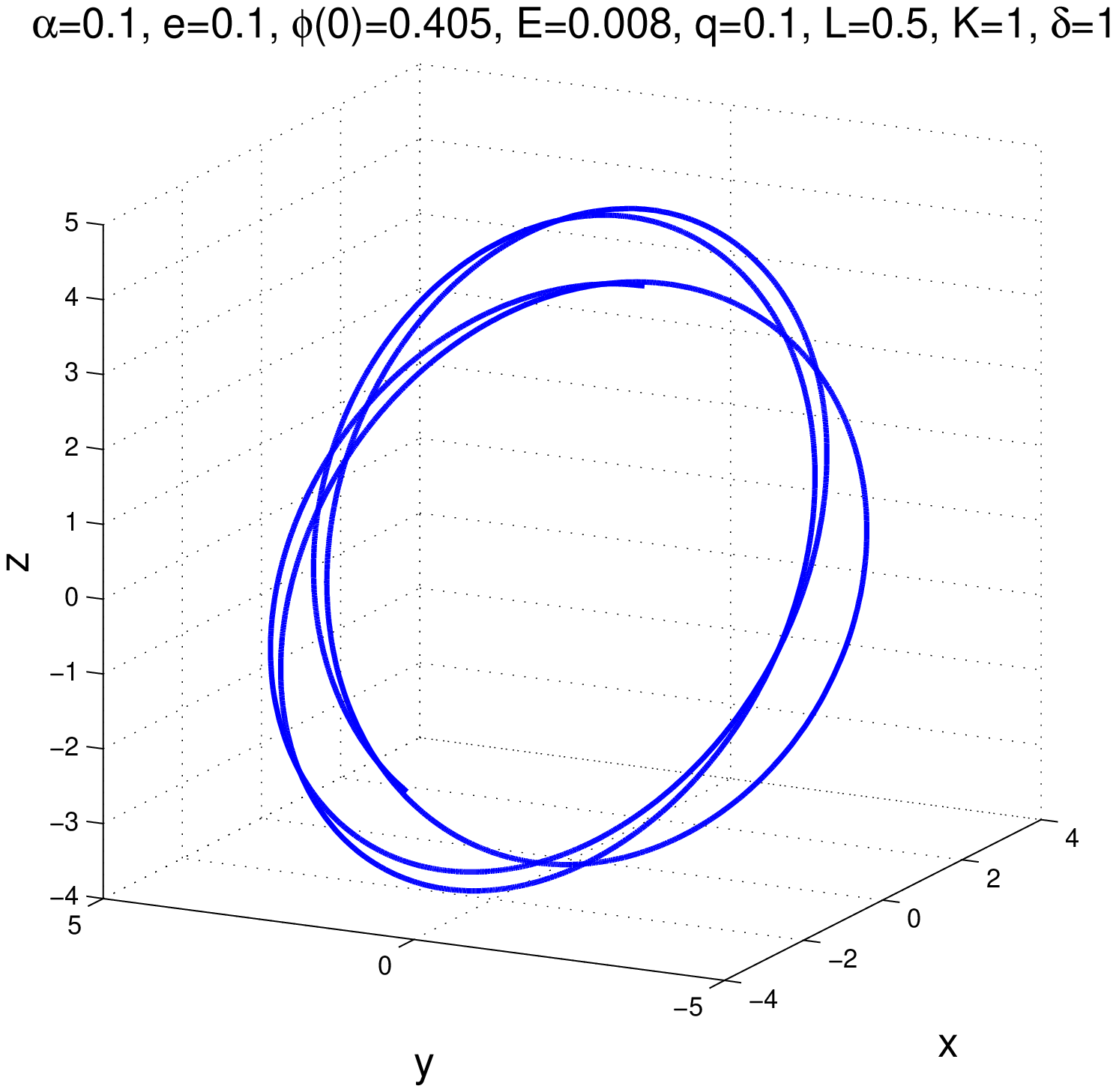}}
\subfigure[][$r$-$\theta$ plot for figure~\subref{orb31}]{\label{orb31-rth}\includegraphics[width=7cm]{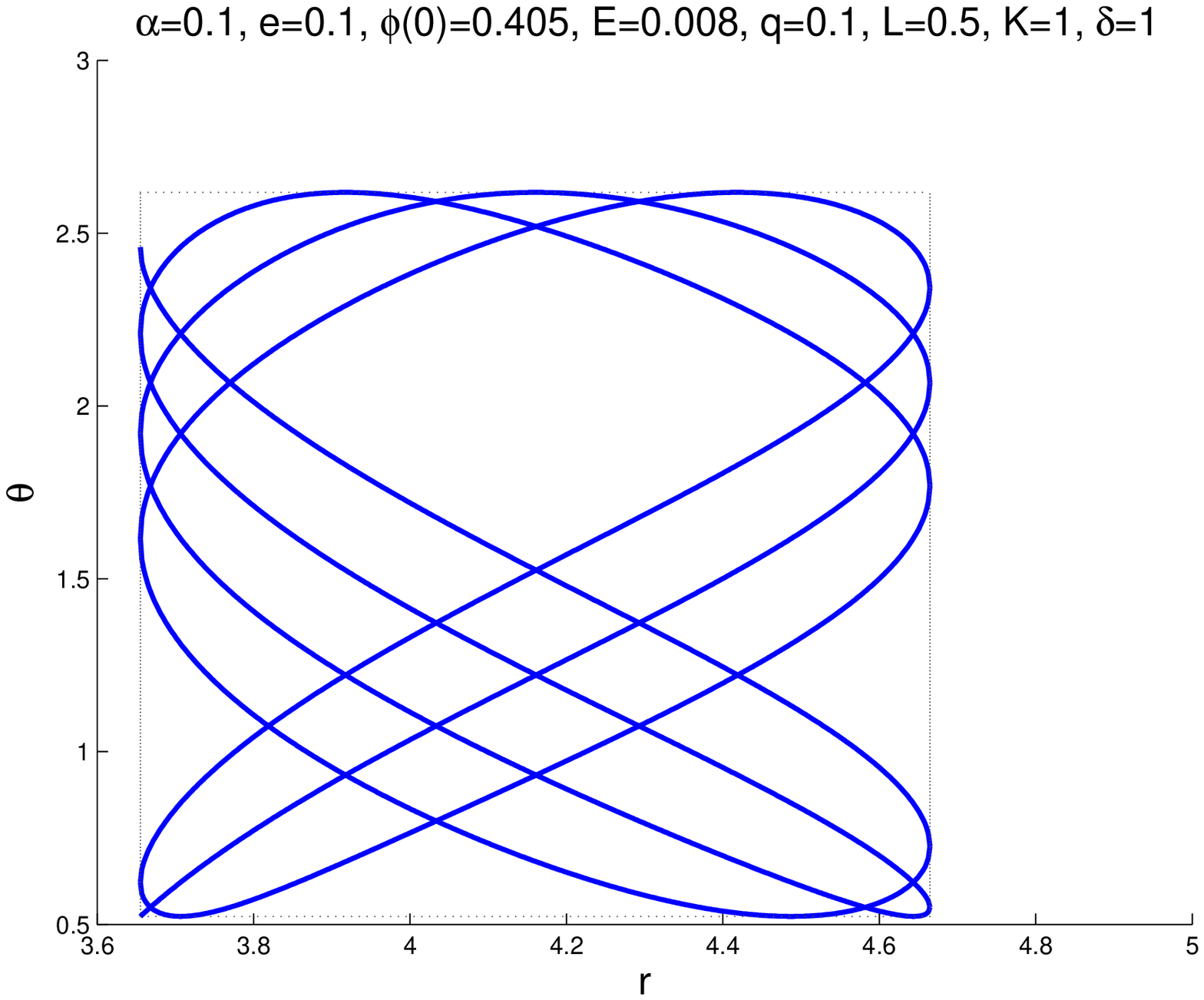}}
\subfigure[][EO for $E=0.5$]{\label{orb32}\includegraphics[width=6cm]{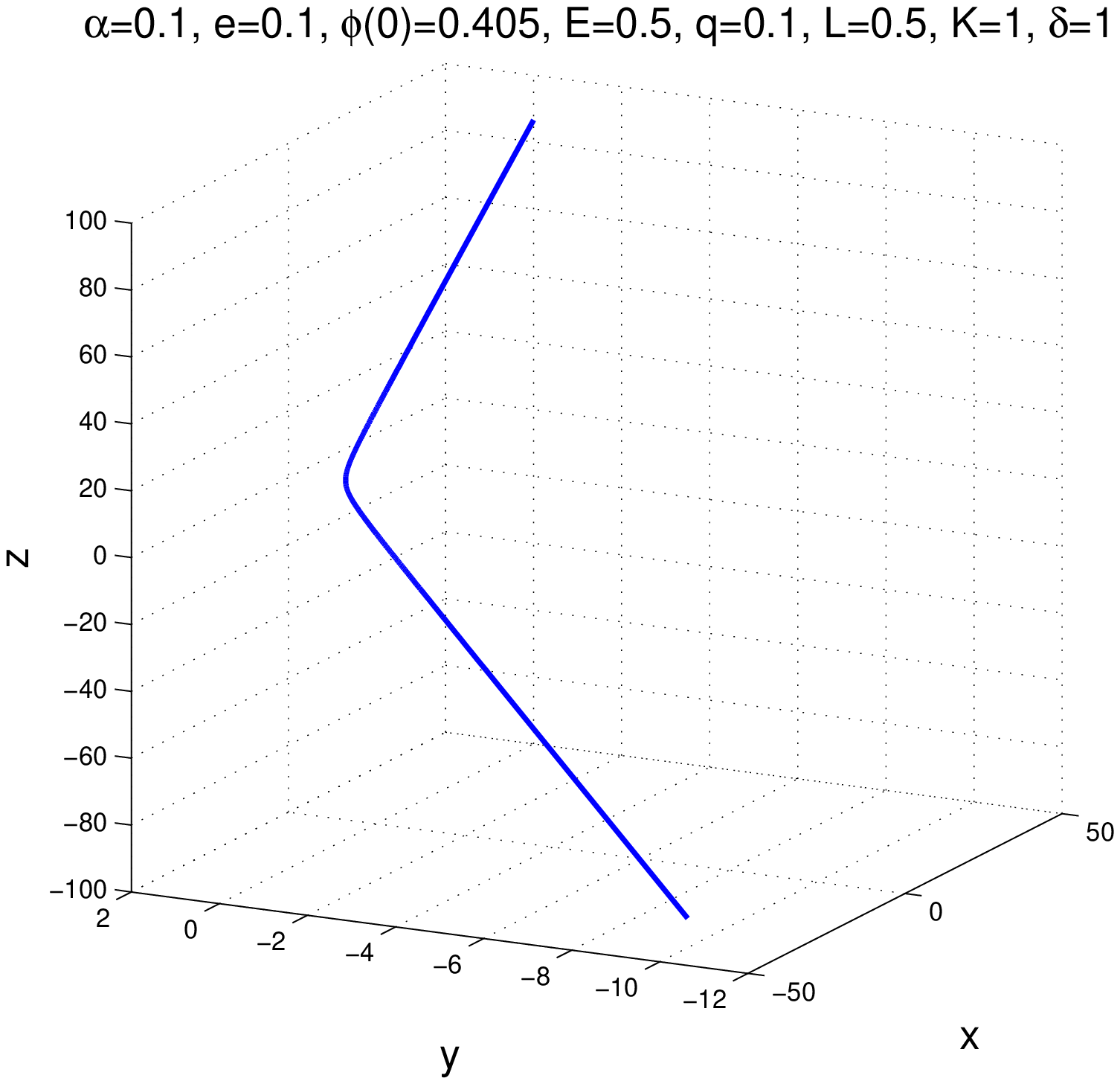}}
\subfigure[][EO for $E=0.065$]{\label{orb33}\includegraphics[width=6cm]{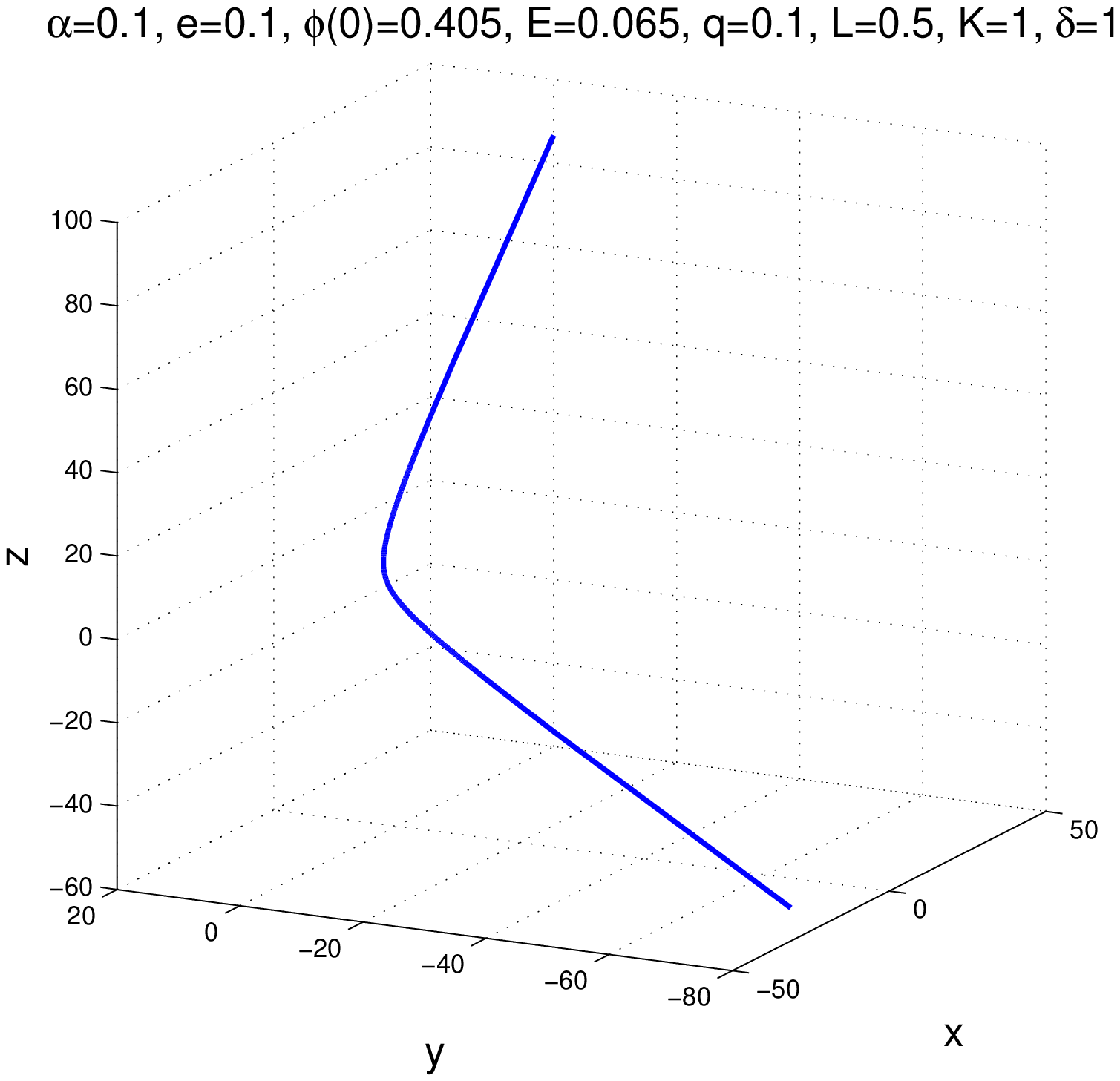}}
\end{center}
\caption{Motion of a charged, massive test particle with $q=0.1$, $L=0.5$ and $K=1.0$ in the space-time
of a boson star with $\alpha=0.1$, $e=0.1$ and $\phi(0)=0.405$.   \label{fig:orb3}}
\end{figure*}

For small value of $\phi(0)$ ($\phi(0)=0.405$) bound orbits are possible 
for very small positive values of the energy $E$ as shown in Fig.~\ref{orb3} and Fig.~\ref{orb31}. 
Figs.~\ref{orb3-rth} and~\ref{orb31-rth} 
illustrate the $r-\theta$ motion of these orbits. 
The escape shown in Fig.~\ref{orb32} and Fig.~\ref{orb33} have parabolic form. 
The corresponding effective potential is shown in Fig.~\ref{potential_massive3}.

\begin{figure*}[h!]
\begin{center}
\subfigure[][BO for $E=0.19$]{\label{orb4}\includegraphics[width=6cm]{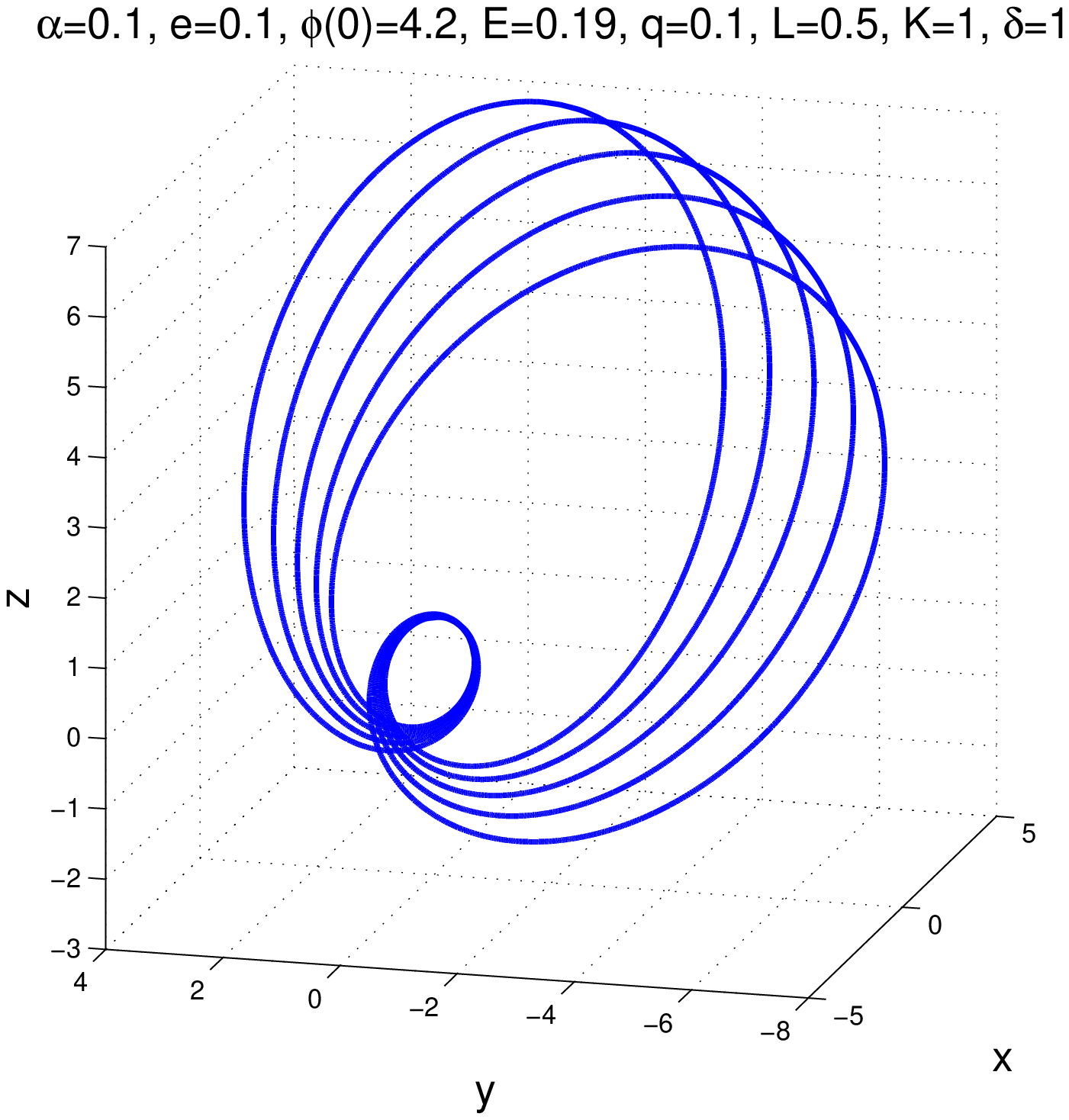}}
\subfigure[][$r$-$\theta$ plot for figure~\subref{orb3}]{\label{orb4-rth}\includegraphics[width=7cm]{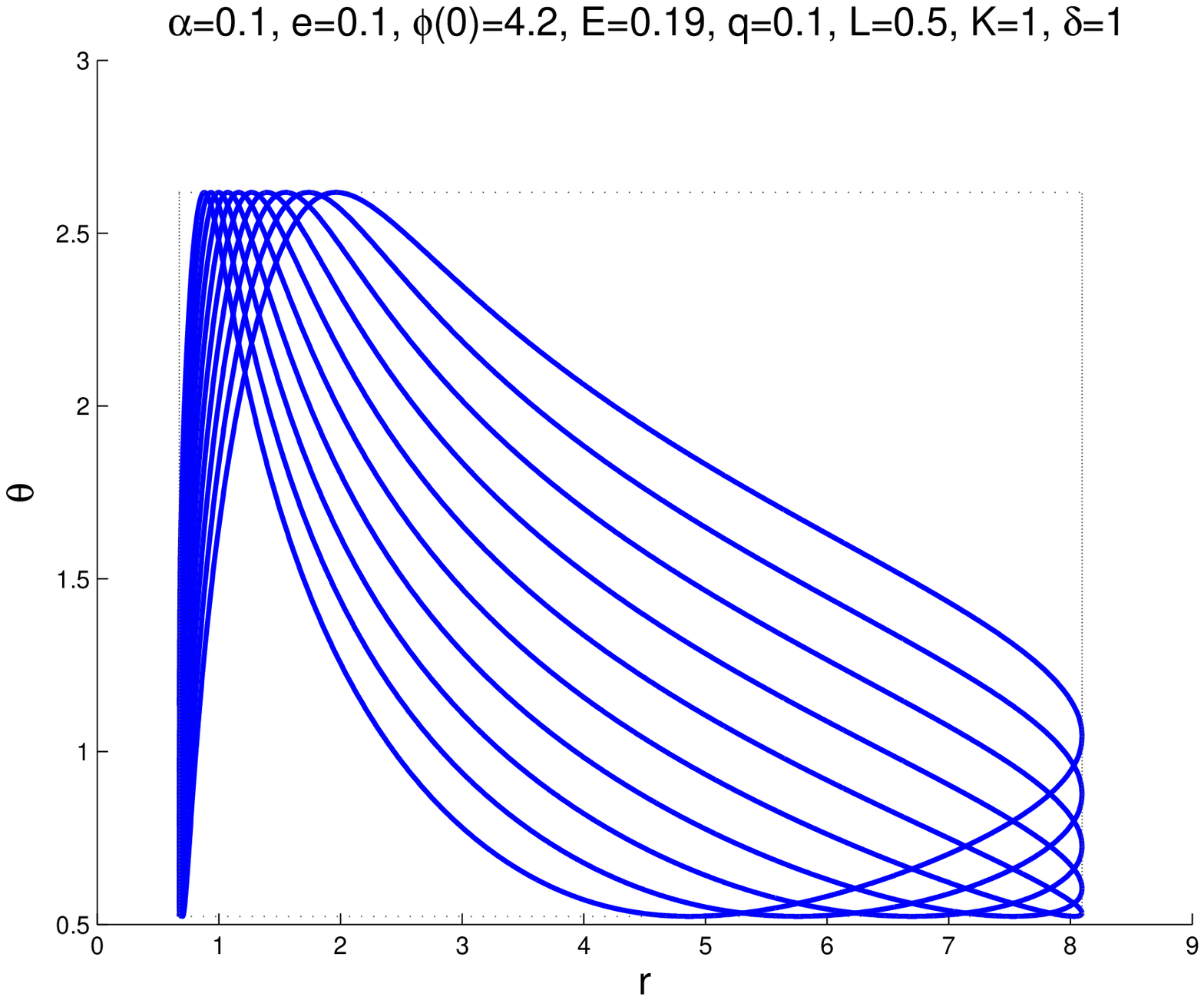}}
\subfigure[][BO for $E=0.1635$]{\label{orb41}\includegraphics[width=6cm]{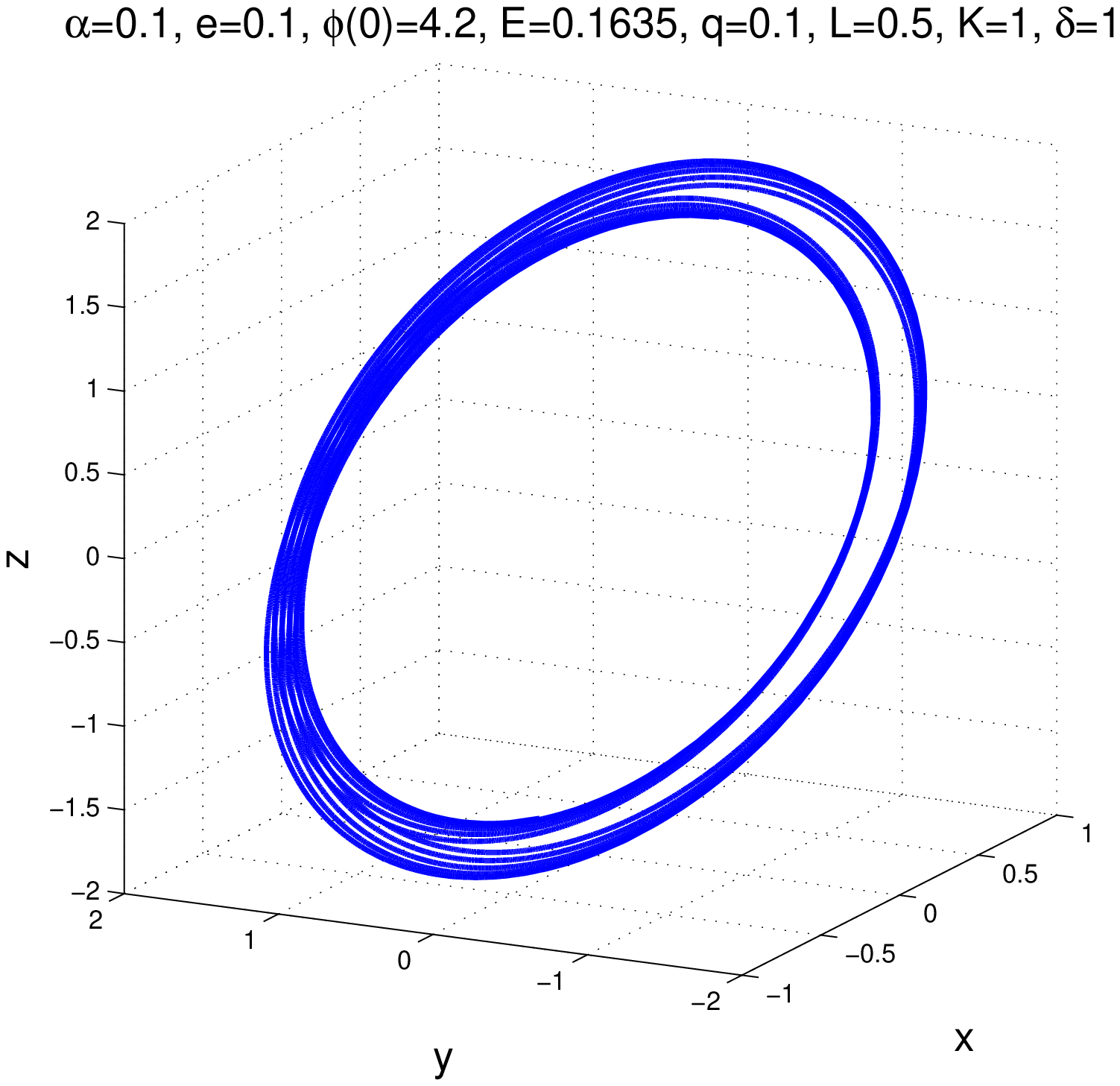}}
\subfigure[][$r$-$\theta$ plot for figure~\subref{orb31}]{\label{orb41-rth}\includegraphics[width=7cm]{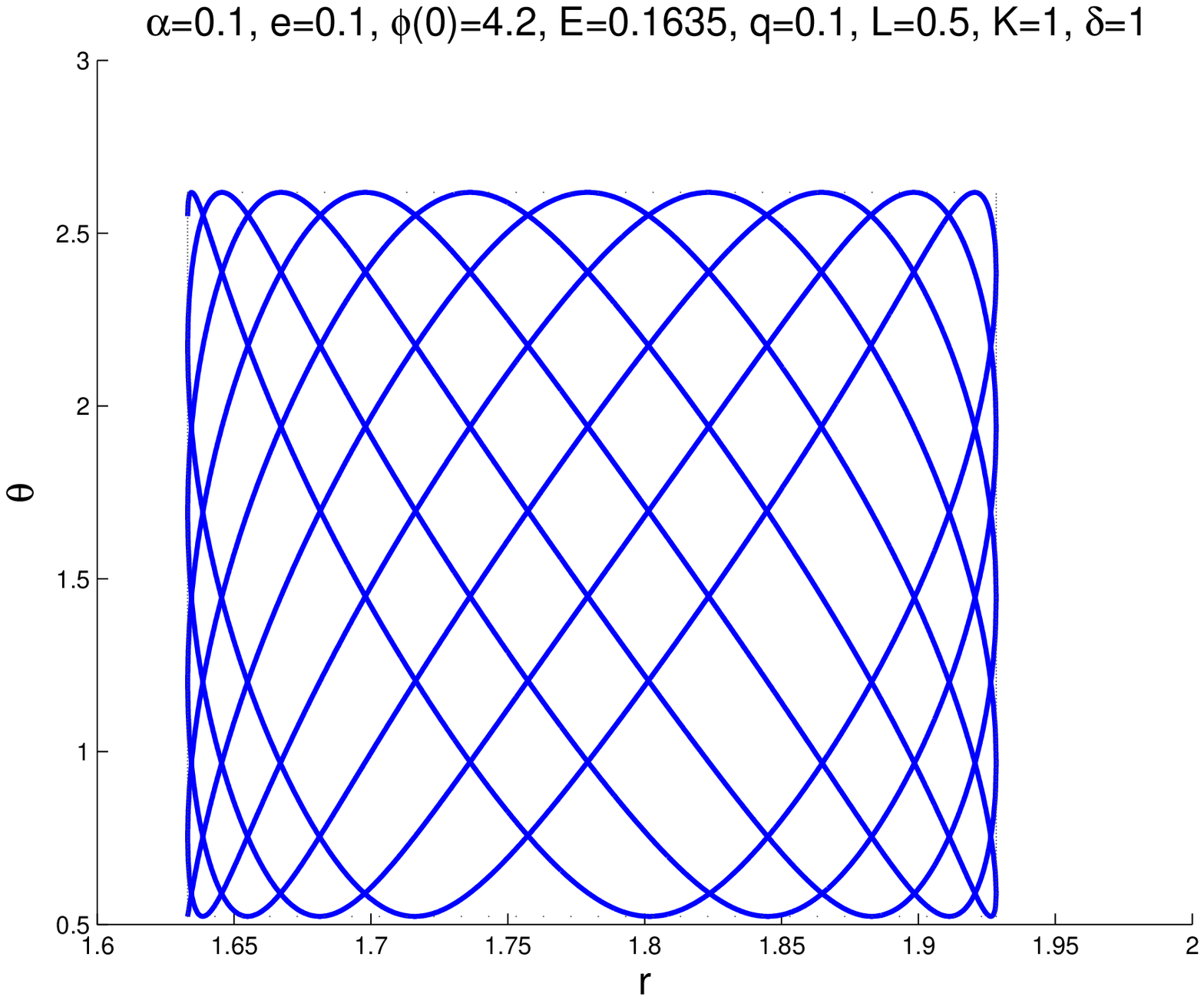}}
\subfigure[][EO for $E=0.4$]{\label{orb42}\includegraphics[width=6cm]{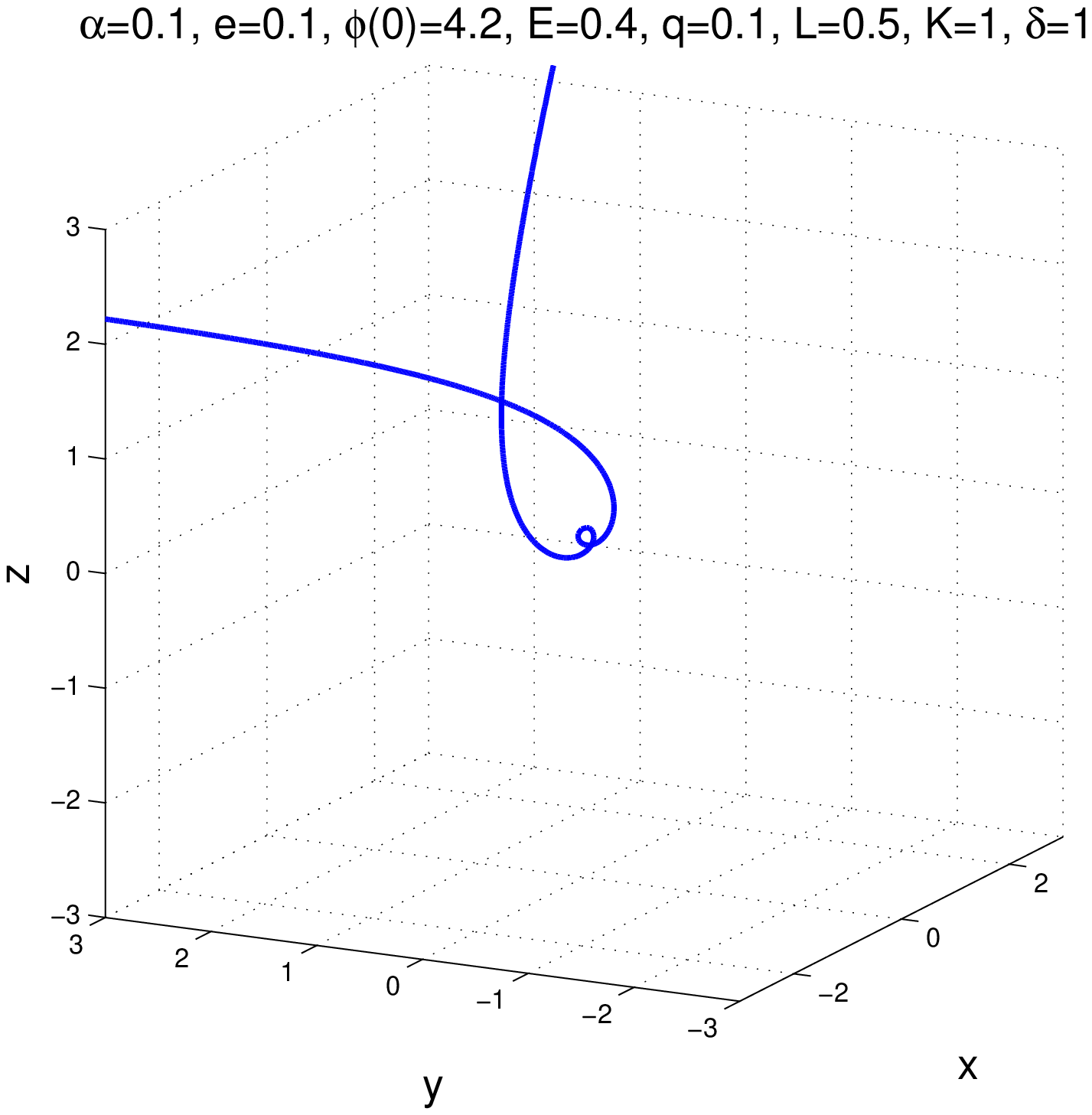}}
\subfigure[][EO for $E=0.3$]{\label{orb43}\includegraphics[width=6cm]{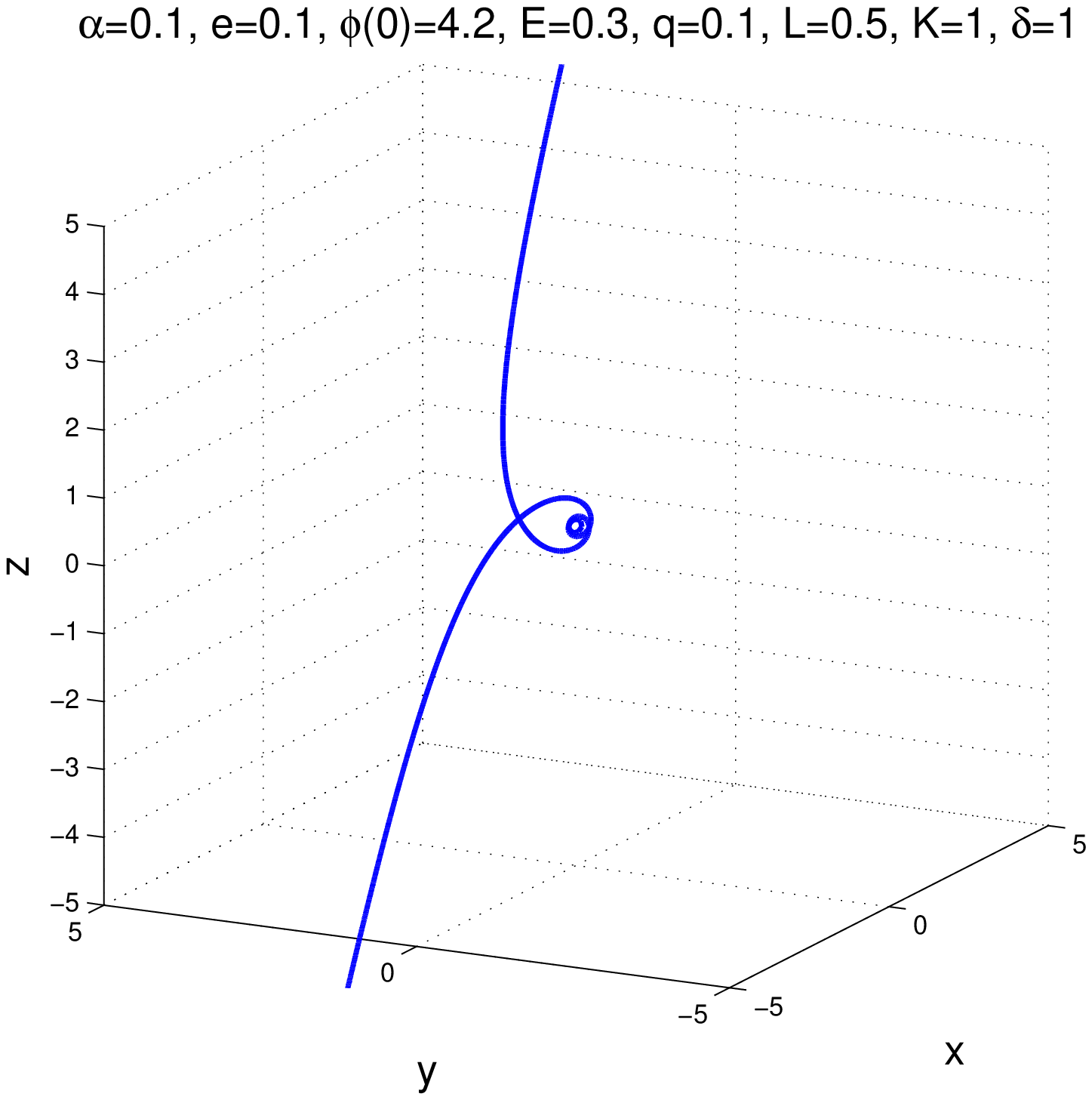}}
\end{center}
\caption{Motion of a charged, massive test particle with $q=0.1$, $L=0.5$ and $K=1.0$ in the space-time
of a boson star with $\alpha=0.1$, $e=0.1$ and $\phi(0)=4.2$.   \label{fig:orb4}}
\end{figure*}

\begin{figure*}[h!]
\begin{center}
\subfigure[][EO for $E=0.001$, $q=1$]{\label{orb42_q1}\includegraphics[width=6cm]{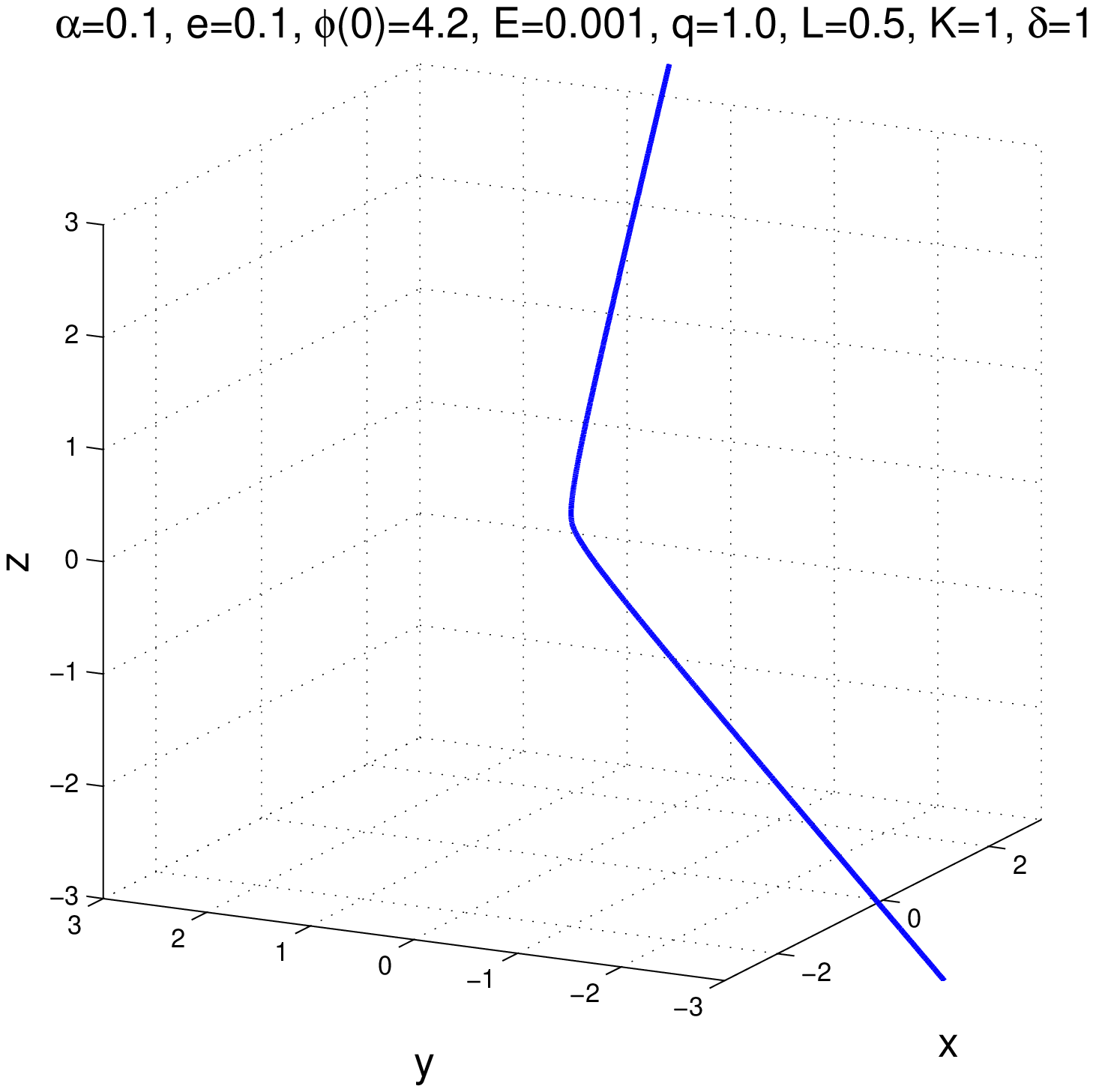}}
\subfigure[][EO for $E=0.001$, $q=3$]{\label{orb43_q3}\includegraphics[width=6cm]{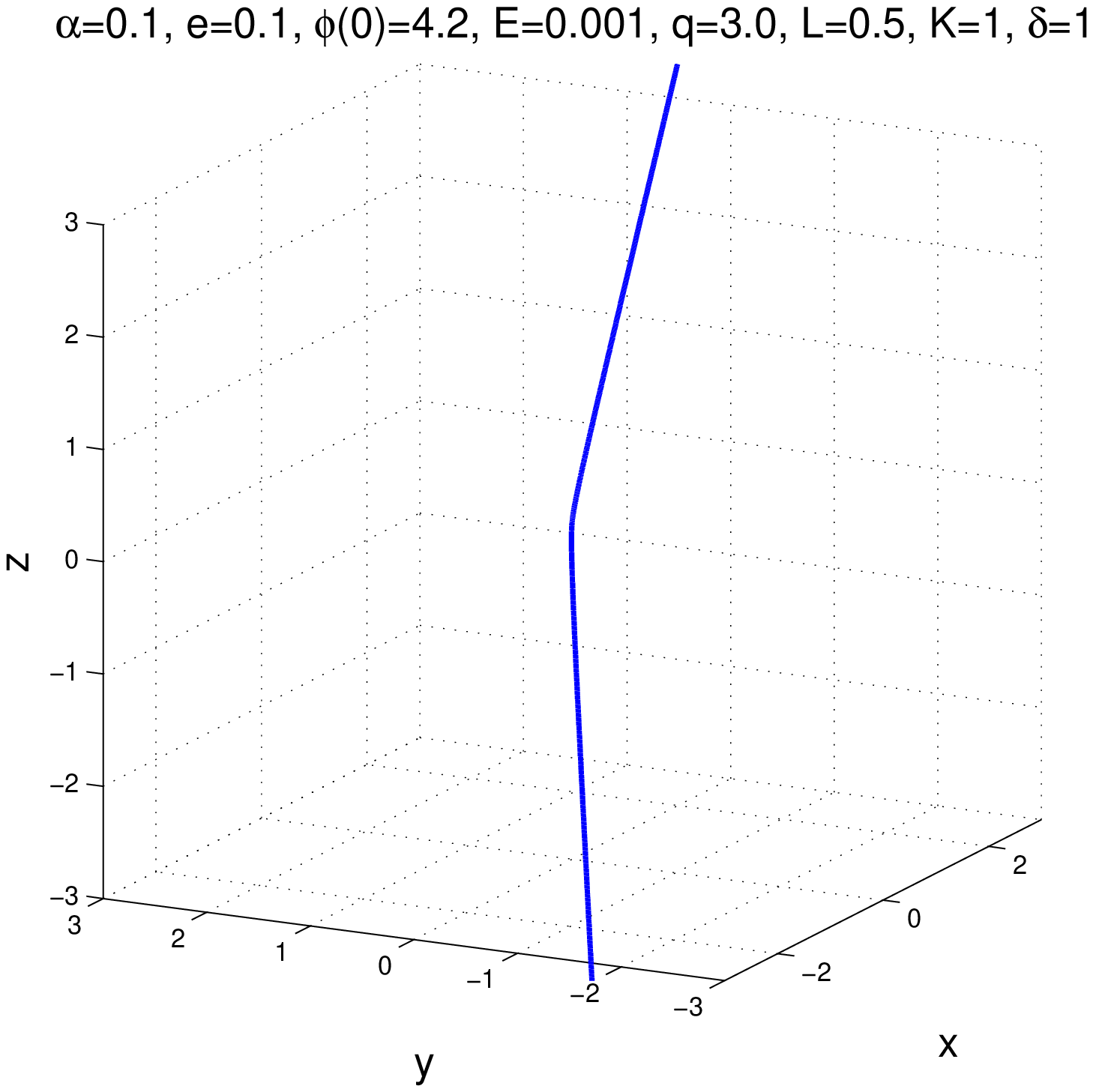}}
\end{center}
\caption{Motion of a charged, massive test particle with $L=0.5$, $K=1.0$ in the space-time
of a boson star with $\alpha=0.1$, $e=0.1$ and $\phi(0)=4.2$.   \label{fig:orb4_q}}
\end{figure*}

For large values of the scalar field $\phi(0)$ (here $\phi(0)=4.2$) 
with effective potential given in Fig.~\ref{potential_massive4} 
the particle trajectories display more structure. This is shown 
in the Fig.~\ref{fig:orb4}. Bound orbits given in Fig.~\ref{orb4} and Fig.~\ref{orb41} are different than 
the trajectories for smaller $\phi(0)$ considered before. 
Here a test particle circles around the boson star a number of times during a single revolution. 
Also the corresponding escape orbits given in Fig.~\ref{orb42} and Fig.~\ref{orb43} 
show this property. For large charges $q$ and 
positive energies of a test particle only escape orbits can be found as we 
have already argued with the help of the effective potentials in Fig.~\ref{effective_qpos}. 
We illustrate this in Fig.~\ref{orb42_q1} where an escape orbit for a test particle with $q=1$ 
and an escape orbit with $q=3$ in Fig.~\ref{orb43_q3}.

In order to understand the influence of the physical properties of the boson stars
on the dynamics of the test particles and make predictions about possible observations
we give the minimal and maximal radii of some of the orbits in Tables~\ref{rminmax} and~\ref{rminmax2}.
Comparing with Table~\ref{table_boson} we notice that there is no obvious correlation between
the properties of the orbits and e.g. the mass of the boson star. This can be seen when e.g. comparing
the minimal radius of EOs with $E=0.5$. When increasing $\phi(0)$ and with it $M$ it seems that the bigger
the mass the closer the test particle can approach the boson star due to stronger gravitational pull.
However, the boson star for $\phi(0)=4.2$ has smaller small than the other examples in the table, however,
the test particle can approach this boson star even further (we have not computed the case of $E=0.5$ here, but
the values for $E=0.4$ indicate that $r_{\rm min}$ becomes very small). Our results suggest rather that
the non-linear interplay between the scalar, gauge and gravitational field is responsible here.
Noting that there seems to be a correlation between the value of $\phi(0)$ and the value of $r_{\rm min}$ this
suggests that it is rather the character of the fields themselves and their extension of the range of $r$ that
matters here and not that much the mass and charge of the boson star which are quantities measured at infinity.
This distinguishes the space-time of the boson star from that of charged, static black holes like the
Reissner-Nordstr\"om solution. It also demonstrates that the presence of the scalar field alters the situation
considerably.

\begin{table}
\begin{center}
  \begin{tabular}{| c | c | c | c |c | c |}
\hline
$E$ & $\phi(0)$ &  $r_{\rm min}$ & $r_{\rm max}$ & $\varepsilon$ & type \\                
\hline\hline
0.008  & 0.405 & 3.65519 & 4.66470   & 0.62128 & BO \\ % 4.664698722
0.03   & 0.405 & 2.32106 & 12.09449  & 0.98141 & BO \\
0.065  & 0.405 & 1.78116 & $\infty$  & 1       & EO \\
0.5    & 0.405 & 0.68978 & $\infty$  & 1       & EO \\ \hline
0.13   & 2.25  & 0.54478 & 0.96771   & 0.82648 & BO \\
0.25   & 2.25  & 0.33427 & 5.11551   & 0.99786 & BO \\
0.33   & 2.25  & 0.28263 & $\infty$  & 1       & EO \\
0.5    & 2.25  & 0.21903 & $\infty$  & 1       & EO \\ \hline
0.13   & 2.42  & 0.54478 & 0.96771   & 0.82648 & BO \\
0.25   & 2.42  & 0.33427 & 5.11551   & 0.99786 & BO \\
0.33   & 2.42  & 0.28263 & $\infty$  & 1       & EO \\
0.5    & 2.42  & 0.21903 & $\infty$  & 1       & EO \\ \hline
0.1635 & 4.2   & 1.63308 & 1.92840   & 0.53183 & BO \\ 
0.19   & 4.2   & 0.67398 & 8.09555   & 0.99653 & BO \\
0.3    & 4.2   & 0.09344 & $\infty$  & 1       & EO \\
0.4    & 4.2   & 0.04654 & $\infty$  & 1       & EO \\ \hline
\hline
\end{tabular}
\end{center}
  \caption{The values of $r_{\rm min}$, $r_{\rm max}$ as well as of the eccentricity $\varepsilon:=\sqrt{(r_{\rm max}^2 - r_{\rm min}^2)/r_{\rm max}^2}$ of 
bound orbits with given $E$ and $L=0.5$ 
in boson star space-times with $\alpha=0.1$, $e=0.1$. The test particle has $q=0.1$ and $K=1.0$.}
\label{rminmax}
  \end{table}

\begin{table}
\begin{center}
  \begin{tabular}{| c | c|  c | c | c |c | c |}
\hline
$E$ & $q$ & $\phi(0)$ &  $r_{\rm min}$ & $r_{\rm max}$ & $\varepsilon$ & type \\                
\hline\hline
14 & -2  & 2.25  & 0.11745 & 2.41749  & 0.99882 & BO \\
28 & -4 & 2.25  & 0.05633 & 6.14704  & 0.99996 & BO \\ \hline
0.001 & 1  & 4.2  & 0.00416 & $\infty$  & 1 & EO \\
0.001 & 3 & 4.2  & 0.00138 & $\infty$  & 1 & EO \\ \hline
\hline
\end{tabular}
\end{center}
  \caption{The values of $r_{\rm min}$, $r_{\rm max}$ as well as of the eccentricity $\varepsilon:=\sqrt{(r_{\rm max}^2 - r_{\rm min}^2)/r_{\rm max}^2}$ of 
bound orbits with given $E$ and $L=0.5$ in boson star space-times with 
$\alpha=0.1$, $e=0.1$. The separation constant is $K=1.0$.}
\label{rminmax2}
  \end{table}

\clearpage

\section{Conclusions}
In this paper, we have studied the properties of charged $Q$-balls and boson stars
in a model with a self-interaction scalar potential motivated from gauge-mediated supersymmetry breaking.
We find that $Q$-balls exist only up to a maximal value of the charge of the scalar field. Furthermore,
we find that in contrast to the uncharged case a minimal, non-vanishing value of the 
number of bosonic scalar constituents forming these objects is necessary. On the other hand, there is also 
an upper limit on the maximal number of such bosonic constituents that can form them. 

We have also studied the motion of charged, massive test particles in the space-time of charged boson stars.
Interestingly, we find that the motion is fundamentally related to the presence of the scalar field.
One important point is that the motion of charged test particles in the space-time 
of charged boson stars is always planar. This is very different to the case of charged test particles
in Reissner-Nordstr\"om space-times \cite{GruKa}. This is on the one hand related to the fact that
boson star space-times do not possess any horizons and are globally regular, but on the other hand
can also be connected to the presence of the scalar field that plays an important and non-trivial r\^ole in
the test particle motion. 

The model presented here can be seen as a toy model for other compact objects such as neutron stars.
Accretion on such objects
is an important and hugely studied topic. However, numerical simulations of these  involve large hydrodynamical
simulations. Recently it has been suggested that within the ballistic limit, i.e.
in the limit in which the particles do not interact with each other, it is possible
to construct simple toy models for the formation of accretion discs. This has been done
in the Newtonian limit \cite{tejeda1} as well as in the Schwarzschild space-time \cite{tejeda2}.
The essential idea here is to let a finite number of particles with initial conditions given
on the surface of a sphere move on particle trajectories. The study done in the present paper could be used to extend
these models providing a more realistic model of accretion discs (which consist mainly of plasma) around compact
objects. 

It would also be interesting to see how test particles move in radially excited boson star space-times 
as well as in the space-time of a rotating boson star. For once this is interesting since it is believed that
the flat space-time counterparts of boson stars, so-called $Q$-balls could originally form in an excited 
state \cite{kusenko}. Furthermore, due to the interaction with surrounding matter, e.g. accretion, boson stars
might be excited. 
If boson stars should act as toy model for neutron stars it would be surely also of interest to study
the rotating counterparts since basically all observed neutron stars seem to rotate, in fact quite
quickly as the example of a radio pulsar (assumed to be a neutron star) spinning at 716 Hz shows 
\cite{Hessels:2006ze}. In our case, a rotating, electrically charged boson star would create a magnetic field.
While we have found in this paper that the electrically charged test particles move in a plane in a charged
boson star space-time, we believe that the presence of a magnetic field would make them move out of the plane.
It would then be interesting to see whether observed phenomena connected to compact objects and plasmas
such as polar jets could be described by our model.\\
\\
\\
{\bf Acknowledgments} We gratefully acknowledge the Deutsche Forschungsgemeinschaft (DFG) for financial support 
within the framework of the DFG Research Group 1620 {\it Models of Gravity}. B.H. would like to acknowledge
the Brazilian CNPq for financial support during part of this work. Y.B. would like to thank
 the Belgian FNRS for financial support.

\end{document}